# The Optimality Principle for MR signal excitation and reception: New physical insights into ideal radiofrequency coil design

Daniel K. Sodickson[1,2], Riccardo Lattanzi[1,2], Manushka Vaidya[1,2], Gang Chen[1,2], Dmitry S. Novikov[1,2], Christopher M. Collins[1,2], Graham C. Wiggins[1,2*]

[1]*Center for Advanced Imaging Innovation and Research (CAI²R) and Bernard and Irene Schwartz Center for Biomedical Imaging, Department of Radiology, New York University Langone Health, New York, NY, USA;* [2]*Sackler Institute of Graduate Biomedical Sciences, New York University School of Medicine, New York, NY, USA*

**Grant support:** National Institutes of Health: P41 EB017183; R01 EB002568; R01 EB024536

National Science Foundation: NSF CAREER 1453675

**Running Head:** The Optimality Principle

Address correspondence to:

Daniel K. Sodickson, MD, PhD

Center for Advanced Imaging Innovation and Research (CAI²R)

Department of Radiology

New York University School of Medicine

660 First Avenue, Fourth Floor, Room 407

New York, NY 10016

Phone: 212-263-4844

Fax: 212-263-4845

Email: Daniel.Sodickson@nyumc.org

*** Special affiliation information:**

**In Memoriam, Graham Charles Wiggins, 1962-2016:** On Tuesday, September 6, 2016, one of our coauthors, Dr. Graham C. Wiggins, passed away suddenly and unexpectedly. Graham's notable accomplishments included not only seminal contributions to magnetic resonance imaging, but also a celebrated musical career with the bands *Outback* and *Dr. Didg*. His groundbreaking RF coil designs, his commitment to clinical translation of basic discoveries in electromagnetics, and his rigorous and rational approach to coil performance evaluation all served as inspirations to countless colleagues in our field. With his passing, the field of magnetic resonance has lost a signature innovator, and the world has lost a unique creative spirit. Farewell, dear friend.

## Abstract

**Purpose:** Despite decades of collective experience, radiofrequency coil optimization for MR has remained a largely empirical process, with clear insight into what might constitute truly task-optimal, as opposed to merely "good," coil performance being difficult to come by. Here, a new principle, the Optimality Principle, is introduced, which allows one to predict, rapidly and intuitively, the form of optimal current patterns on any surface surrounding any arbitrary body.

**Theory:** The Optimality Principle, in its simplest form, states that the surface current pattern associated with optimal transmit field or receive sensitivity at a point of interest (per unit current integrated over the surface) is a precise scaled replica of the tangential electric field pattern that would be generated on the surface by a precessing spin placed at that point. A more general perturbative formulation enables efficient calculation of the pattern modifications required to optimize signal-to-noise ratio in body-noise-dominated situations. A full derivation of the Optimality Principle, in both unperturbed and perturbative formulations, is provided here.

**Methods and Results:** The unperturbed principle is validated numerically, and convergence of the perturbative formulation is explored in simple geometries. Predictions of both formulations are then explored in a variety of concrete case studies, spanning a range of operating frequencies and body geometries. Current patterns and corresponding field patterns derived using the Optimality Principle are used to separate signal and noise effects in coil optimization, to understand the emergence of electric dipoles as strong performers at high frequency, and to highlight the importance of surface geometry in coil design.

**Conclusion:** Like the Principle of Reciprocity from which it is derived, the Optimality Principle offers both a conceptual and a computational shortcut. In addition to providing quantitative targets for coil design, in the form of parsimoniously-computed ideal current patterns for arbitrary bodies and surfaces, the Optimality Principle affords direct physical insight into the fundamental determinants of coil performance, including, for example, current topology, surface geometry, and conservative vs non-conservative electric fields.

**Dedication:** *To Graham, who brought out the best in all of us.**

## Introduction

The interaction of electromagnetic fields with nuclear or electron spins defines the magnetic resonance phenomenon. The dynamics of MR signal generation and reception, in particular, are determined by the distribution of time-varying magnetic ($\mathcal{B}_1$) fields, whereas the corresponding time-varying electric ($\mathcal{E}$) fields tend to dictate the behavior of deleterious effects such as noise and deposited energy. Consequently, the process of shaping $\mathcal{B}_1$ and $\mathcal{E}$ fields to yield desired behaviors has evolved into a refined art. A wide range of MR transmitter and receiver designs have been proposed in the past several decades, spanning a similarly wide range of operating frequencies, element numbers, and conductor topologies. That said, despite this rich investigative activity, a much smaller subset of designs is used in routine practice, based on time-honored building blocks such as birdcage resonators (1) or arrays of loop coils (2). Meanwhile, coil optimization has remained a largely empirical process, in which prospective new designs are tested against highly variable benchmark designs, with no common standard or guarantee of theoretical optimality. The advent of accurate electrodynamic simulation tools (3,4) has greatly facilitated the testing of design hypotheses; however, the complexity of Maxwell's equations generally prevents all but the most seasoned experts from gaining clear intuition about what might constitute a truly task-optimal, as opposed to a merely "good," coil performance.

In 2011-12, we introduced a method to determine ideal current patterns for MR signal reception and transmission, as a concrete target for coil optimization, and as a source of physical intuition about the determinants of coil performance (5). An ideal current pattern constitutes the best possible arrangement of currents, on a specified surface encircling a body, that will receive (or create) MR signal from that body with the minimum possible associated noise (or deposited power). In the case of signal reception, these ideal patterns are associated with the ultimate intrinsic signal-to-noise ratio (UISNR), i.e. the maximum SNR consistent with the constraints of Maxwell's equations (6-15). Whereas UISNR may be used as an absolute reference for coil performance in practice (16,17), ideal current patterns provide a visual guide for high-performance coil design. Ideal patterns have been demonstrated for simple surface and body geometries using a dyadic Green's function (DGF) formalism to determine $\mathcal{B}_1$ and $\mathcal{E}$ fields associated with a complete basis of surface currents. The results yield insight into why birdcage or surface quadrature coils have stood the test of time (they approximate the ideal current patterns for central or superficial regions of a body, respectively). Perturbations in the ideal patterns at high field strength also highlight the importance of coil types newer to MR, such as electric dipole elements, which have increasingly significant contributions to UISNR at increasing frequency (5). Since our initial explorations of the behavior of ideal current patterns, we have also investigated means of approaching ideal patterns with practical coil designs, e.g. using combinations of current loops with electric dipoles (18-20).

One problem with ideal pattern calculation is that, until recently, it has only been practical in simple geometries, for which analytic solutions to Maxwell's equations are available. For more complicated -- and more anatomically realistic -- geometries, a prohibitive number of calls to electrodynamic solver algorithms have been required for convergence. Recent advances in the speed of electrodynamic computation, e.g. using the MARIE integral equation software suite (21,22), have facilitated the computation of UISNR for more general body models and coil surface geometries (23). Only in the past year have related methods at last been used for generalized ideal current pattern calculations in realistic body models (24). Nevertheless, such an approach still requires the careful design of numerical simulations with large numbers of time-consuming operations.

Perhaps more importantly, there is an ongoing need for enhanced physical intuition about the shapes of ideal patterns, which, even in our analytic DGF formalism, arise out of a numerical optimization process. One case in point involves the electric dipole elements mentioned earlier. Ever since favorable SNR performance was demonstrated for radiative dipole antennas at high frequency (25), centers around the world have deployed arrays of electric dipoles for ultra-high-field MR applications at 7 Tesla and above. The question of *why* such designs might be preferentially effective at high frequency remains unresolved, however, despite various speculations. Evocations of antenna theory have been brought to bear, suggesting that propagating field components associated with radiative elements such as electric dipoles might play a beneficial role at high frequencies, but such arguments do not directly address the detailed balance of electric and magnetic fields that might make any particular configuration optimal for a particular surface and body geometry. Thoughtful appeals have also been made to 'near-field' versus 'far-field' behavior, but these distinctions tend not to be entirely illuminating in what is known to be a mixed regime associated with typical radiofrequencies and corresponding wavelengths in tissue. All of these arguments, moreover, tend to be based on the behavior of individual elements with comparatively simple structure (such as straight filamentary electric dipoles), whereas ideal current patterns generally exhibit a more complex distributed structure that can only be approximated by combinations of numerous simple individual elements. It is true that DGF-derived ideal current patterns on a cylinder do show a preponderance of z-directed non-closed currents at high frequency (as opposed to the closed loop configurations they adopt at low frequency), indicating that 'electric-dipole-like' elements should be substantial contributors to high-field SNR (5). However, the simulations in themselves are silent on the question of why the balance of signal and noise might shift in this way.

The goal of this paper is to describe, and to illustrate implications of, a new principle that provides direct insight into the physical origin of ideal current patterns. We call this principle the "Optimality Principle" (OP), since it allows one to predict the form of signal-optimizing current patterns using a simple intuition. Indeed, like the principle of reciprocity from which it is derived, the Optimality Principle offers both a conceptual and a computational shortcut. Just as reciprocity enables rapid calculation of distributed receive sensitivity patterns for any particular conductor arrangement by appealing to a single reciprocal transmission process, application of OP allows one to arrive immediately at ideal conductor patterns for arbitrary surface and body geometries with a minimum of computation, once again by switching perspective to a simple transmission process.

After deriving a particularly simple formulation of the Optimality Principle which applies in coil-noise-dominated situations, we show that the effects of body-derived noise may be captured using a straightforward perturbative framework. We then provide numerical validation of the simple formulation, and explore conditions for convergence of the perturbative framework, at least in simple geometries. Both the simple formulation and the more general perturbative approach promote physical intuition about, and enable efficient computation of, optimal current patterns for arbitrary geometries, without appealing to any cumbersome basis of surface currents, and without requiring a prohibitive number of calls to an electrodynamic solver. Once we have laid out the theoretical groundwork for application of the Optimality Principle, we explore concrete changes in optimal current and field patterns as a function of magnetic field strength / operating frequency, of coil surface configuration, and of body geometry. Based on these studies, we are able to highlight certain key determinants of coil performance, including current topology (e.g., closed loops vs non-closed electric dipoles), surface geometry, and conservative vs non-conservative electric fields. For example, we show some surprising results regarding the role of closed versus non-closed current contributions in approaching ultimate performance. We demonstrate that OP enables natural separation of signal-optimizing from noise-minimizing contributions to ideal patterns, and this separation provides new insight into the functioning of traditional coil designs at both low and high frequencies. We also use OP to explain, with new rigor, the reasons for emergence of electric dipoles as strong performers at high frequency, and show that OP enables a particularly simple prediction of ideal current patterns at ultra-high field strength. For simple geometries, we compare OP-derived signal-optimizing patterns with traditionally-computed ideal patterns, defining new "dark mode" patterns whose function is to cancel noise without affecting signal. We explore the electric field patterns associated with signal-optimizing and SNR-optimizing current patterns, and highlight the role of conservative electric fields at low frequency, and of propagating electric fields at high frequency, in shaping ideal currents. We use an OP perspective to highlight the importance of current-carrying surface geometry, rather than body shape or individual coil element design per se, in determining coil performance. Finally, we derive optimal current patterns for an exemplary complex surface and body geometry. Early versions of some of this work were presented at the Annual Meetings of the International Society for Magnetic Resonance in Medicine in 2016 (26) and 2017 (27).

## Theory

From an electrodynamic point of view, the design of efficient MR coils generally involves maximizing the following quantity:

$$\frac{\mathcal{B}_1^{(\pm)}(\mathbf{r}_0)}{\left(\int_V d^3\mathbf{r}\,\sigma(\mathbf{r})\boldsymbol{\mathcal{E}}(\mathbf{r})\cdot\boldsymbol{\mathcal{E}}^*(\mathbf{r})\right)^{\frac{1}{2}}} \qquad [1]$$

Here, $\mathcal{B}_1^{(\pm)} \equiv \left(\mathcal{B}_{1,x} \pm i\mathcal{B}_{1,y}\right)/2$ represents the right- or left-circularly polarized component of the complex $\mathcal{B}_1$ field, defined according to conventions which will be laid out in detail momentarily. By careful application of the principle of reciprocity, Hoult has shown (28) that the strength of excitation (i.e., the flip angle) produced by a transmit coil at any given position of interest $\mathbf{r}_0$ is proportional to $\mathcal{B}_1^{(+)}(\mathbf{r}_0)$, whereas the strength of signal captured by a receive coil from precessing magnetization at position $\mathbf{r}_0$ is proportional to the oppositely polarized component $\mathcal{B}_1^{(-)}(\mathbf{r}_0)$ of the field which would be produced by the same coil used as a putative transmitter driven with a unit current. (A new, compact derivation of these results is provided in the next section.) The denominator in Eq. [1] integrates the complex electric field vector $\boldsymbol{\mathcal{E}}(\mathbf{r})$ associated with $\mathcal{B}_1$, dotted with its complex conjugate $\boldsymbol{\mathcal{E}}^*(\mathbf{r})$, against the conductivity $\sigma(\mathbf{r})$ of the imaged body. If the integral is performed over the entire volume $V$ of the body, this quantity represents the global Specific Absorption Rate (SAR) in the transmit case, or, by reciprocity, the noise transferred from the body to the coil in the receive case. In other words, Eq. [1] is a measure of either transmit efficiency or received intrinsic signal-to-noise ratio (SNR).

In what follows, we will focus on the receive case, using SNR as our performance metric. Close analogies with and equivalent derivations for the transmit case (e.g., to maximize transmit efficiency) will be identified in the Discussion and Appendices. Note that Eq. [1] assumes body noise dominance. Other noise sources, such as conductor and circuit noise, may be incorporated by including additional terms in the noise denominator (29,30). Likewise, cable losses and related terms may be included in the transmit case. Note also that, for typical static magnetic field strengths associated with biomedical MR, our time-varying $\mathcal{B}_1$ and $\mathcal{E}$ fields lie in the radiofrequency (RF) band, and we will sometimes use the term 'RF coils': however, except where otherwise noted, our results are generalizable to other frequency bands.

In addition to the particular form of the fields in Eq. [1], one general feature that distinguishes MR coils from antennas designed for other purposes is that MR coils must balance magnetic fields at particular points of interest against electric fields integrated over extended volumes, often including the entire imaged body. (There are some modest exceptions, such as in cases where local SAR becomes the limiting factor, but these are special cases specific to transmit coils.) We shall see later that this general feature dictates much of the behavior of ideal current patterns for MR.

Our calculations of ideal current patterns to date have relied upon a Dyadic Green's Function (DGF) framework, as defined for cylindrical geometries in a seminal paper by Schnell et al (9), and subsequently modified for spherical geometries as well (5). We shall review the fundamentals of the DGF framework in Methods to follow, but, in brief, it begins by defining a basis of surface current modes $\boldsymbol{J}_n$ whose combinations are capable of describing any arbitrary current pattern on the chosen cylindrical or spherical surface. Electric and magnetic fields associated with these modes are then determined (analytically in the simple cylindrical and spherical geometries) by integrating each mode against Green's functions appropriate to the geometry and boundary conditions. Combinations of these field modes that maximize the metric in Eq. [1] are then formed, yielding ultimate intrinsic SNR or SAR. Combinations of the associated current basis elements with the same optimal weights yield ideal current patterns.

Green's functions, of course, play a far more general role in electrodynamics. Indeed, we will next show that the symmetry of Green's functions associated with Maxwell's equations may be used to construct an intuitive proof of the principle of reciprocity as it is used in MR. Like the extensive and rigorous derivation provided by Hoult (28), our compact proof identifies signal sensitivity directly with $\mathcal{B}_1^{(-)}$, and excitation strength with $\mathcal{B}_1^{(+)}$, without appealing to the complexities of rotating and counter-rotating frames. We provide our compact proof here in order to inform the subsequent, and related, proof of the Optimality Principle.

*Compact proof of the principal of reciprocity for MR excitation and reception, based on Green's Functions*

We begin by assuming time-harmonic electromagnetic fields driven at the Larmor frequency, $\omega$. We use traditional complex notation such that all time-dependent quantities (e.g., voltages, currents or fields) in the laboratory frame are expressed as

$$\mathbf{A}(\mathbf{r},t) = \mathrm{Re}\left(\boldsymbol{\mathcal{A}}(\mathbf{r})e^{i\omega t}\right). \tag{2}$$

**Figure 1a** illustrates the geometry of an exemplary current loop (shown in yellow) on a surface outside a body, with total current $I_{\text{loop}}$ passing through any given cross-section of the loop in a local direction $\mathbf{dl}_{\text{loop}}$. A precessing spin with magnetic moment $\boldsymbol{\mathcal{M}}_{\text{spin}}$ is also shown (in blue) at a location within the body. The electromotive force (EMF), or the voltage along the path of the loop, as a result of the precessing magnetic dipole of the spin is

$$\boldsymbol{\mathcal{V}} = \oint \boldsymbol{\mathcal{E}}_{\text{spin}}(\mathbf{r}) \cdot \mathbf{dl}_{\text{loop}} \quad . \tag{3}$$

For an ideal filamentary loop of infinitesimal cross-section, with length sufficiently small to ensure vanishing phase delay around the path of the loop, we can express the current density induced in the loop in response to this EMF as

$$\boldsymbol{\mathcal{J}}_{\text{loop}}(\mathbf{r}) = I_{\text{loop}} \mathbf{dl}_{\text{loop}} \delta^3\left(\mathbf{r} - \mathbf{r}_{\text{loop}}\right) \quad . \tag{4}$$

Substituting Eq. [4] into Eq. [3], and taking advantage of the delta function to extend the path integral to a volume integral, yields

$$\boldsymbol{\mathcal{V}} = \frac{1}{I_{\text{loop}}} \int d^3\mathbf{r} \boldsymbol{\mathcal{E}}_{\text{spin}}(\mathbf{r}) \cdot \boldsymbol{\mathcal{J}}_{\text{loop}}(\mathbf{r}) \quad . \tag{5}$$

Note that multiplying both sides of Eq. [5] by $I_{\text{loop}}$ yields a power balance expression, in which the power delivered to the loop by the spin is equal to a product of voltage and current. This expression holds true when loop current is constant, i.e. when the cross-sectional current distribution is uniform, and the current $I_{\text{loop}}$ integrated over the cross-section is the same at all positions around the loop. For physical conductive structures with nonuniform current distribution across a finite cross-section, and for structures in which the cross-sectional current varies along the path (such as large self-resonant loops at high frequency, or non-closed conductor paths), the equality in Eq. [5] must be replaced by a proportionality, with the scaling factor (i.e., the effective loop current that balances the power) depending upon the details of port placement and circuit composition.

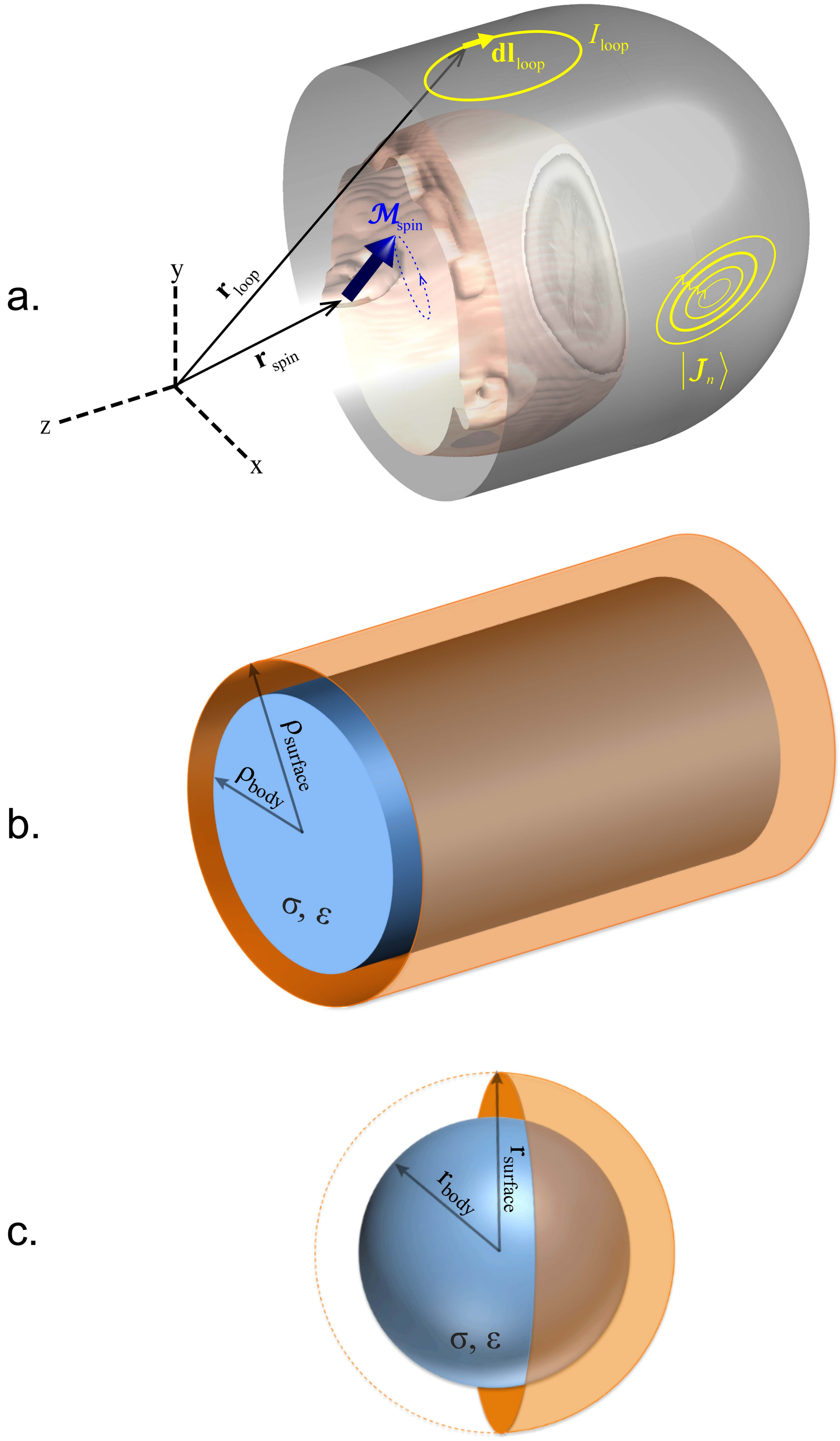

**Figure 1:** Geometry of the bodies and current-carrying surfaces used for proofs, and for numerical computations. a) Arbitrary body and surface geometry, defining terms used in proofs of the Principle of Reciprocity and the Optimality Principle. b) Cylindrical geometry used for calculations of current and field patterns. c) Spherical geometry used for similar calculations.

With a suitably scaled version of Eq. [5] in hand, we may now use fundamental symmetries underlying the principle of reciprocity to connect, and exchange, electric field and current density. Maxwell's equations relate time-harmonic electric fields to corresponding current densities in a material of interest as follows:

$$\left(\nabla\times\nabla\times-\omega^2\mu\varepsilon\right)\boldsymbol{\mathcal{E}}\equiv\bar{\mathbf{G}}^{-1}\boldsymbol{\mathcal{E}}=-i\omega\mu\boldsymbol{\mathcal{J}}\ , \qquad [6]$$

where $\mu$ is the magnetic permeability and $\varepsilon$ the electrical permittivity of the material. The differential operator $\bar{\mathbf{G}}^{-1}$ may be represented as a real symmetric matrix with elements indexed by continuous spatial coordinates. Inverting this matrix, and expressing the product $\bar{\mathbf{G}}\boldsymbol{\mathcal{J}}$ as a spatial integration, we arrive at the traditional Green's function expression for electric field:

$$\boldsymbol{\mathcal{E}}(\mathbf{r})=-i\omega\mu\int d^3\mathbf{r}'\bar{\mathbf{G}}(\mathbf{r},\mathbf{r}')\cdot\boldsymbol{\mathcal{J}}(\mathbf{r}')\,. \qquad [7]$$

We may now apply Eq. [7] to Eq. [5] to yield

$$\begin{aligned}\boldsymbol{\mathcal{V}}&=\frac{1}{I_{\text{loop}}}\int d^3\mathbf{r}\boldsymbol{\mathcal{E}}_{\text{spin}}\left(\mathbf{r}\right)\cdot\boldsymbol{\mathcal{J}}_{\text{loop}}\left(\mathbf{r}\right)\\&=\frac{-i\omega\mu}{I_{\text{loop}}}\iint d^3\mathbf{r}'d^3\mathbf{r}\bar{\mathbf{G}}(\mathbf{r},\mathbf{r}')\cdot\boldsymbol{\mathcal{J}}_{\text{spin}}\left(\mathbf{r}'\right)\cdot\boldsymbol{\mathcal{J}}_{\text{loop}}\left(\mathbf{r}\right)\\&=\frac{-i\omega\mu}{I_{\text{loop}}}\iint d^3\mathbf{r}'d^3\mathbf{r}\boldsymbol{\mathcal{J}}_{\text{spin}}\left(\mathbf{r}'\right)\cdot\bar{\mathbf{G}}(\mathbf{r}',\mathbf{r})\cdot\boldsymbol{\mathcal{J}}_{\text{loop}}\left(\mathbf{r}\right)\\&=\frac{1}{I_{\text{loop}}}\int d^3\mathbf{r}'\boldsymbol{\mathcal{J}}_{\text{spin}}\left(\mathbf{r}'\right)\cdot\boldsymbol{\mathcal{E}}_{\text{loop}}\left(\mathbf{r}'\right)\ ,\end{aligned} \qquad [8]$$

where $\boldsymbol{\mathcal{E}}_{\text{loop}}$ is the electric field which would be produced within the body by exciting the loop with current $I_{\text{loop}}$, and $\boldsymbol{\mathcal{J}}_{\text{spin}}$ is the effective current density associated with the spin's magnetic dipole moment (we will clarify this association presently). In the third line of Eq. [8], we have made use of the property that the Green's function is symmetric under exchange of the primed with the unprimed coordinate system (since $\mathbf{r}$ and $\mathbf{r}'$ may be understood as continuous indices of the symmetric operator $\bar{\mathbf{G}}$). Eq. [8] represents the classic form of the principle of reciprocity as it is generally used in MR, allowing transmitting and receiving structures to be exchanged. It holds true under the usual conditions of linear materials with time-invariant properties, and compactly-supported currents with vanishing incoming or outgoing energy at infinity.

To complete a proof of the particular form of the reciprocity principle relating MR transmit and receive sensitivities to $\mathcal{B}_1^{(\pm)}$, we identify the current density associated with the precessing dipole at any given instant as an effective infinitesimal current loop (which we will connect in a moment to the familiar dipole moment):

$$\boldsymbol{\mathcal{J}}_{\text{spin}}(\mathbf{r}) = I_{\text{spin}}\mathbf{dl}_{\text{spin}}\delta^3(\mathbf{r}-\mathbf{r}_{\text{spin}}) \ . \qquad [9]$$

We then have

$$\begin{aligned}\boldsymbol{\mathcal{V}} &= \frac{1}{I_{\text{loop}}}\int d^3\mathbf{r}'\boldsymbol{\mathcal{E}}_{\text{loop}}(\mathbf{r}')\cdot\boldsymbol{\mathcal{J}}_{\text{spin}}(\mathbf{r}') = \frac{I_{\text{spin}}}{I_{\text{loop}}}\oint\boldsymbol{\mathcal{E}}_{\text{loop}}(\mathbf{r})\cdot\mathbf{dl}_{\text{spin}} \\ &= \frac{I_{\text{spin}}}{I_{\text{loop}}}\int\left(\nabla\times\boldsymbol{\mathcal{E}}_{\text{loop}}(\mathbf{r})\right)\cdot\mathbf{da}_{\text{spin}} = \frac{-i\omega}{I_{\text{loop}}}\int\boldsymbol{\mathcal{B}}_{\text{loop}}(\mathbf{r})\cdot I_{\text{spin}}\mathbf{da}_{\text{spin}} \\ &= \frac{-i\omega}{I_{\text{loop}}}\int\boldsymbol{\mathcal{B}}_{\text{loop}}(\mathbf{r})\cdot\boldsymbol{\mathcal{M}}_{\text{spin}}(\mathbf{r}) \ .\end{aligned} \qquad [10]$$

In Eq. [10], we have used Stokes' theorem together with the Maxwell-Faraday equation to identify the curl of $\boldsymbol{\mathcal{E}}_{\text{loop}}$ with the time-derivative, $-i\omega\boldsymbol{\mathcal{B}}_{\text{loop}}$. Note that this is the form of reciprocity identified by Hoult and Richards in their original 1976 paper deriving the SNR of the nuclear magnetic resonance experiment (31).

Finally, we identify the positively precessing dipole as

$$\begin{aligned}\mathbf{M}_{\text{spin}}(\mathbf{r},t) &= M_0\delta^3(\mathbf{r}-\mathbf{r}_{\text{spin}})(\cos\omega t\hat{\mathbf{x}}+\sin\omega t\hat{\mathbf{y}}) \\ &= \text{Re}\left(M_0\delta^3(\mathbf{r}-\mathbf{r}_{\text{spin}})(\hat{\mathbf{x}}-i\hat{\mathbf{y}})e^{i\omega t}\right) \\ &\equiv \text{Re}\left(\boldsymbol{\mathcal{M}}_{\text{spin}}(\mathbf{r})e^{i\omega t}\right) \ .\end{aligned} \qquad [11]$$

(As a minor note, this convention of right-handed spin precession, which was chosen for historical reasons to be consistent with conventions used elsewhere in treatments of RF magnetic field components, would technically require $\mathbf{B}_0$ to be oriented in the negative $z$ direction, since nuclear precession occurs in a left-handed sense about $\mathbf{B}_0$ (32,33).) Inserting the definition of $\boldsymbol{\mathcal{M}}_{\text{spin}}(\mathbf{r})$ from Eq. [11] into Eq. [10] yields the desired result:

$$\begin{aligned}\boldsymbol{\mathcal{V}} &= \frac{-i\omega}{I_{\text{loop}}}\int \boldsymbol{\mathcal{B}}_{\text{loop}}(\mathbf{r})\cdot\boldsymbol{\mathcal{M}}_{\text{spin}}(\mathbf{r}) = \frac{-i\omega}{I_{\text{loop}}}\int \mathrm{d}^3\mathbf{r}\,\boldsymbol{\mathcal{B}}_{\text{loop}}(\mathbf{r})\cdot M_0\delta^3\left(\mathbf{r}-\mathbf{r}_{\text{spin}}\right)\left(\hat{\mathbf{x}}-i\hat{\mathbf{y}}\right)\\ &= \frac{-i\omega M_0}{I_{\text{loop}}}\boldsymbol{\mathcal{B}}_{\text{loop}}\left(\mathbf{r}_{\text{spin}}\right)\cdot\left(\hat{\mathbf{x}}-i\hat{\mathbf{y}}\right) = \frac{-i\omega M_0}{I_{\text{loop}}}\left(\mathcal{B}_{\text{loop,x}}\left(\mathbf{r}_{\text{spin}}\right)-i\mathcal{B}_{\text{loop,y}}\left(\mathbf{r}_{\text{spin}}\right)\right)\\ &= \frac{-2i\omega M_0}{I_{\text{loop}}}\mathcal{B}_{1,\text{loop}}^{(-)}\left(\mathbf{r}_{\text{spin}}\right)\ .\end{aligned} \qquad [12]$$

Note the scaling by $I_{\text{loop}}$ in this familiar result, indicating that the equivalent transmitter in the reciprocity framework must indeed be excited with unit cross-sectional current. Note also that, as for Eq. [5], the equality in Eq. [12] must be replaced by a proportionality for physical structures with nonuniform current distributions.

Meanwhile, by expressing the RF magnetic field of the loop as

$$\begin{aligned}\boldsymbol{\mathcal{B}}_{1,\text{loop}}(\mathbf{r}) &= \mathcal{B}_{1x,\text{loop}}(\mathbf{r})\hat{\mathbf{x}}+\mathcal{B}_{1y,\text{loop}}(\mathbf{r})\hat{\mathbf{y}}+\mathcal{B}_{1z,\text{loop}}(\mathbf{r})\hat{\mathbf{z}}\\ &= \mathcal{B}_{1,\text{loop}}^{(+)}(\mathbf{r})\left(\hat{\mathbf{x}}-i\hat{\mathbf{y}}\right)+\mathcal{B}_{1,\text{loop}}^{(-)}(\mathbf{r})\left(\hat{\mathbf{x}}+i\hat{\mathbf{y}}\right)+\mathcal{B}_{1z,\text{loop}}(\mathbf{r})\hat{\mathbf{z}}\ ,\end{aligned} \qquad [13]$$

we can immediately recognize that the component of the transmit field co-rotating with the dipole (see Eq. [11]) is the $\left(\hat{\mathbf{x}}-i\hat{\mathbf{y}}\right)$ component, whose coefficient is $\mathcal{B}_{1,\text{loop}}^{(+)}(\mathbf{r})$. (Here and in Eq. [12], as for Eq. [1], we have used the definition $\mathcal{B}_{1,\text{loop}}^{(\pm)} = \left(\mathcal{B}_{1x,\text{loop}} \pm i\mathcal{B}_{1y,\text{loop}}\right)/2$.) Thus, the transmit sensitivity is proportional to $\mathcal{B}_{1,\text{loop}}^{(+)}\left(\mathbf{r}_{\text{spin}}\right)$, while the receive sensitivity is proportional to $\mathcal{B}_{1,\text{loop}}^{(-)}\left(\mathbf{r}_{\text{spin}}\right)$, as shown by Eq. [12].

No reference to any particular rotating or counter-rotating frame is required in this simple derivation. Instead, the distinct expressions for transmit and receive sensitivities arise from the fact that the transmit field component is the coefficient of the vector $\left(\hat{\mathbf{x}}-i\hat{\mathbf{y}}\right)$, whereas the received voltage is, by reciprocity, proportional to the dot product of the field with that same vector. (Starting with Eq. [3] above, we have shown that this dot product derives directly from the inductive character of signal detection: in signal reception, we are measuring the work done by the field of the precessing spin, whereas what matters for signal excitation is the field experienced by the spin itself (34).) If we take the original reciprocity expression of Hoult and Richards as a starting point (Eq. [10] or Ref. (31)), then all the key results may be derived in a total of three equations: Eq.'s [11], [12], and [13].

*The Optimality Principle: Derivation, and A Simple Form for Signal-Optimizing Current Patterns*

So far, we have used the principle of reciprocity to characterize the sensitivities of particular coils or current distributions. Let us now return to the question of what constitutes an "ideal" current distribution for MR signal reception. Here we may be aided by the qualitative intuition that, just as an effective transmit coil creates a $B_1$ field that rotates with excited spins, an effective receive coil should also be associated with a $B_1$ field that "tracks" the spins as they precess. Certainly, the example of the birdcage resonator bears out this intuition, as does the case of any quadrature coil. One might, therefore, suppose that the distribution of current induced by a precessing spin on a surface surrounding that spin might, if it were played out as a transmit current, constitute an effective coil on that surface. The $B_1$ field created by such a current will not reproduce the singularity of a dipole field, but it will rotate with the dipole. It will also generally be strong where the dipole field is strong, suggesting a matched filter character for signal reception.

Here, we show that **the surface current pattern associated with optimal $\mathcal{B}_1^{(-)}$ at a point of interest (per unit current integrated over the surface) is in fact a precise scaled replica of the tangential electric field pattern generated on the surface by a precessing spin placed at that point**. This is the simplest form of the Optimality Principle, and it applies for any arbitrary surface surrounding any arbitrary body.

Our proof is based directly upon the reciprocity principle. Once again, **Figure 1a** illustrates the geometry for the derivation, showing a current-bearing surface surrounding a spin within a dielectric body to be imaged.

For convenience, in what follows we will use a bra-ket formulation familiar from quantum mechanics, in order to prevent a proliferation of integrals. We define spatial integration as an inner product of bra and ket vectors:

$$\int d^3\mathbf{r}\boldsymbol{J}^*(\mathbf{r})\cdot\boldsymbol{\mathcal{E}}(\mathbf{r}) \equiv \left\langle \boldsymbol{J}(\mathbf{r}) \middle| \boldsymbol{\mathcal{E}}(\mathbf{r}) \right\rangle \ . \qquad [14]$$

(Note the complex conjugation of $\boldsymbol{J}$ on the left hand side of Eq. [14], which is required for a true inner product of complex quantities. The bra vector $\left\langle \boldsymbol{J}(\mathbf{r}) \right|$ is, thus, a conjugate transpose of the ket vector $\left| \boldsymbol{J}(\mathbf{r}) \right\rangle$.) The effect of a Green's function operator in the fundamental Green's function equation (Eq. [7]

above) is represented in bra-ket notation as follows, where integration occurs over the repeated spatial coordinate:

$$-i\omega\mu\int d^3\mathbf{r}'\overline{\mathbf{G}}(\mathbf{r},\mathbf{r}')\cdot\boldsymbol{\mathcal{J}}(\mathbf{r}') \equiv -i\omega\mu\overline{\mathbf{G}}(\mathbf{r},\mathbf{r}')\left|\boldsymbol{\mathcal{J}}(\mathbf{r}')\right\rangle = \left|\boldsymbol{\mathcal{E}}(\mathbf{r})\right\rangle \ . \tag{15}$$

Now let us assume that we have a complete basis of surface current patterns,

$$\left|\boldsymbol{\mathcal{J}}_n(\mathbf{r})\right\rangle = \left|\boldsymbol{\mathcal{K}}_n(\mathbf{r}_{\text{surface}})\right\rangle \delta^3(\mathbf{r}-\mathbf{r}_{\text{surface}}) \ , \tag{16}$$

where *n* is an integer index and $\left|\boldsymbol{\mathcal{K}}_n(\mathbf{r}_{\text{surface}})\right\rangle$ represents a current pattern existing only on the surface. For simple geometries, these basis functions may take the form of vector spherical or cylindrical harmonics, as in (5). For any surface, all possible surface current distributions may be represented by suitable linear combinations of basis elements.

If the $\left|\boldsymbol{\mathcal{J}}_n(\mathbf{r})\right\rangle$ do indeed form a complete basis on the surface, then projection onto the surface may be accomplished by the operator

$$\mathbf{P} = \sum_n \frac{\left|\boldsymbol{\mathcal{J}}_n(\mathbf{r})\right\rangle\left\langle\boldsymbol{\mathcal{J}}_n(\mathbf{r})\right|}{\left\langle\boldsymbol{\mathcal{J}}_n(\mathbf{r})\middle|\boldsymbol{\mathcal{J}}_n(\mathbf{r})\right\rangle} \tag{17}$$

Projection of the electric field produced by a precessing spin and its point magnetic dipole onto the surface yields the following expression for tangential dipole field (with a complex conjugation whose role will become clear presently):

$$\left|\boldsymbol{\mathcal{E}}^*_{\text{spin, tangential}}(\mathbf{r})\right\rangle = \mathbf{P}\left|\boldsymbol{\mathcal{E}}^*_{\text{spin}}(\mathbf{r})\right\rangle = \sum_n \frac{\left|\boldsymbol{\mathcal{J}}_n(\mathbf{r})\right\rangle\left\langle\boldsymbol{\mathcal{J}}_n(\mathbf{r})\middle|\boldsymbol{\mathcal{E}}^*_{\text{spin}}(\mathbf{r})\right\rangle}{\left\langle\boldsymbol{\mathcal{J}}_n(\mathbf{r})\middle|\boldsymbol{\mathcal{J}}_n(\mathbf{r})\right\rangle} \ . \tag{18}$$

Following a similar line of argument as for Eq.'s [10] through [12] for the reciprocity derivation above, we may write

$$\begin{aligned}\left\langle\boldsymbol{\mathcal{J}}_n(\mathbf{r})\middle|\boldsymbol{\mathcal{E}}^*_{\text{spin}}(\mathbf{r})\right\rangle &= \left\langle\boldsymbol{\mathcal{E}}_n(\mathbf{r})\middle|\boldsymbol{\mathcal{J}}^*_{\text{spin}}(\mathbf{r})\right\rangle = i\omega\left\langle\boldsymbol{\mathcal{B}}_n(\mathbf{r})\middle|\boldsymbol{\mathcal{M}}^*_{\text{spin}}(\mathbf{r})\right\rangle \\ &= i\omega M_0\left(\boldsymbol{\mathcal{B}}^*_n(\mathbf{r}_{\text{spin}})\cdot(\hat{\mathbf{x}}+i\hat{\mathbf{y}})\right) = 2i\omega M_0\left(\mathcal{B}^{(-)}_{1,n}(\mathbf{r}_{\text{spin}})\right)^* \ ,\end{aligned} \tag{19}$$

so that

$$\left|\boldsymbol{\mathcal{E}}^*_{\text{spin, tangential}}(\mathbf{r})\right\rangle = 2i\omega M_0 \sum_n \frac{\left(\mathcal{B}^{(-)}_{1,n}(\mathbf{r}_{\text{spin}})\right)^*}{\left\langle\boldsymbol{\mathcal{J}}_n(\mathbf{r})\middle|\boldsymbol{\mathcal{J}}_n(\mathbf{r})\right\rangle}\left|\boldsymbol{\mathcal{J}}_n(\mathbf{r})\right\rangle \tag{20}$$

The ideal current pattern (i.e., the current pattern resulting in optimal SNR) may be written in terms of the same surface current basis functions, as follows:

$$\begin{aligned}\left|\boldsymbol{J}^*_{\text{ideal}}(\mathbf{r})\right\rangle &= \sum_n W_{n,\text{ideal}}\left|\boldsymbol{J}_n(\mathbf{r})\right\rangle = \sum_n \left(\mathbf{S}^H\boldsymbol{\Psi}^{-1}\mathbf{S}\right)^{-1}\left(\mathbf{S}^H\boldsymbol{\Psi}^{-1}\right)_n\left|\boldsymbol{J}_n(\mathbf{r})\right\rangle \\ &= \left(\mathbf{S}^H\boldsymbol{\Psi}^{-1}\mathbf{S}\right)^{-1}\sum_n\sum_{n'} S^*_{n'}\Psi^{-1}_{n'n}\left|\boldsymbol{J}_n(\mathbf{r})\right\rangle\end{aligned} \qquad [21]$$

(Once again, we will discuss the role of the complex conjugation in the ideal current pattern momentarily.) The ideal weights $W_{n,\text{ideal}}$ in Eq. [21] are determined by treating each surface current element as an effective coil in a large coil array, and generating an SNR-optimizing combination according to Roemer (2) and/or Pruessmann (35), with $S_n$ representing the receive sensitivity of the $n^{\text{th}}$ element at the position of interest, and $\Psi_{n'n}$ representing the noise covariance between elements *n'* and *n*. In the absence of parallel imaging acceleration, $S_n$ is simply a scalar quantity, as is the product $\left(\mathbf{S}^H\boldsymbol{\Psi}^{-1}\mathbf{S}\right)$. We already know from reciprocity that the receive sensitivity takes the form

$$S_n \propto \mathcal{B}^{(-)}_{1,\text{n}}\left(\mathbf{r}_{\text{spin}}\right) \ . \qquad [22]$$

With this substitution, and making use of Eq. [19], we may write

$$\left|\boldsymbol{J}^*_{\text{ideal}}(\mathbf{r})\right\rangle \propto \sum_n\sum_{n'}\left|\boldsymbol{J}_n(\mathbf{r})\right\rangle \Psi^{-1*}_{nn'}\left\langle \boldsymbol{J}_{n'}(\mathbf{r})\middle|\boldsymbol{\mathcal{E}}^*_{\text{spin}}(\mathbf{r})\right\rangle \ , \qquad [23]$$

where we have taken advantage of the fact that the noise covariance matrix is Hermitian: $\Psi_{n'n} = \Psi^*_{nn'}$ . **This expression relating the ideal current pattern to the electric field of a precessing spin is the general form of the Optimality Principle.**

We may immediately recognize that

$$\left|\boldsymbol{J}^*_{\text{ideal}}(\mathbf{r})\right\rangle \propto \left|\boldsymbol{\mathcal{E}}^*_{\text{spin, tangential}}(\mathbf{r})\right\rangle \ \text{ if } \ \Psi_{n'n} \propto \left\langle \boldsymbol{J}_n(\mathbf{r})\middle|\boldsymbol{J}_n(\mathbf{r})\right\rangle\delta_{n'n} \qquad [24]$$

In other words, our earlier intuition is correct, and the SNR-optimizing ideal current pattern $\left|\boldsymbol{J}_{\text{ideal}}(\mathbf{r})\right\rangle$ is an identical replica of the tangential dipole electric field pattern $\left|\boldsymbol{\mathcal{E}}_{\text{spin, tangential}}(\mathbf{r})\right\rangle$ (up to an overall scaling factor), if the noise correlation condition in Eq. [24] is satisfied.

What is the physical meaning of this condition on noise? First, it specifies that noise is not correlated between distinct basis elements. Second, it requires that the relative amplitude of the noise in each mode is determined only by the integrated amplitude $\left\langle \boldsymbol{J}_n(\mathbf{r}) \middle| \boldsymbol{J}_n(\mathbf{r}) \right\rangle$ of current density in that mode. (Taken together, these requirements yield a noise covariance matrix which, when normalized by the integrated current density in each mode, is just a constant multiple of the identity matrix.) Both of these requirements are generally met in the regime of conductor noise dominance, for which noise is not shared among distinct conductors, and noise power is proportional to the quantity of current traveling along the length of those conductors. For conductive material with fixed resistivity $\rho$, Johnson noise power is in fact proportional to $\rho\left\langle \boldsymbol{J}_n(\mathbf{r}) \middle| \boldsymbol{J}_n(\mathbf{r}) \right\rangle$, which increases with increasing conductor length (via the spatial integration implicit in the bra-ket product) as well as with cross-sectional current amplitude. In other words, the simple form of the Optimality Principle in Eq. [24] (which we shall call “simple OP” or “unperturbed OP”) identifies the most parsimonious way to lay conductor on a surface in order to achieve any given receive sensitivity at an interior point of interest. We can therefore identify simple OP patterns as “signal-optimizing” patterns, in the sense that they optimize signal per unit conductor length, or, for straightforward Ohmic conductor loss mechanisms, signal per unit coil noise. These signal-optimizing patterns do not yet take into account body-derived noise, which may both shift diagonal elements and introduce off-diagonal noise correlations into the noise covariance matrix. The perturbing effects of body noise will be addressed in the next section.

Let us return now to the complex conjugation of the patterns in Eq.’s [23] and [24]. The optimal weights $W_{n,\text{ideal}}$ are designed to create a unit net sensitivity, proportional to $\mathcal{B}_1^{(-)}$, at the point of interest (while simultaneously minimizing noise). If the optimally-weighted currents were to be played out in a transmit scenario, they would therefore create a $\mathcal{B}_1$ field with a dominant component precessing in the opposite sense from the spins. Thus, in order to represent receive current patterns that capture spin precession, we must reverse the precession sense by taking a complex conjugate. This conceptual observation has also been validated numerically (see Results to follow). Note that conjugating the receive pattern is not the same as optimizing for transmission, since $\mathcal{B}_1^{(+)} \neq \left(\mathcal{B}_1^{(-)}\right)^*$.

Eq. [24] represents a very simple and general result. It indicates that one may define signal-optimizing current patterns for any arbitrary body and surface of interest, with a single call to an electrodynamic solver, whether based on finite difference time domain (FDTD), finite element (FEM), or other methods. One need only place a precessing magnetic dipole at the point of interest and calculate its electric field on the surface of interest. Despite the fact that a surface current basis was used in intermediate steps of the preceding proof, the result is basis-independent. There is no need to form any particular basis of surface currents, or to compute any explicit fields associated with such a basis.

As is the case for the principle of reciprocity, the Optimality Principle involves a shift in perspective between reception and transmission. Reciprocity asserts a relationship between the sensitivity of a receive coil to a precessing spin and the sensitivity of that spin to unit transmission in the coil. OP relates the current pattern which produces the best signal reception from a spin to the electric field pattern created by that precessing spin. In other words, the effective transmitter in OP is the spin, rather than the coil. The full benefits of this shift in perspective, including both computational efficiency and physical intuition, will be explored presently.

We next generalize to cases involving body noise, which is the dominant noise source in many *in vivo* human MR settings.

*Ideal Current Patterns in the Presence of Body Noise as Perturbations of Signal-Optimizing Patterns*

For the case of significant body noise contributions, we return to the full expression in Eq. [23]. In this case, the noise covariance matrix (or, to be precise, its inverse) rescales and mixes contributions from various current basis elements, perturbing the result away from a strict surface projection of the dipole electric field. Let us therefore approach the problem from the perspective of perturbation theory.

For convenience of notation, we first renormalize our surface current basis functions such that $\left\langle \boldsymbol{J}_n(\mathbf{r}) \middle| \boldsymbol{J}_n(\mathbf{r}) \right\rangle = 1$. This scaling does not affect generality, and simply converts the conductor-noise-dominated covariance matrix in Eq. [24] to the identity matrix $\mathbf{1}$. If we then perform a Taylor series expansion of the full noise covariance matrix inverse around the identity matrix, we find that

$$\boldsymbol{\Psi}^{-1*} = \left(\mathbf{1} - \left(\mathbf{1} - \boldsymbol{\Psi}^*\right)\right)^{-1} \equiv \left(\mathbf{1} - \mathbf{V}\right)^{-1} = \mathbf{1} + \mathbf{V} + \mathbf{V}^2 + \mathbf{V}^3 + \ldots \qquad [25]$$

This series converges as long as the (suitably scaled) eigenvalues of $\mathbf{V} \equiv \mathbf{1} - \boldsymbol{\Psi}^*$ have absolute magnitude less than one. We may immediately recognize the simple Optimality Principle in Eq. [24] as the result of inserting into Eq. [23] only the zeroth-order unity term in this expansion. In order to compute the first-order term involving $\mathbf{V}$, we identify the body-noise-dominated noise covariance matrix as

$$\Psi^*_{nn'} = \int d^3\mathbf{r}\,\sigma(\mathbf{r})\,\boldsymbol{\mathcal{E}}^*_n(\mathbf{r})\,\boldsymbol{\mathcal{E}}_{n'}(\mathbf{r}) = \left\langle \boldsymbol{\mathcal{E}}_n(\mathbf{r}) \middle| \bar{\boldsymbol{\sigma}}(\mathbf{r},\mathbf{r}') \middle| \boldsymbol{\mathcal{E}}_{n'}(\mathbf{r}') \right\rangle , \qquad [26]$$

where we have defined the conductivity operator

$$\bar{\boldsymbol{\sigma}}(\mathbf{r},\mathbf{r}') = \sigma(\mathbf{r})\delta^3(\mathbf{r}-\mathbf{r}') \quad [27]$$

If we now insert the perturbation series expansion of Eq. [25] into the Optimality Principle master equation, Eq. [23], the first-order term becomes

$$\begin{aligned}
&\sum_n \sum_{n'} \left| \boldsymbol{J}_n(\mathbf{r}) \right\rangle V_{nn'} \left\langle \boldsymbol{J}_{n'}(\mathbf{r}) \middle| \boldsymbol{\mathcal{E}}^*_{\text{spin}}(\mathbf{r}) \right\rangle \\
&= \sum_n \sum_{n'} \left| \boldsymbol{J}_n(\mathbf{r}) \right\rangle \left( \delta_{nn'} - \Psi^*_{nn'} \right) \left\langle \boldsymbol{J}_{n'}(\mathbf{r}) \middle| \boldsymbol{\mathcal{E}}^*_{\text{spin}}(\mathbf{r}) \right\rangle \\
&= \sum_n \sum_{n'} \left| \boldsymbol{J}_n(\mathbf{r}) \right\rangle \left( \delta_{nn'} - \left\langle \boldsymbol{\mathcal{E}}_n(\mathbf{r}) \middle| \bar{\boldsymbol{\sigma}}(\mathbf{r},\mathbf{r}') \middle| \boldsymbol{\mathcal{E}}_{n'}(\mathbf{r}') \right\rangle \right) \left\langle \boldsymbol{J}_{n'}(\mathbf{r}) \middle| \boldsymbol{\mathcal{E}}^*_{\text{spin}}(\mathbf{r}) \right\rangle \\
&= \sum_n \sum_{n'} \left| \boldsymbol{J}_n(\mathbf{r}) \right\rangle \left( \delta_{nn'} - \left\langle \boldsymbol{J}_n(\mathbf{r}) \middle| \left( i\omega\mu \bar{\mathbf{G}}(\mathbf{r},\mathbf{r}') \right) \bar{\boldsymbol{\sigma}}(\mathbf{r},\mathbf{r}') \left( -i\omega\mu \bar{\mathbf{G}}(\mathbf{r}',\mathbf{r}) \right) \middle| \boldsymbol{J}_{n'}(\mathbf{r}) \right\rangle \right) \left\langle \boldsymbol{J}_{n'}(\mathbf{r}) \middle| \boldsymbol{\mathcal{E}}^*_{\text{spin}}(\mathbf{r}) \right\rangle \\
&= \left( \mathbf{P} - \mathbf{P}\tilde{\mathbf{G}}^* \bar{\boldsymbol{\sigma}} \tilde{\mathbf{G}} \mathbf{P} \right) \left| \boldsymbol{\mathcal{E}}^*_{\text{spin}}(\mathbf{r}) \right\rangle
\end{aligned}$$

[28]

Here, $\mathbf{P}$ is just the projection operator for the renormalized surface current basis, and $\tilde{\mathbf{G}} \equiv -i\omega\mu\bar{\mathbf{G}}$ is the operator that converts current to electric field. The product $\mathbf{P}\tilde{\mathbf{G}}^*\bar{\boldsymbol{\sigma}}\tilde{\mathbf{G}}\mathbf{P}$ in Eq. [28], when applied to an electric field vector, corresponds to the following set of operations:

1. Project the electric field onto the surface of interest (just as the dipole electric field $\left| \boldsymbol{\mathcal{E}}^*_{\text{spin}}(\mathbf{r}) \right\rangle$ is initially projected onto the surface to yield the simple OP current pattern).
2. Treat the resulting tangential field as an effective current pattern, and compute (via an electrodynamic solver of interest) the electric field throughout the volume that results from that applied current pattern.
3. Multiply the resulting volumetric electric field by the conductivity distribution, yielding a conduction current density in the body.
4. Compute (with a solver of interest) the conjugate of the electric field resulting from the distribution of conduction currents in the body.
5. Project that electric field back onto the surface.

It may easily be shown that inserting higher powers of $\mathbf{V}$ simply results in repeat applications of the above operations, i.e.

$$\sum_n \sum_{n'} \left| \boldsymbol{J}_n(\mathbf{r}) \right\rangle V^m_{nn'} \left\langle \boldsymbol{J}^*_{n'}(\mathbf{r}) \right| = \left( \mathbf{P} - \mathbf{P}\tilde{\mathbf{G}}^* \bar{\boldsymbol{\sigma}} \tilde{\mathbf{G}} \mathbf{P} \right)^m = \left( \mathbf{P}\left( \mathbf{1} - \tilde{\mathbf{G}}^* \bar{\boldsymbol{\sigma}} \tilde{\mathbf{G}} \right) \right)^m \mathbf{P} \quad [29]$$

(In Eq. [29], we have used the fact that projection operators are idempotent, i.e., $\mathbf{P}^m = \mathbf{P}$.) Inserting this into Eq. [23], and noting that both surface projection and field calculation may be accomplished without reference to any particular basis of current vectors, we have **a simple basis-independent expression for the ideal current pattern**:

$$\left|\boldsymbol{\mathcal{J}}^*_{\text{ideal}}(\mathbf{r})\right\rangle \propto \left\{\sum_{m=0}^{\infty}\left[\mathbf{P}\left(\mathbf{1}-\tilde{\mathbf{G}}^*\overline{\boldsymbol{\sigma}}\tilde{\mathbf{G}}\right)\right]^m\right\}\mathbf{P}\left|\boldsymbol{\mathcal{E}}^*_{\text{spin}}(\mathbf{r})\right\rangle \qquad [30]$$

*Applying the Optimality Principle*

According to the generalized form of the Optimality Principle in Eq. [30], in order to compute ideal current patterns on an arbitrary surface for an arbitrary body, we need only make an iterative series of calls to a traditional electrodynamic solver, adding terms for each order of iteration together as specified in the equation. There is no need to compute fields for a large basis of surface currents, which would vary in form depending upon the geometry of the surface. The only modification required to standard electrodynamic solver platforms is the capability to compute external fields resulting from a defined set of internal conduction currents – a capability that is straightforward to implement, but that is not generally exploited in typical RF coil simulations.

Note that we may also recognize the problem in Eq. [30] as a scattering problem, identifying the series expansion of the noise covariance matrix inverse as a Born series (36). We may therefore bring to bear a wide range of known techniques for robust solution, estimation of convergence, etc. (see Appendix A).

One key benefit of the Optimality Principle is evident even from its unperturbed form in Eq. [24]. It is comparatively easy to form mental projections of precessing dipole fields onto a surface of interest, and thereby to predict the general form of ideal current patterns from basic physical principles. Physical intuition about what constitutes optimality can serve as a helpful guide to coil designers, particularly in frequency regimes typical of high-field MR applications, for which quasistatic near-field approximations break down, but far-field approximations do not yet apply robustly. Moreover, as will be elaborated in detail in the Results section to follow, the ability to compare intuitive unperturbed signal-optimizing current patterns with full SNR-optimizing patterns provides additional insight, and enables separation of signal-maximizing from noise-minimizing effects. These effects are otherwise inherently difficult to separate, and can easily be conflated or confused in common explanations of coil performance.

## Methods

*Dyadic Green's Function calculations for simple geometries*

Direct computations of ideal current patterns, either neglecting or including body noise contributions, are possible in simple geometries using a Dyadic Green's Function (DGF) formalism (5). Therefore, DGF calculations were used to validate the Optimality Principle, and to explore some of its predictions. Figure 1 illustrates the cylindrical (**Fig 1b**) and spherical (**Fig 1c**) geometries used for these calculations. In all cases, a uniform dielectric cylindrical or spherical body is surrounded by a concentric cylindrical or spherical surface positioned at some small distance from the surface of the body. Simulations were performed for cases spanning a range of parameters, including magnetic field strength and corresponding operating frequency, object size, and position of interest for SNR optimization. Unless otherwise specified, simulations for cylindrical geometries used the following default parameters (chosen to approximate an average human torso): conductivity $\sigma$ = 0.4 S/m, relative permittivity $\varepsilon_{rel}$ = 39, cylinder length 125cm, cylinder radius $\rho_{body}$ = 20cm, and cylindrical current-carrying surface radius $\rho_{surface}$ = 21cm. For spherical geometries, default parameters were as follows: $\sigma$ = 0.4 S/m, $\varepsilon_{rel}$ = 39, dielectric sphere radius $\rho_{sphere}$ = 20cm, and spherical current-carrying surface radius $\rho_{surface}$ = 21cm.

In brief, the DGF framework operates by positing a complete basis of surface currents, then using Green's functions subject to appropriate boundary conditions to compute analytic expressions for the electromagnetic fields associated with each surface current element. Such simple analytic expressions only exist for simple object and surface geometries. For more general cases, a DGF approach would require calling a full electrodynamic solver for each current basis element – a computationally expensive proposition for element numbers typically running into the tens of thousands. In any case, once the EM fields are known, one can compute, again analytically, the contribution of each mode *n* to MR signal at the point of interest (i.e., $\mathcal{B}_{1,n}^{(-)}(\mathbf{r}_0)$), and also to noise in the conductive body (i.e., $\Psi_{nn'} = \int d^3\mathbf{r}\sigma(\mathbf{r})\boldsymbol{\mathcal{E}}_n(\mathbf{r})\boldsymbol{\mathcal{E}}_{n'}^*(\mathbf{r})$). An SNR-optimizing matched filter combination of mode contributions, as indicated in Eq. [21], results in the ultimate intrinsic SNR. A similar combination in which the noise covariance matrix is set to the identity matrix results in ultimate intrinsic signal-to-coil-noise-ratio, the "signal-optimizing" limit neglecting body noise. The same combination weights applied to current basis elements yields the ideal current pattern, $\left|\boldsymbol{J}_{\text{ideal}}^*(\mathbf{r})\right\rangle$ (or, with unit noise covariance matrix, the "signal-optimizing" unperturbed OP pattern). Likewise, the same weights may be applied to electric and magnetic fields for each mode, in order to construct the net effective fields associated with the ideal pattern. By construction, the ideal magnetic field at the point of interest will always be circularly polarized with unit amplitude, i.e. $\mathcal{B}_{1,n}^{(-)}(\mathbf{r}_0)=1$.

There is, of course, some flexibility in choosing the particular form of current basis functions spanning a chosen surface. While this choice is rendered largely unnecessary in basis-free Optimality Principle calculations, it can be instructive in simple geometries. For cylindrical geometries, we used cylindrical harmonic surface currents, as in (5). In order to represent the full range of possible local vector directions on the surface, we used either pairs of divergence-free and curl-free elements at each mode order (5), or else locally orthogonal $\phi$-directed and z-directed elements, which more efficiently capture z-directed electric dipole contributions (since a filamentary electric dipole may be shown to be neither divergence-free nor curl-free, but rather a well-defined combination (17,26)). This choice is inconsequential if a full basis is used, since the same ideal current pattern and UISNR will result from any complete basis. If we wish to determine the best possible coil using a constrained set of current types, however, we can optimize using, e.g., only divergence-free elements (for the best array of closed loop coils), or only z-directed elements (for the best array of z-directed electric dipole coils). For spherical geometries, we used vector spherical harmonics on the current surface, which separate naturally into divergence-free and curl-free elements (5). Note that a closed sphere is not, of course, a realistic model for actual coils, which must have an opening to insert a body part of interest. Spheres are useful, however, as models of the cap of a close-fitting helmet, and for general exploration of geometrical dependencies.

In the results to follow, optimizations with and without body noise are labeled "unperturbed OP (neglecting body noise)" and "ideal (considering body noise)," respectively. The conductor noise contributions identified in (5) were included in all cases. Since these contributions are uncorrelated between current modes, and scale linearly with current density, they fulfill the conditions identified earlier for unperturbed OP.

DGF calculations were performed using custom-designed Matlab code (The Mathworks, Natick, MA), which will be made available for download at http://www.cai2r.net/resources/software. The accuracy of field calculations using this DGF code has been tested extensively against experiment, and also against other electrodynamic solver software.

*Validations, and calculations for complex geometries, using general-purpose electrodynamic solver software*

Numerical calculations were performed using CST Microwave Studio 2016 (Computer Simulation Technology, Darmstadt, Germany). A beta version of a magnetic dipole source macro provided by CST was used to model a synthetic precessing spin by two oscillating magnetic sources mutually orthogonal and 90 degrees apart in phase at the position of interest. (This magnetic dipole source code was also checked against more cumbersome CST simulations using two small orthogonal current loops driven independently in quadrature at the interior position of interest, which yielded equivalent results.) To validate the proportionality between unperturbed OP patterns and tangential E fields on surfaces of interest, the synthetic precessing spin was modeled within a spherical geometry matching the geometry used in the DGF calculations. In order to approximate ideal current patterns for complex heterogeneous bodies and more general surface geometries, numerical simulations with the synthetic precessing spin were also carried out in a body model ("Duke," Virtual Population, IT'IS Foundation, Zurich, Switzerland) (37). Complex volumetric electric fields obtained from these simulations were then imported into Matlab, and were projected onto surfaces of interest by evaluating the fields on a grid of surface locations, then subtracting the inner product of those fields against local surface normal vectors.

*Curl-free filter for conservative electric fields*

Some of the results and discussion to follow explore the role of conservative electric fields in shaping ideal current patterns. In order to separate conservative fields from other field components, we used a curl-free filter, introduced here.

First, let us define conservative electric fields. If the current vector field, $\boldsymbol{J}$, is not divergence-free (i.e. not closed) on the current-carrying surface, then the continuity equation requires a time-varying surface charge density, $\rho$:

$$\nabla \cdot \boldsymbol{J} = -\frac{\partial \rho}{\partial t} = -i\omega\rho(\mathbf{r})e^{i\omega t} \quad [31]$$

Charge separation over time in turn results in a "quasistatic" electric field contribution, satisfying Gauss' Law, $\nabla \cdot (\varepsilon \boldsymbol{\mathcal{E}}) = \rho$, that oscillates with the separated charge. This so-called conservative electric field decreases with increasing distance from the current source, and also decreases with increasing frequency $\omega$, since local unbalanced charge density also clearly decreases with increasing frequency for a given current distribution (see Eq. [31] above).

This conservative field is mixed in a nontransparent fashion with the magnetically induced field (satisfying the Maxwell-Faraday equation, $\nabla\times\boldsymbol{\mathcal{E}} = -\partial\mathcal{B}/\partial t = -i\omega\mathcal{B}$ ) to form the total field. To separate out the conservative component of the electric field, we must therefore find the field component with a vanishing curl, corresponding to a vanishing induced field. In order to accomplish this task, we first take a Fourier transform of the field:

$$\nabla\times\boldsymbol{\mathcal{E}}(\mathbf{r}) = \int d^3\mathbf{k}\left(i\mathbf{k}\times\hat{\boldsymbol{\mathcal{E}}}(\mathbf{k})\right)\exp(i\mathbf{k}\cdot\mathbf{r}) \ . \tag{32}$$

We then project the transformed field onto the direction of the spatial frequency vector **k**:

$$\hat{\boldsymbol{\mathcal{E}}}(\mathbf{k})^{\text{curl-free}} = \frac{\left(\hat{\boldsymbol{\mathcal{E}}}(\mathbf{k})\cdot\mathbf{k}\right)\mathbf{k}}{\mathbf{k}\cdot\mathbf{k}} \ . \tag{33}$$

An inverse Fourier transform of this projected field results in the conservative electric field,

$$\boldsymbol{\mathcal{E}}(\mathbf{r})^{\text{curl-free}} = \int d^3\mathbf{k}\hat{\boldsymbol{\mathcal{E}}}(\mathbf{k})^{\text{curl-free}}\exp(i\mathbf{k}\cdot\mathbf{r}) = \boldsymbol{\mathcal{E}}(\mathbf{r})^{\text{conservative}} \ , \tag{34}$$

which satisfies the desired condition that $\nabla\times\boldsymbol{\mathcal{E}}(\mathbf{r})^{\text{conservative}} = 0$. In evaluating conservative electric fields in practice, we applied the previous three steps numerically to the full electric fields derived from DGF calculations.

## Results

The figures that follow illustrate applications of the Optimality Principle, first in simple geometries, then for more complex bodies and surfaces. They are intended to validate the principle, and to provide intuition about why ideal current patterns take the form that they do in a variety of potential experimental conditions.

**Figure 2** shows successive temporal snapshots of a precessing spin at the center of a dielectric cylinder (left) or sphere (right). The precession frequency of 4.26 MHz corresponds to the proton Larmor frequency at 0.1T, and three sample snapshots are shown, at time zero (bottom), at one eighth of a full cycle (middle), and at one quarter of a full cycle (top). Each spin is surrounded by a translucent cylindrical or spherical surface, on which the tangential electric field induced by spin precession is rendered as a quiver plot with superimposed color map. Arrows indicate the field direction, and field amplitude is indicated both by arrow length and by color map (with yellow being high and blue low). The dielectric object – either a cylinder or a sphere concentric with the current surface and 1 cm smaller in radius – is omitted from the plots for ease of visualization of the surface and the magnified interior spin. According to the Optimality Principle, these tangential electric field patterns correspond to the signal-optimizing current pattern for each central point, unperturbed by considerations of body noise. Note first that these patterns precess with the spin over time, thus replicating the general features of a quadrature-driven volume coil. Note also key differences between the signal-optimizing current patterns for cylinders and spheres, resulting from differences in surface geometry. Between each cylinder and sphere, the spin is shown surrounded by selected magnetic field lines (in red) and electric field lines (in blue). The magnetic field lines trace out a classic magnetic dipole pattern, emerging from the top of the dipole and looping around to converge again at the bottom of the dipole. When the dipole precesses at a fixed angular frequency, its time-varying magnetic field results in an induced electric field that curls around the transverse component of the dipole with circular axial symmetry. Projection of these electric field lines onto the current-carrying surface in each case yields the signal-optimizing current pattern. In a sphere, the resulting pattern is closed, since all points on the surface of a sphere are equidistant from the central spin position, and, therefore, the electric field lines are naturally fully tangential. By contrast, not all points on a cylindrical surface are equidistant from the central spin position. Thus, the circular symmetry of the pattern is lost. In particular, the tangential electric field, and the corresponding signal-optimizing current, is attenuated at $|z|>0$, since these points are further from the spin than are points at $z=0$. In other words, circumferential return currents that would otherwise complete the loop are attenuated, and the pattern is naturally dominated by z-directed currents. This feature is not immediately tied to the frequency of operation, but is rather a direct result of the surface geometry.

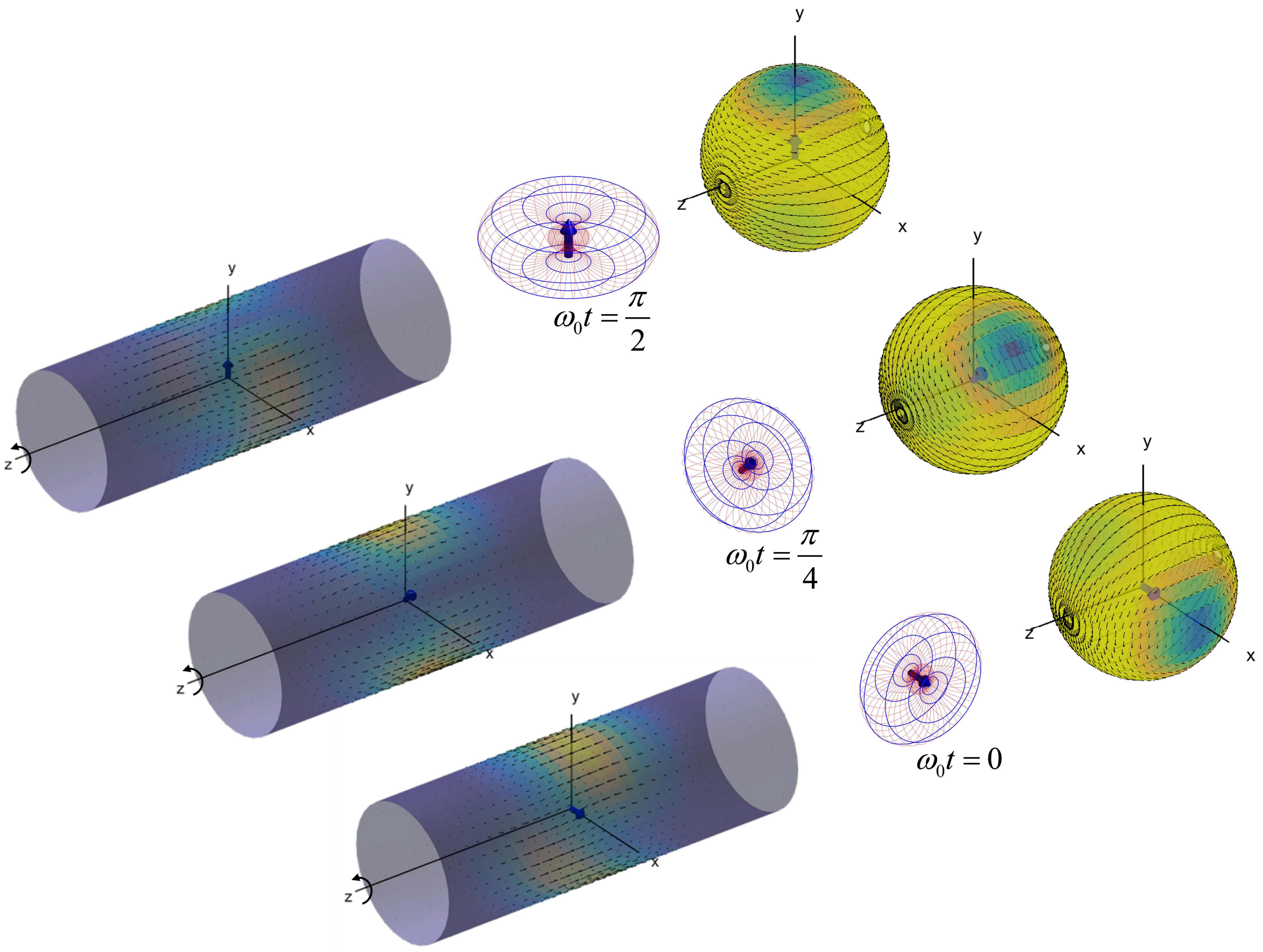

**Figure 2:** Schematic illustration of the Optimality Principle. Successive temporal snapshots of a magnified precessing central spin (thick blue arrow) are shown surrounded by translucent cylindrical (left) and spherical (right) surfaces on which the spin's tangential electric field patterns are superimposed (with arrows on quiver plots indicating tangential field direction, and with field amplitude indicated both by arrow length and by color map, with yellow being high and blue low). These patterns precess with the spin, here at a frequency of 4.26 MHz, corresponding to the proton Larmor frequency at 0.1T. For ease of visualization, the enclosed dielectric cylinder or sphere is not shown. According to the Optimality Principle, these tangential electric field patterns correspond to the signal-optimizing current pattern for each central point, unperturbed by considerations of body noise. The three sample temporal snapshots correspond to time zero (bottom), one eighth of a full precession cycle (middle), and one quarter of a full cycle (top). Note that time advances here from the bottom to the top of the figure, corresponding to a right-handed spin precession about the positive *z* axis. Between each cylinder and sphere, the spin is shown along with selected magnetic field lines (in red) and electric field lines (in blue). (These field lines are schematic, and will be perturbed somewhat in practice by the dielectric properties and boundaries of the objects in question.) Projection of the electric field lines onto the current-carrying surface in each case yields the signal-optimizing current pattern. Note strong differences in characteristics of the resulting current patterns associated with different surface projections (e.g. closed patterns for the sphere, and non-closed patterns for the cylinder).

**Supplementary Movie S1** contains an animation of the precessing central spin and surrounding signal-optimizing current pattern for the sphere from Fig. 2. Likewise, **Supplementary Movie S2** contains an animation of the precessing central spin and surrounding signal-optimizing current pattern for the cylinder from Fig. 2. **Supplementary Movie S3** shows a similar animation for a spin location intermediate between the center and the surface of the cylinder, and **Figure 3** shows two snapshots, at time zero (bottom) and at one quarter of a full cycle (top), respectively. In this case, the z-directed character of the signal-optimizing pattern is still dominant when the spin points along the x direction, but the pattern takes on a more closed character when the spin points along y. At all times, the pattern is now localized to the side of the cylinder closest to the spin. The animation in **Supplementary Movie S4** illustrates the case for a central position in the cylinder at a frequency of 298.1 MHz, corresponding to 7T field strength. **Figure 4** shows two snapshots, once again at time zero (bottom) and at one quarter of a full cycle (top), respectively. At this comparatively high precession frequency, propagation delay results in an apparent lag in tangential electric field patterns with increasing axial distance from the center: by the time the fields have reached the surface, the spin has already moved along in its precession, and, since the distance from the center to the surface is larger at $|z|>0$ than at $z=0$, the pattern is distorted as compared to the low-frequency pattern. The z-directed character of ideal currents is maintained, but the entire pattern is warped by differing field arrival times. This is the origin of the characteristic chevron shape observed in high-field ideal current patterns in previous studies (5,38), and also demonstrated in subsequent figures here. (Note that Reykowski et. al. correctly attributed the V-shaped character of cylindrical ideal current patterns to phase delay (38).) Once again, this feature is a direct outcome of surface geometry, in a setting of finite propagation speed. No propagation-related warping of signal-optimizing patterns is expected for a central point in a uniform sphere, since all points on a concentric spherical surface are equidistant from the origin, and there is no differential propagation delay.

**Figure 5** compares current patterns derived from direct signal optimization (i.e., matched filter combination of mode contributions neglecting body noise) using a DGF formalism in a spherical basis (left) with patterns derived using the Optimality Principle via surface projection of fields computed with a commercial electrodynamic solver (right). As was summarized in more detail in the Methods section, a synthetic precessing dipole was created for the commercial solver by combining, in quadrature, electric fields generated in a dielectric sphere by two mutually orthogonal magnetic dipole sources oscillating at 42.57MHz (corresponding to 1.0T proton Larmor frequency) and driven 90 degrees apart in phase. The pattern generated by projecting this combined electric field onto the spherical surface precisely matches the pattern generated by DGF signal optimization (up to a constant overall scaling factor, which was removed by normalizing each pattern by its maximum value). This constitutes a numerical validation of the Optimality Principle. The animation in **Supplementary Movie S5** shows that the match is preserved at all time points. A similar match was obtained for cylindrical geometries (data not shown).

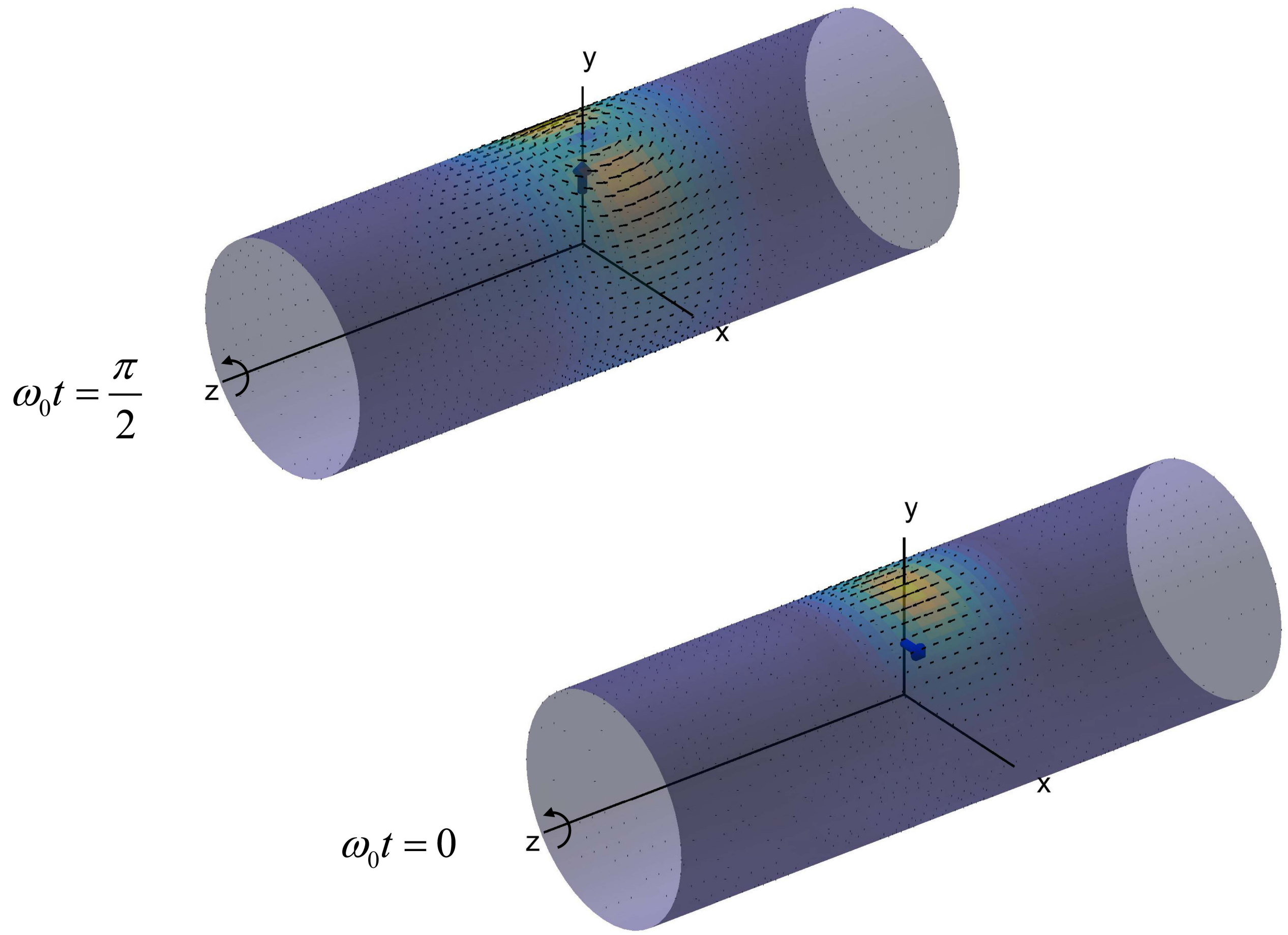

**Figure 3:** Successive temporal snapshots, as for the left-hand side of Fig. 2, but for a spin location (and a corresponding point of interest for signal optimization) intermediate between the center and the surface of a cylinder, i.e. at position (x, y, z) = (0, 10, 0) cm within the 20cm-radius cylinder. Note that the current pattern has a pronounced z-directed character at time zero (bottom), but some more circulatory behavior appears at one quarter of a full cycle (top). At both phases, the pattern is localized to the side of the cylinder closest to the spin. A full animation is provided in Mov. S3.

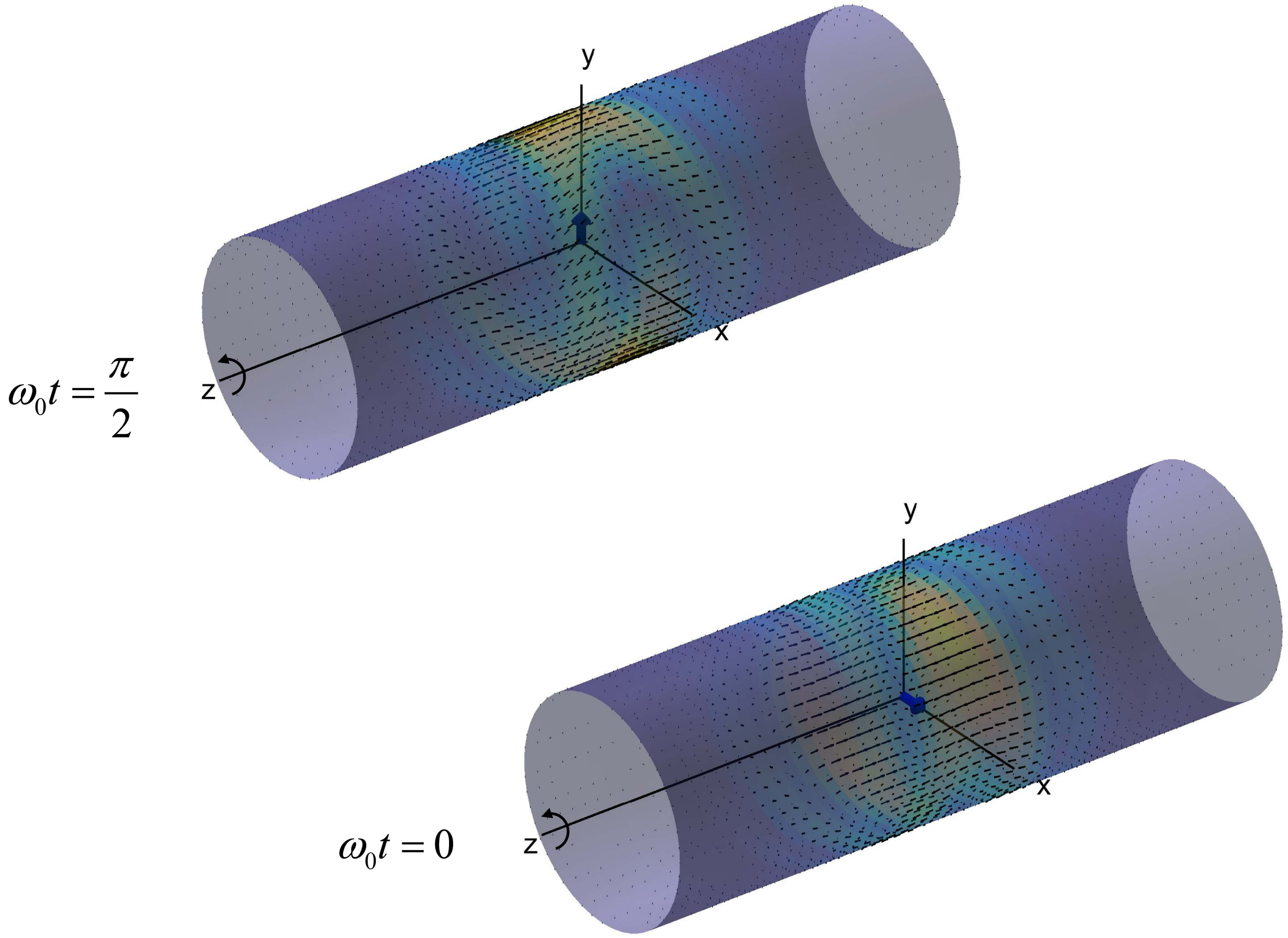

**Figure 4:** Successive temporal snapshots, as for Fig. 3, but for a central position in the 20cm-radius cylinder at 7T Larmor frequency. z-directed character is maintained, but note the distortion of the pattern at ±z, as compared with low-field patterns, resulting in a characteristic "chevron" structure. This distortion is a result of differential relativistic time delay associated with the spin's field reaching different locations on the surface. At the high frequency (298.1 MHz) associated with 7T field strength, the spin has already precessed appreciably by the time the electric field reaches the more distant regions on the surface. A full animation is provided in Mov. S4.

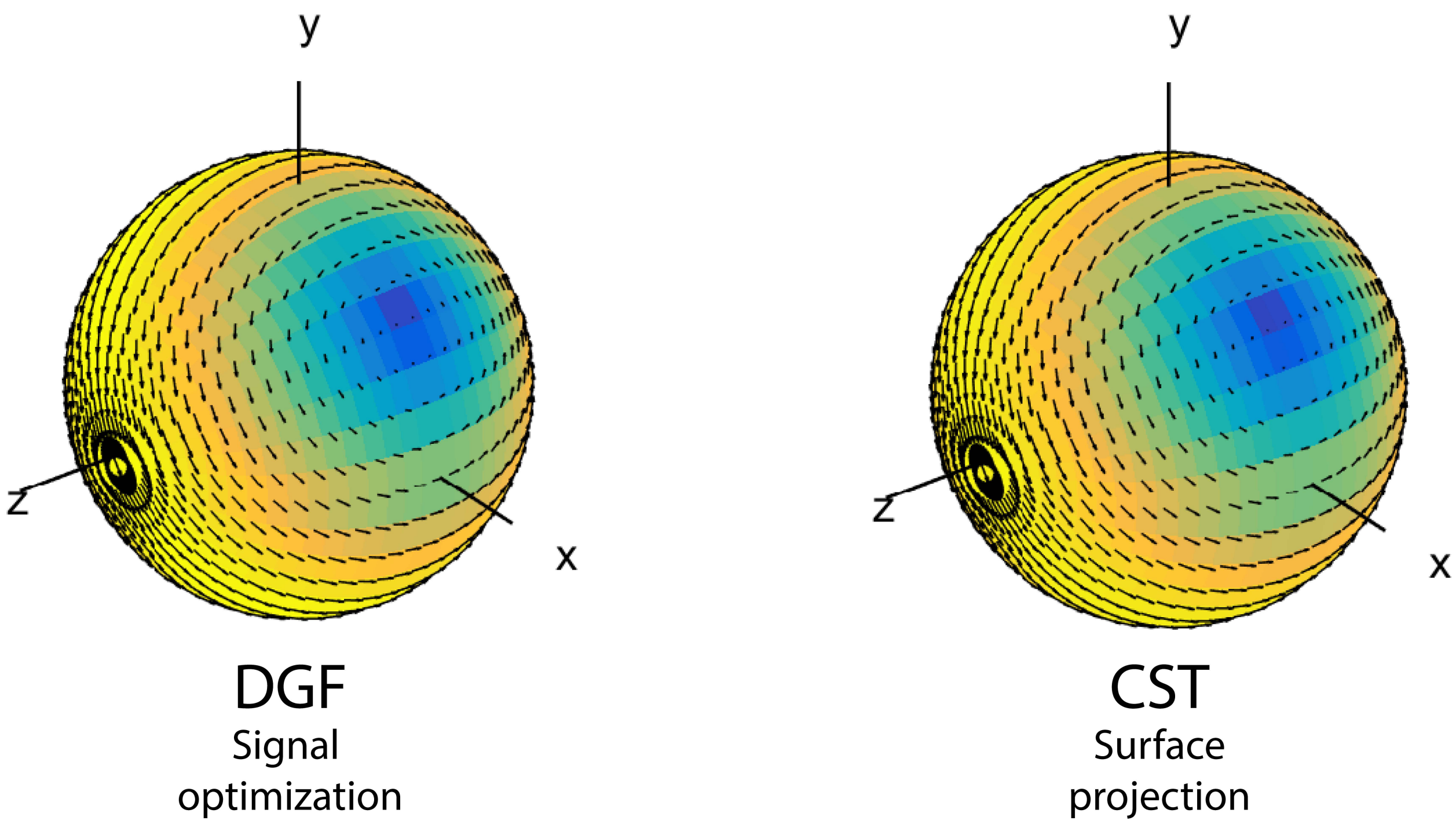

**Figure 5:** Validation of the Optimality Principle by comparing direct optimization of signal in our DGF framework (left) with field calculation in CST (right), for a spherical geometry. Sample temporal snapshots of current patterns for a central position in a dielectric sphere ($\sigma$ = 0.4 S/m, $\varepsilon_r$ = 39) with 20cm radius at 1.0T are shown (no longer translucent, and hence hiding the central precessing spin). Patterns on the left were derived in the DGF framework by direct optimization of central signal per unit current on a concentric 21cm-radius spherical surface. Patterns on the right were derived in CST by computing the electric field created by a synthetic precessing magnetic dipole (see details in text), and projecting this field onto the spherical current-carrying surface. The two patterns are, as expected, identical.

Validation for body-noise-dominated cases, and/or for cases with more complex geometries, is a more challenging proposition (though, of course, the derivations in the Theory Section are sufficiently general that we would expect them to hold for all such cases). On the one hand, determination of ideal current patterns for arbitrary surfaces surrounding inhomogeneous bodies is a computationally-intensive task when the Optimality Principle is not used, given the large number of current basis functions required. Computation of ultimate intrinsic SNR for realistic body models has only recently been accomplished using advanced integral-equation methods (23), and associated ideal current patterns have not yet been demonstrated. On the other hand, use of the Optimality Principle to determine SNR-optimizing ideal current patterns when the effects of body noise are significant would require a full perturbative OP calculation, as outlined in the text leading up to Eq. [30]. Such a full perturbative calculation requires iterative computation of the field outside the body produced by defined eddy currents within the body – a feature not commonly included in most commercially available electrodynamic solvers. Development and dissemination of full perturbative code will be the subject of future work. (The Marie software suite (21,22), for example, uses so-called polarization currents as an intermediate step, and should therefore be relatively straightforward to modify for our purposes.) In the meantime, though, we can validate the expansion in Eq. [25], on which the perturbative formulation is based, for simple geometries, using analytically computed noise covariance matrix elements for a well-defined surface current basis in the DGF formalism. **Figure 6a** illustrates that the expansion does indeed converge for realistic noise covariance matrix elements in the 20cm-radius dielectric cylinder. Plotted in the figure are signal contributions, for a point of interest at the center of the cylinder, as a function of expansion order. Here we have defined signal contribution at expansion order *m* as the following convenient scalar quantity, which approaches unity as the matrix expansion converges:

$$\left(\mathbf{S}^H \mathbf{\Psi}^{-1}\mathbf{S}\right)^{-1}\left(\mathbf{S}^H\left(\mathbf{1}+\left(\mathbf{1}-\mathbf{\Psi}\right)+\left(\mathbf{1}-\mathbf{\Psi}\right)^2+\ldots+\left(\mathbf{1}-\mathbf{\Psi}\right)^m\right)\mathbf{S}\right). \qquad [35]$$

Elements of the signal sensitivity vector $\mathbf{S}$ and the (block-diagonal) noise covariance matrix $\mathbf{\Psi}$ were computed for each cylindrical harmonic surface current mode, as previously described (5). At 7T Larmor frequency (solid blue line), the rate of convergence is comparatively rapid, reaching 90% at an expansion order of approximately 50. At 0.1T frequency (solid red line), the perturbation associated with body noise is far more significant, and the signal contribution series has only reached 0.2% of its final value at order 100. Rate of convergence is a key consideration for practical perturbative OP calculations, since this rate sets the number of time-consuming solver calls required for an accurate calculation. Observe, however, that convergence is extremely rapid (passing 94% at an expansion order of only 3) if we only use divergence-free modes in the expansion at 0.1T (dashed red line). This suggests a practical strategy for limiting the number of solver calls in perturbative OP calculations:

1. Initialize the calculation with subspace projections corresponding to use of a suitable subset of modes – for example, preserve only the circumferential or the axial components of surface projections (corresponding to use of $\phi$-directed or z-directed modes only), or perform a divergence-free or a curl-free filter on surface projections (corresponding to use of divergence-free or curl-free modes only). Then,
2. Sum the resulting current patterns for different projections, and
3. Use the composite pattern as a starting point for subsequent iterations.

This separate initialization corresponds to ignoring correlations between different subsets of modes or projections, but, for suitably selected subsets, these correlations represent a small perturbation away from the full answer, requiring only a small number of iterations to correct. **Figure 6b** illustrates this strategy in the uniform dielectric cylinder. Following ten iterations using separate divergence-free and curl-free contributions to the noise covariance matrix (dashed and fine dotted lines leading up to the vertical grey dotted line in Fig. 6a), the resulting diagonal matrix was used as a starting point for a second-pass iteration, whose convergence is shown in Fig. 6b. Signal contributions in the second pass exceed 99% after only 10 additional iterations, both at low and at high field strength. While complex body models and tortuous surfaces may slow convergence somewhat in practice, these results suggest that, when suitably initialized, full perturbative basis-free OP calculations of ideal current patterns may require only a small number of solver calls, and may therefore be far more parsimonious and convenient than methods employing any particular current basis.

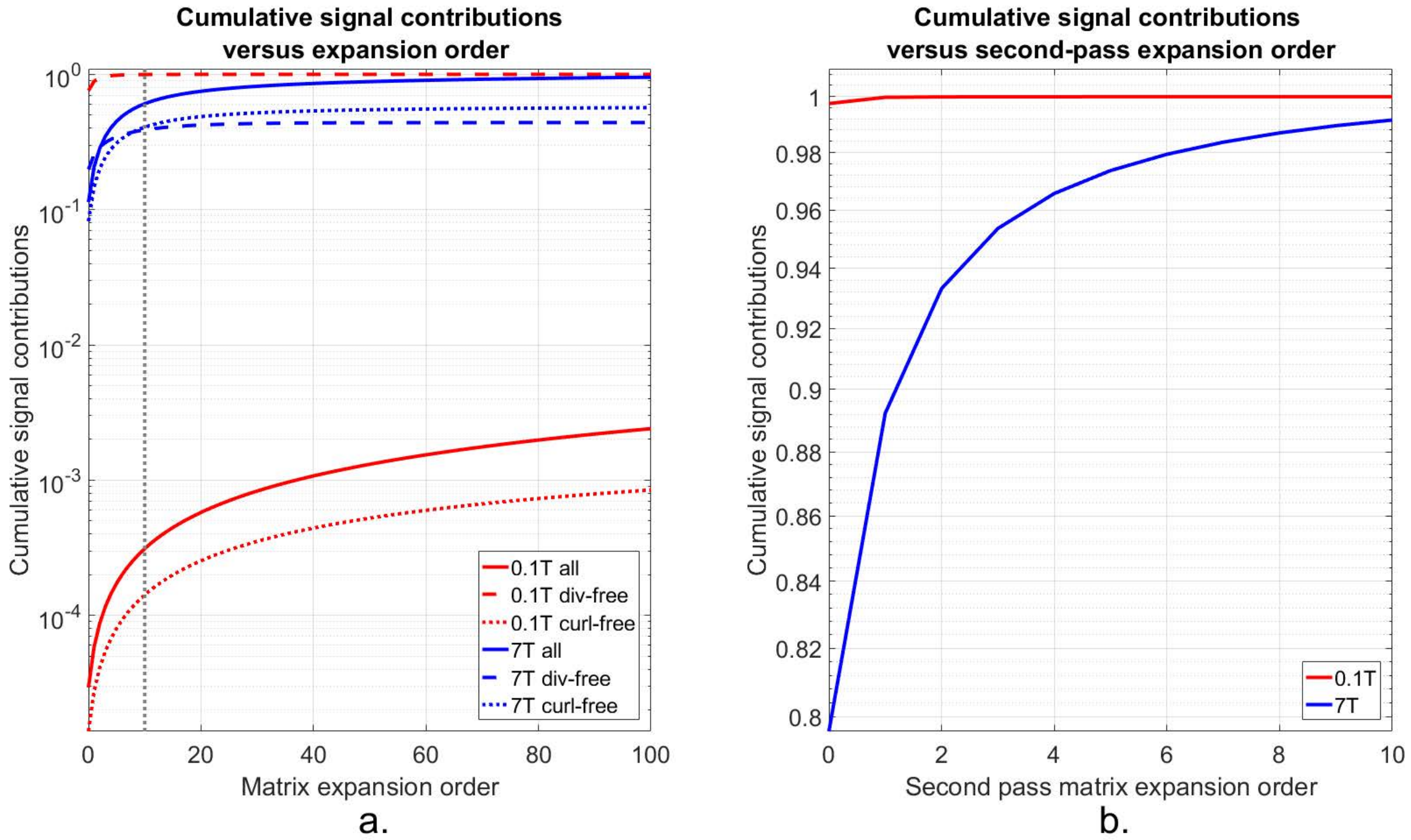

**Figure 6:** Convergence of the matrix expansion in Eq. [25], on which the perturbative formulation of the Optimality Principle is based, for the 20cm cylinder, with noise covariance matrix elements computed analytically for a basis of surface current modes in the DGF formalism. a) Cumulative signal contributions, defined in Eq. [35], as a function of matrix expansion order. At 7T Larmor frequency (solid blue line), the rate of convergence is comparatively rapid, whereas at 0.1T frequency (solid red line), the perturbation associated with body noise is far more significant, and convergence is far slower. Rapid convergence is restored at 0.1T, however, if we only use divergence-free modes in the expansion (dashed red line). b) Illustration of convergence at 0.1T (solid red line) and 7T (solid blue line) following an initialization procedure, in which 10 iterations are performed using separate divergence-free and curl-free current contributions (dashed and fine dotted lines leading up to the vertical grey dotted line in a), and the resulting diagonal matrix is used as a starting point for a second-pass iteration. With this initialization, convergence is rapid for both low and high field strengths, suggesting that full perturbative basis-free OP calculations of ideal current patterns may be feasible using only a comparatively small number of solver calls.

Returning now to the inspection of current patterns of interest, **Figure 7** illustrates the "unwrapped" view of cylindrical current patterns which will be used henceforward. (Once again, relative current amplitude is represented both by arrow length and by colormap.) With the circumferential direction unrolled onto the vertical axis, the z-directed character of a central-signal-optimizing pattern snapshot at 0.1T (matching the case shown in Fig. 2) may easily be appreciated as a circulating pattern attenuated along z. In such an unwrapped view, rotation of the pattern over time, matching the precession of a centrally located spin, appears as cyclical progress along the vertical direction within the 2D frame, with the topmost structures reappearing at the bottom. This is illustrated in **Supplementary Movie S6** (central position at 0.1T, matching Mov. S2), **Supplementary Movie S7** (intermediate position at 0.1T, matching Mov. S3), and **Supplementary Movie S8** (central position at 7T, matching Mov. S4). Also making use of the unwrapped 2D view, **Supplementary Movie S9** illustrates convergence of the signal-optimizing current pattern for the same central position at 7T, as the weighted contributions of different surface current modes are added together. Application of the Optimality Principle circumvents, in a single calculation, the iteration associated both with computation of weights and with accumulation of the final pattern.

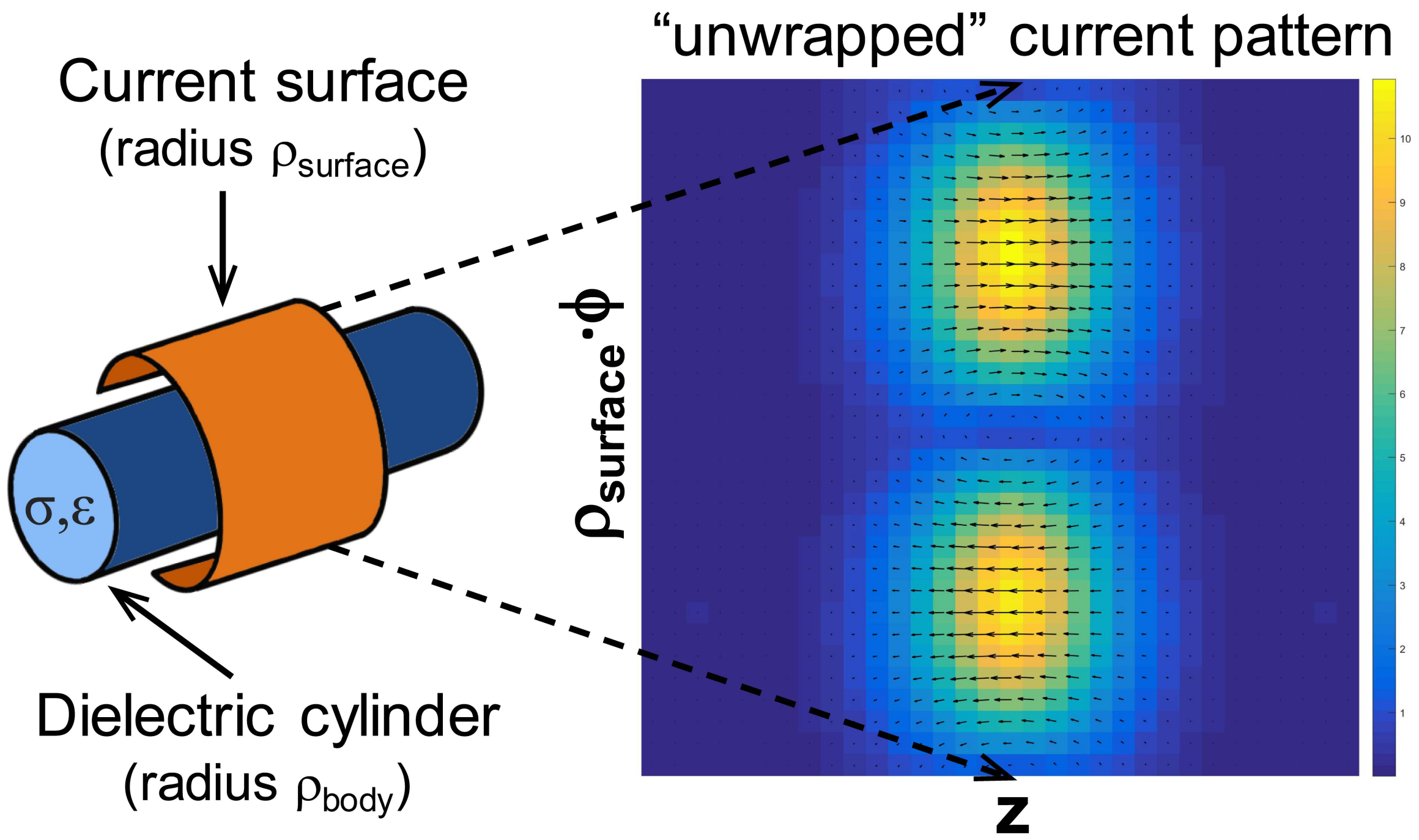

**Figure 7:** Unwrapped view of cylindrical current patterns, which will be used in subsequent figures for ease of visualization. A snapshot of the central 0.1T signal-optimizing current pattern from Fig. 2 and Supplementary Movie S2 is shown with the circumferential direction unrolled onto the vertical axis. As before, relative current amplitude is represented both by arrow length and by colormap.

All of the signal-optimizing patterns shown so far have neglected the perturbing effects of body-derived noise. We know, however, that full SNR-optimizing patterns may also be computed, either using perturbation theory, as described earlier in the Theory section, or else using a DGF formalism in simple geometries (5). While the need to account separately for body noise may be seen as a nuisance, it may also be turned to advantage, for by comparing signal-optimizing patterns with full SNR-optimizing ideal patterns, we may isolate the effects of body noise. Such a separation of signal-optimizing and noise-minimizing effects has not been possible before now, and it provides noteworthy insights into the determinants of effective coil structures. In particular, ideal current patterns considering body noise may be decomposed as a sum of unperturbed OP patterns neglecting body noise plus patterns made up of what might be called "dark modes," which have no contribution to signal at the point of interest, but which reduce body noise by canceling some of the electric field associated with the unperturbed OP patterns. (Note that "dark modes" have been identified in transmit coil designs (39,40), and we use the term here in an analogous fashion for the case of signal reception.) Indeed, if we subtract unperturbed OP patterns from full ideal patterns considering body noise, the unit signal at the point of interest in each case cancels exactly, leaving a difference pattern that only affects noise.

**Figure 8** shows illustrative snapshots of unwrapped full ideal (left), unperturbed OP (center), and dark mode (right) patterns, for a position at the center of a 20cm dielectric cylinder at frequencies associated with various field strengths, including a) 0.1T, b) 1.5T, c) 3.0T and d) 7.0T. In each case, the dark mode pattern was computed as described above, by subtracting the unperturbed OP pattern neglecting body noise from the full ideal pattern considering body noise, resulting in no signal at the location of interest. A number of noteworthy features become apparent when one compares these patterns as a function of field strength. Observe that the propagation-related distortions mentioned earlier increase noticeably as field strength increases. Whereas the unperturbed OP pattern is highly symmetric along both axial and circumferential directions at 0.1T, the corresponding 7T pattern is warped into a distinctive chevron, with z=0 defining the leading edge of precession, and with progressive delays appearing for increasing |z|. Second, note that full ideal patterns considering body noise are markedly different from unperturbed OP patterns at low field strength. For the 0.1T case, in fact, the full ideal pattern closes on itself in a complete circuit, whereas the unperturbed OP pattern remains distinctly non-closed. This difference corresponds to a high-amplitude dark mode on the right. As field strength increases, however, full ideal and unperturbed OP patterns begin to resemble one another more closely, and the dark mode contribution diminishes in amplitude and significance. In other words, unperturbed OP patterns capture much of the behavior of full ideal patterns at high field strength. As we shall discuss further to follow, this is why z-directed electric dipoles perform so well in cylindrical geometries at high field strength: they are effective at optimizing signal for central locations of interest.

But unperturbed OP patterns are also z-directed at low frequency; in fact, without propagation effects to contend with, these signal-optimizing patterns resemble a relatively straightforward array of straight z-directed dipoles (albeit with varying length). What, then, is the reason for SNR-optimizing ideal current patterns to be closed at low frequency? Such closed patterns clearly differ from the patterns that optimize central signal – so the differences must be required to minimize noise. This hypothesis is borne out by inspection of the electric fields associated with ideal current patterns.

**Figure 9** shows snapshots of electric field patterns associated with selected current patterns from Fig 8. Electric fields are shown in three-plane slices of a 3D volume containing the dielectric cylinder. The absolute magnitude of the field at each point (in arbitrary but consistent units) is indicated by color, according to colormaps to the right of each plot. Field direction and magnitude are also indicated by the direction and magnitude of small purple arrows superimposed at each grid point on the slices. Fields corresponding to the ideal pattern considering body noise are shown in the left-hand column, and fields corresponding to the unperturbed OP pattern neglecting body noise are shown in the right-hand column. Colormaps are matched for direct comparison of field magnitude within each row, with the maximum value of the field in each plot indicated with an asterisk on the shared colormap for each row, whenever that maximum lies below the top of the colormap range. At 0.1T (a, top row), the electric field associated with the unperturbed OP pattern is clearly much larger at all points in the volume than the corresponding field for the ideal case, indicating that body-derived noise would indeed overwhelm the signal benefits of the unperturbed pattern. At 7T (b, second row), on the other hand, the electric fields are similar both in amplitude and in spatial pattern for the ideal and the unperturbed OP current patterns. Note the spiral character of the electric field as a function of radius at 7T frequency. This field pattern results from delays associated with propagation out from the central location of the precessing spin. **Supplementary Movies S10 and S11** show animations of the unperturbed OP electric field patterns at 0.1T and 7T, respectively. The field appears to rotate as a unit at 0.1T (Mov. S10), since precession is slow enough that field propagation is effectively instantaneous. (Likewise, as shown in **Supplementary Movie S12**, the magnetic field rotates as a unit, with comparatively simple circular polarization, consistent with a value of $\mathcal{B}_1^{(-)} = 1$ at the central point of interest.) At 7T (Mov. S11), on the other hand, the electric field clearly radiates outward from the spin position, twisting around itself as the spin advances while the field propagates.

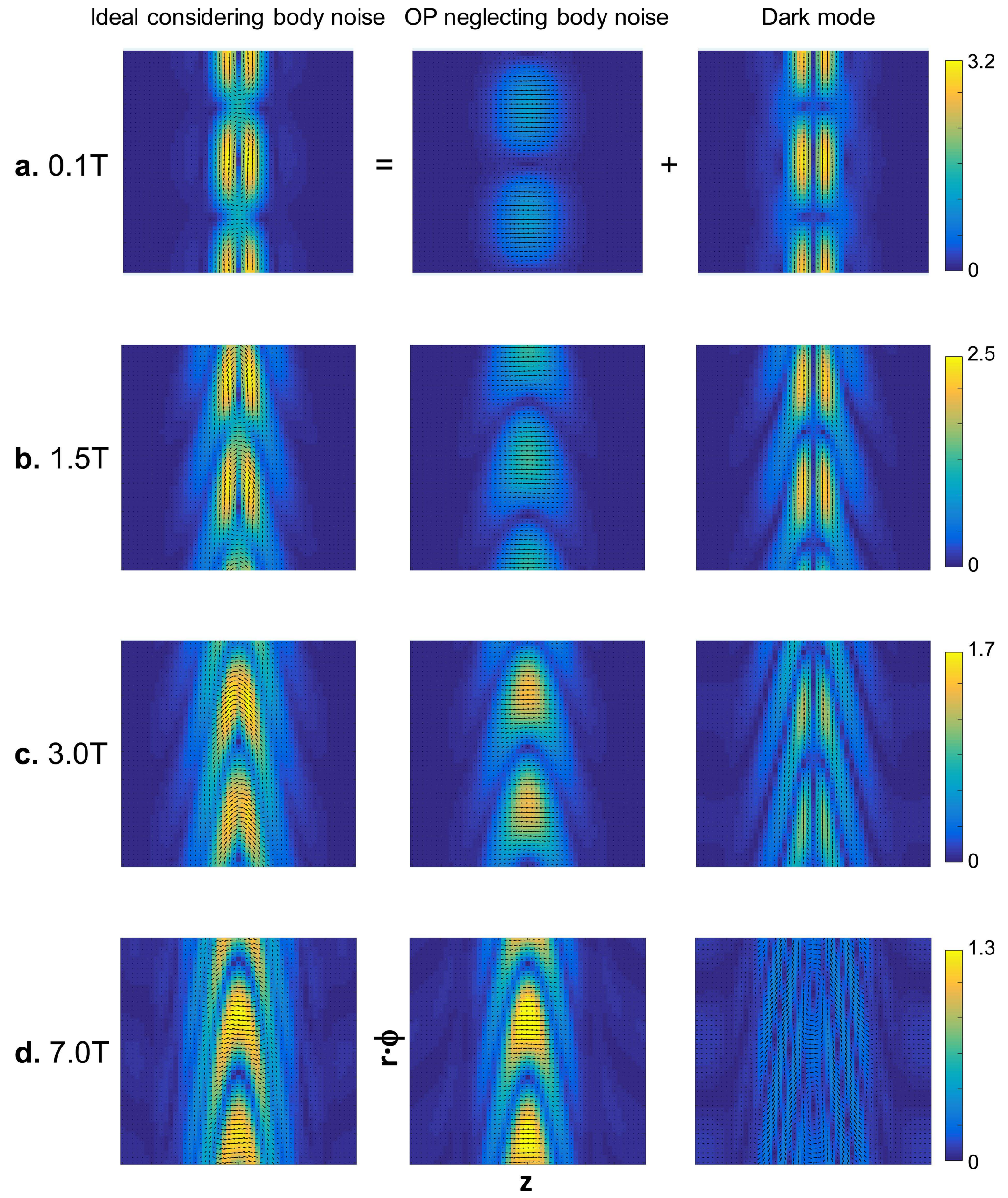

**Figure 8:** Ideal current patterns considering body noise (left), decomposed into a sum of unperturbed OP patterns neglecting body noise (center) plus dark mode patterns (right), which have no contribution to signal at the point of interest, but which reduce body noise by canceling some of the electric field associated with the unperturbed OP patterns. Snapshots of unwrapped patterns optimized for the center of a 20cm dielectric cylinder are shown at various field strengths: 0.1T (a), 1.5T (b), 3.0T (c), and 7.0T (d). Each row uses the same colormap, in order to compare the relative contributions of unperturbed OP and dark mode patterns to the ideal pattern, but colormaps differ between rows. (In this figure and henceforward, on the other hand, arrow length is autoscaled for maximum visibility of current direction in each individual pattern. In other words, arrows indicate relative amplitude within a pattern, but not absolute amplitude compared with other patterns.) Note the progression in the shape of ideal patterns from fully closed at low field strength to strongly non-closed at high field strength. Note also the largely z-directed non-closed character of unperturbed OP patterns at all field strengths, and the decreasing contribution of dark mode patterns with increasing field strength.

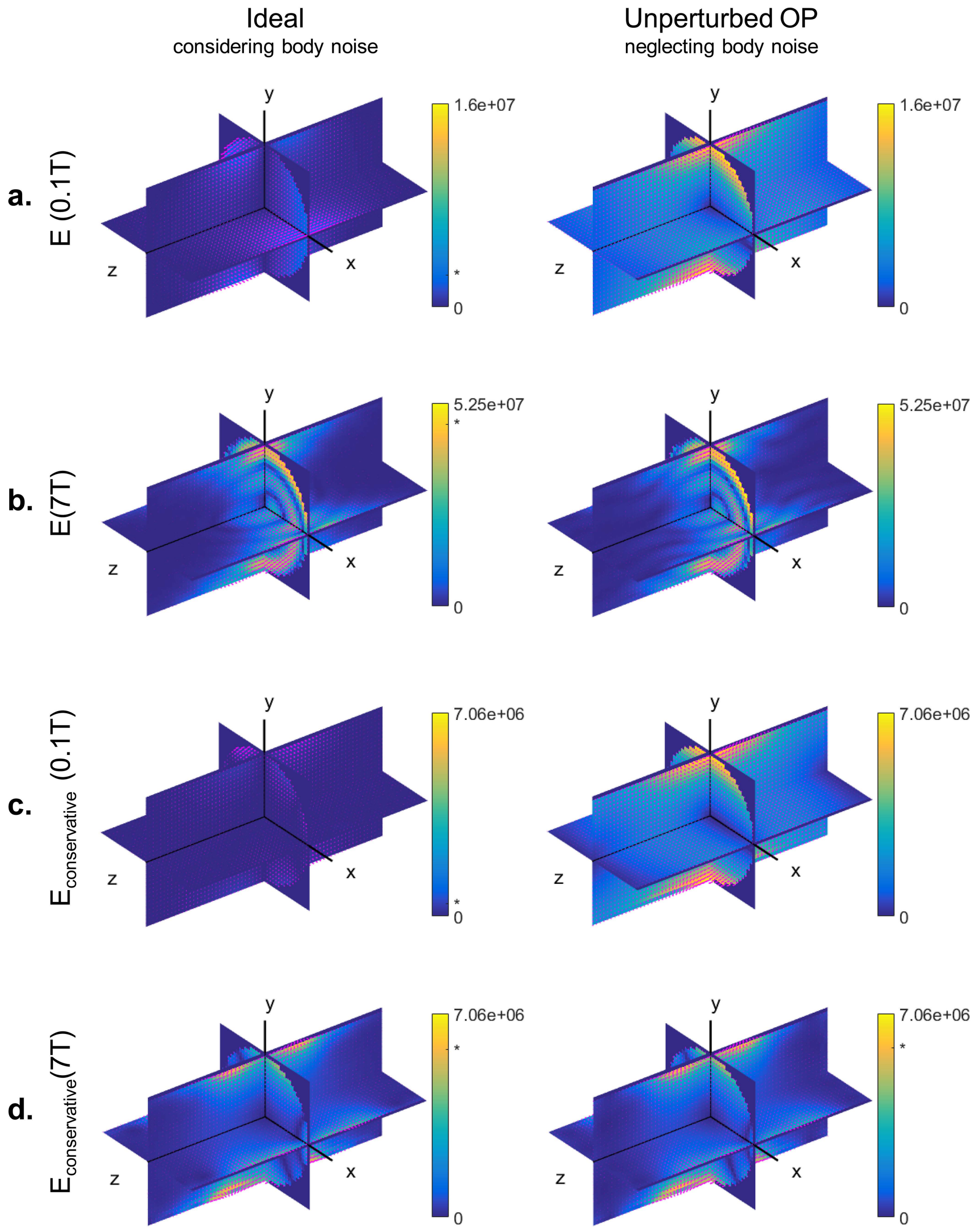

**Figure 9:** Snapshots of electric field patterns associated with selected current patterns from Fig. 8. Electric fields are shown in three-plane slices of a 3D volume containing the dielectric cylinder. The absolute magnitude of the field at each point (in arbitrary but consistent units) is indicated by color, according to colormaps to the right of each plot. Field direction and magnitude are also indicated by the direction and magnitude of small purple arrows superimposed at each grid point on the slices. Fields corresponding to the ideal pattern considering body noise are shown in the left-hand column, and fields corresponding to the unperturbed OP pattern neglecting body noise are shown in the right-hand column. Colormaps are matched for direct comparison of field magnitude within each row, with the maximum value of the field in each plot indicated with an asterisk on the shared colormap for each row, whenever that maximum lies below the top of the colormap range. Full electric fields in the top two rows (a: 0.1T and b: 7.0T) are compared with the conservative components of those fields (computed using the curl-free filter described in Methods) in the bottom two rows (c: 0.1T and d: 7.0T, all shown with a common colormap). Note the marked reduction in integrated electric field for ideal versus unperturbed OP patterns at 0.1T, as compared with a much more modest reduction at 7.0T. Note also the dominance of conservative electric field components at low field/frequency for non-closed OP current patterns, as opposed to the minor contribution of conservative components to the full electric field at high field/frequency.

Why are the electric fields associated with unperturbed OP patterns so much larger than those associated with full ideal patterns at low frequency, but not at high frequency? For large dielectric bodies, like the 20cm-radius-cylinder used for illustrations here, one key answer relates to the role of quasi-static conservative electric fields, as opposed to the propagating field components which are clearly significant at high frequency. As was mentioned earlier, any non-closed (i.e. non-divergence-free) current pattern necessarily implies charge separation, and an associated conservative electric field which tracks that separated charge. **Supplementary Movie S13** illustrates a dominant conservative electric field pattern for the case of a simple filamentary z-directed electric dipole with a sinusoidal current distribution (indicated with red arrows in the animated figure) oscillating at the 0.1T Larmor frequency. At each instant of time, the electric field vectors clearly conform to a static electric dipole pattern, emerging from an instantaneous "source" on one end of the dipole, and proceeding to a "sink" on the other end. The pattern is quasi-static, in that it simply oscillates in sync with the current. The same current dipole at the 7T Larmor frequency, however, is associated with a markedly different electric field pattern, as shown in **Supplementary Movie S14**. In this case, no discrete sources or sinks are evident, the field amplitude tracks the sinusoidal distribution of current along the z direction, and the pattern is clearly that of a propagating wavefront. (The fact that the field appears to be radiating towards rather than away from the dipole results from the complex conjugation that converts transmit to receive, as discussed in the Theory section.) In other words, the current dipole at high frequency is acting as a well-known radiative antenna. When we apply the conservative electric field filter described in the Methods section to the high-frequency electric field in Mov. S14, we recover a quasi-static dipole field pattern, as shown in **Supplementary Movie S15**, but with an amplitude nearly an order of magnitude lower than that of the full field pattern. In other words, though the conservative electric fields which dominate at low frequency do not vanish altogether at high frequency, they become minor contributors to a total electric field pattern increasingly dominated by propagating components.

Returning to **Figure 9**, the third and fourth rows (c and d) show the filtered conservative components of the electric fields from the first and second rows (a and b), respectively. At 0.1T (c), the conservative component clearly dominates the field for the non-closed unperturbed OP current pattern, whereas the closed ideal current pattern has a vanishing conservative component (likely reflecting the numerical precision of the curl-free filter). Thus, it stands to reason that an SNR-optimizing algorithm will go to some lengths to close current patterns, sacrificing signal strength per unit current in order to reduce noise which would result from conservative electric fields. At 7T (d), on the other hand, the non-closed ideal and OP current patterns both maintain a residual conservative component, but with an amplitude approximately an order of magnitude smaller than the amplitude of the total electric fields in (b). At high frequency, we may therefore posit, there is no incentive to close current patterns, since doing so would only introduce additional propagating field components, which would overwhelm the benefit of eliminating the comparatively small conservative field components. Indeed, we can test this hypothesis by artificially applying the closed low-frequency ideal current pattern not only at low frequency (**Supplementary Movie S16**) but also at high frequency (**Supplementary Movie S17**), and comparing the resulting electric field patterns. When a closed 0.1T ideal current pattern is applied at 7T (Mov. S17), propagating electric field components dominate, and propagation actually extends farther along the cylinder axis than for the non-closed pattern optimized for high frequency (Mov. S11). One may also note a high concentration of electric field around the circumferential return current legs. In other words, the noise benefits of closing the current pattern are indeed lost at high frequency. This is why the ideal current pattern closely resembles the unperturbed OP pattern at high frequency (Fig. 8d).

At sufficiently high frequencies, the cylindrical RF shield of the scanner bore itself is known to support traveling wave modes, which may be used to excite and receive MR signal (41). These axially-propagating cavity modes deviate from the predominantly radial propagation seen in field plots so far. Nevertheless, the Optimality Principle may easily be applied to evaluate the SNR efficiency of such modes. All of the simulated results presented so far were computed in the absence of a surrounding shield. It is straightforward, however, to place a conductive shield around the object, whether in a DGF framework (5) or in a more general electrodynamic solver. When we add such a shield, at a radius of 37.5cm, matching typical commercial configurations for a 60cm bore diameter, we observe somewhat exotic behavior at 7T frequency, as shown in **Supplementary Movie S18**. The unperturbed OP current pattern (Mov. S18a) looks generally similar to the unshielded case (Mov. S8 and Fig. 8d), since OP electric fields at the current-carrying surface are only mildly perturbed by the presence of the remote shield, spreading out further along z as a result of axial propagation. There is a marked difference, however, between the unperturbed OP pattern and the SNR-optimizing ideal pattern when the shield is present. The ideal pattern considering body noise has as strong repeating structure along z, and even loses the characteristic chevron structure associated with propagation delay (Mov. S18a). Inspection of the corresponding electric fields (Mov. S18b) reveals that the additional currents flanking the central axial position actually serve to *cancel* the axially-propagating traveling wave modes supported by the bore. For the ideal pattern, electric fields are more confined along z, and propagation is dominantly radial, whereas appreciable axial propagation is evident for the unperturbed OP pattern without traveling-wave cancellation. In other words, for this central position of interest, traveling wave modes are sufficiently SNR-inefficient that the optimization sacrifices signal in order to arrange for their active cancellation. (One caveat is that the case considered here involves a long cylindrical bore, which requires traversal of a large volume of dielectric tissue to reach the point of interest from the sides, and which includes no coils that might efficiently generate traveling waves from the end caps, in the manner of patch antennas. To explore the role of patch antennas in shorter bores, one need only place a current-carrying surface on one or both ends of the bore, in addition to the radially encircling surface used so far. Optimality Principle calculations – with a general-purpose solver rather than DGF in this case – will replicate patch-antenna-like cap current distributions if and only if they have an appreciable role to play in SNR optimization.)

**Figure 10** summarizes the SNR contributions of various types of current patterns as a function of field strength / frequency in a cylindrical geometry. Plotted on the vertical axis is the fraction of ultimate intrinsic SNR (for a central point in the 20cm-radius cylinder surrounded by a 21cm-radius surface) captured by the unperturbed OP current pattern (solid blue line and stars), the optimized divergence-free current pattern corresponding to the best possible loop array (dashed red line and circles), and the optimized z-directed current pattern corresponding to the best array composed of z-directed electric dipoles (dashed yellow line with triangles), at field strengths ranging from 0.5T to 20T. These quantities were computed in the DGF framework by optimizing signal per unit current, SNR for divergence-free current modes only, and SNR for z-directed current modes only, respectively, and comparing with a full SNR optimization for all current modes. Divergence-free (i.e. closed) current patterns capture nearly all of the UISNR at low field, but their efficacy drops below 50% at the highest field strengths tested. z-directed currents are SNR-inefficient at low field, become equal to divergence-free currents in efficiency around 3T, and maintain high performance at higher field strengths. Unperturbed OP patterns, as we have seen, are dominated by z-directed currents due to the geometry of the cylinder, and they perform similarly to optimal z-directed patterns at all field strengths. Note that these results (and the results in all other figures except Mov. S18) omit the scanner's RF shield, for simplicity of interpretation. They also use a fixed relative permittivity ($\varepsilon_{rel}$ = 39) and conductivity (0.4 S/m) independent of frequency, once again for ease of interpretation. In actuality, tissue electrical properties are known to vary with frequency, but this variation is trivial to incorporate into simulations, and results in no substantial changes in overall trends.

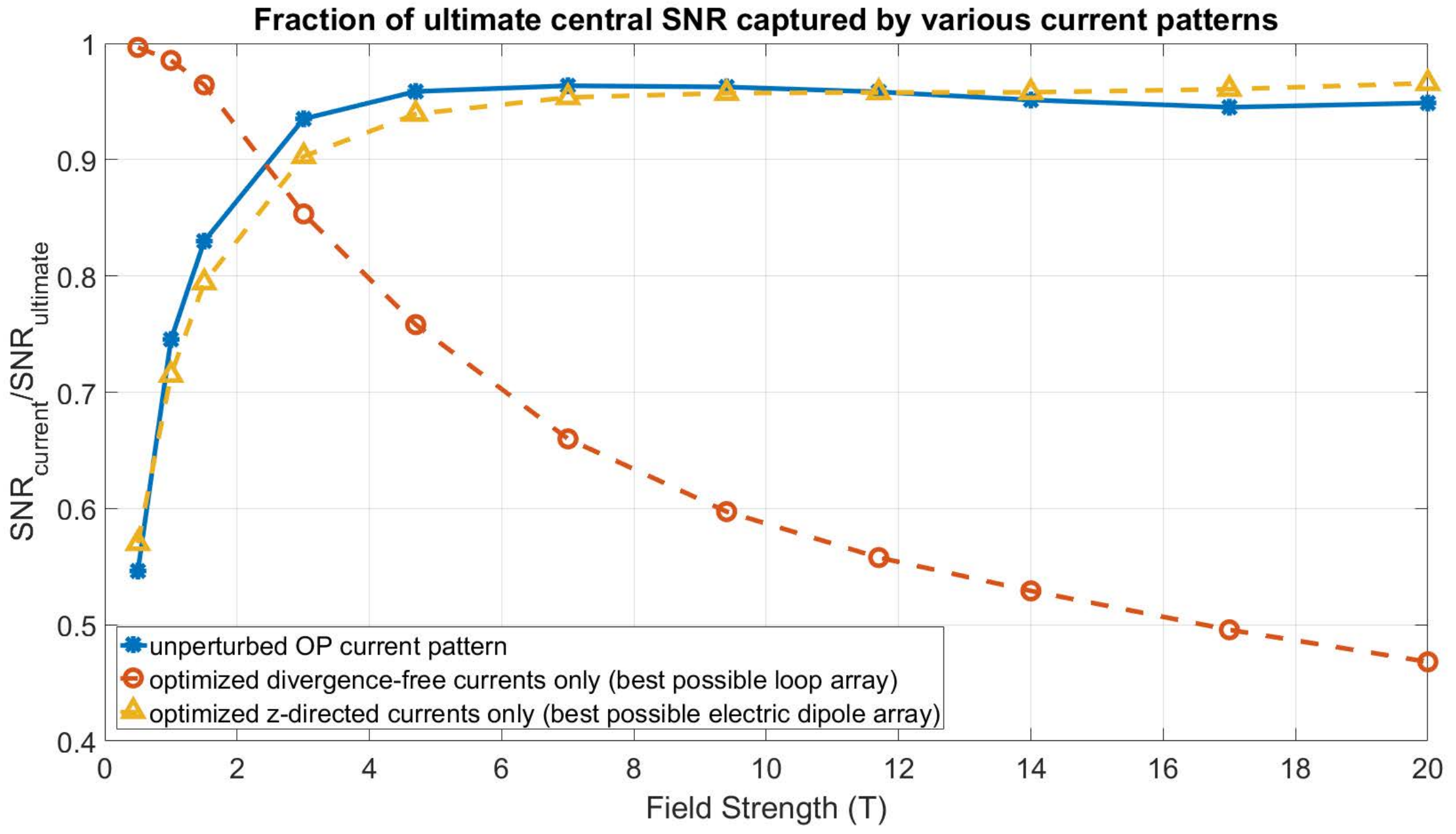

**Figure 10:** Fraction of the ultimate intrinsic SNR (for a central point in the 20cm-radius cylinder, surrounded by a 21cm-radius surface without an encircling shield) captured by the unperturbed OP current pattern (solid blue line with stars), the optimized divergence-free current pattern corresponding to the best possible loop array (dashed red line with circles), and the optimized z-directed current pattern corresponding to the best array composed of z-directed electric dipoles (dashed yellow line with triangles), at field strengths ranging from 0.5T to 20T. Divergence-free (i.e. closed) current patterns capture nearly all of the UISNR at low frequency, but their efficacy drops below 50% above 16T. z-directed currents are equally efficacious as divergence-free currents at capturing UISNR in the vicinity of 3T, and they maintain high performance at higher field strengths. Unperturbed OP patterns are dominated by z-directed currents due to the geometry of the cylinder, and they perform similarly to optimal z-directed patterns at all field strengths.

**Figure 11** shows current patterns (a, c) and associated electric field patterns (b,d) optimized at 1.5T for a point intermediate between the center and the edge of the cylinder, at coordinates (x,y,z) = (0,10,0) cm. Temporal snapshots are included at two distinct phases, 110 degrees and 200 degrees into a cycle at the 1.5T Larmor frequency (displaced from 0 and 90 degrees to account for propagation delay). The ideal current pattern considering body noise assumes a figure-eight pattern at 18 degrees, and a tight loop pattern at 108 degrees, in a distributed version of a traditional surface quadrature coil. The unperturbed OP current pattern has larger spatial extent, and alternates between predominantly z-directed and more circulatory configurations. (Corresponding animations may be found in **Supplementary Movie S19**.) As for previous cases, the unperturbed OP electric fields display a distinctive conservative character, at least in some phases. However, in this case, the tightly looped concentration of ideal current results in electric fields with a maximum amplitude nearly as large as for the unperturbed OP current. The noise minimization strategy in this case, then, is less a matter of reducing conservative electric fields, per se, than it is about limiting the extent of the fields away from the point of interest – a proposition that is more difficult for a deep central position than for a more superficial one. Ideal currents are concentrated in order to minimize the extent of associated electric fields, and they travel in complete circuits not only to reduce conservative electric field but also to minimize total electric field by symmetry.

As shown in **Figure 12**, similar strategies are at work for the center of sufficiently small cylinders. The same is true for intermediate voxels in a sphere (**Supplementary Movie S20**). When frequency increases enough for propagation to play a significant role, however, constructive and destructive interference comes into play, and symmetry-derived electric field cancellations become more difficult to coordinate across the body. At that point, the ideal current pattern begins to resemble the unperturbed OP pattern once again (see **Supplementary Movie S21**). **Figure 13** encapsulates some of these trends by plotting ultimate SNR efficiency as a function of field strength for an intermediate voxel (a, (x,y,z) = (0,10,0) cm) and an edge voxel (b, (x,y,z) = (0,19,0) cm) in our canonical 20 cm cylinder.

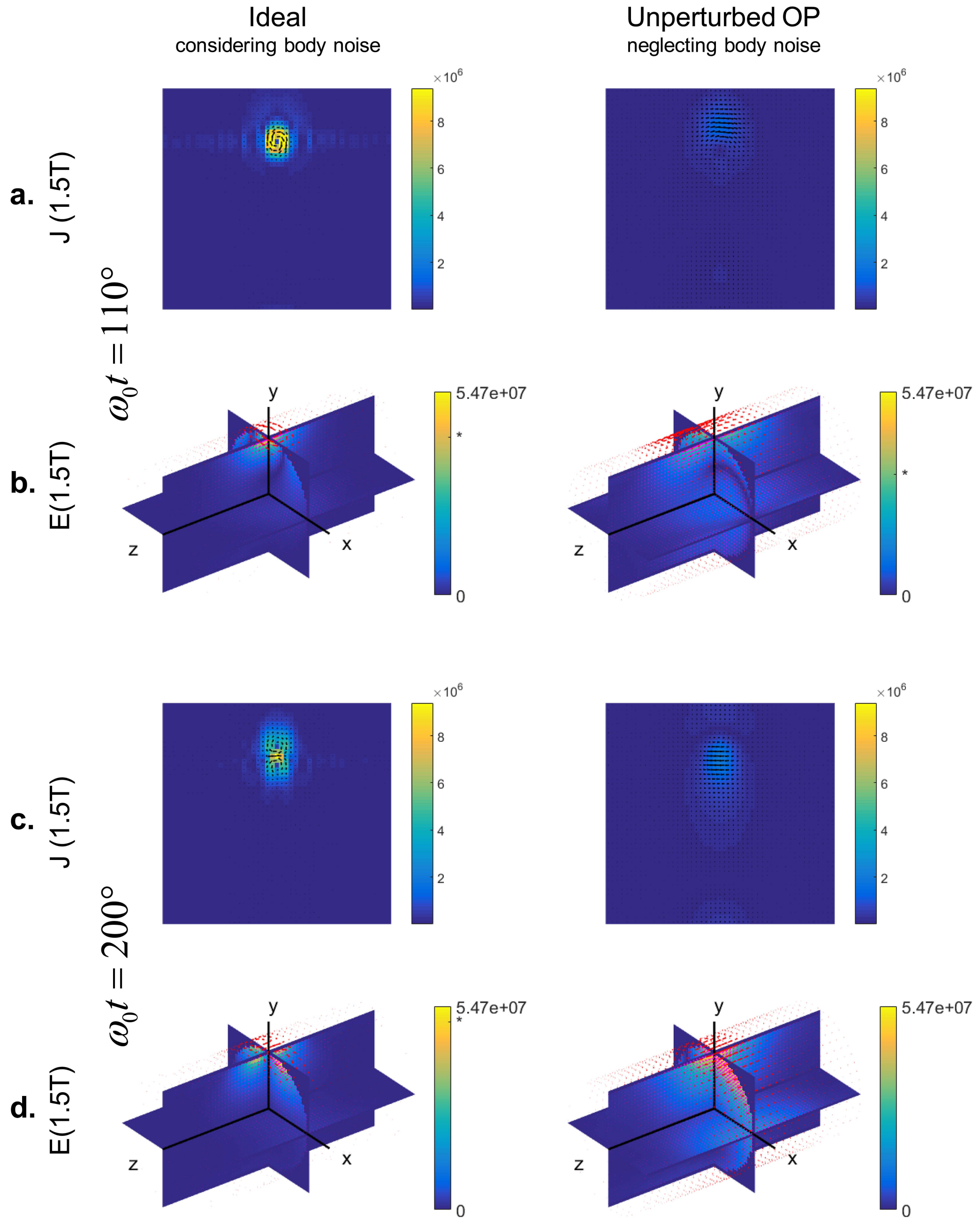

**Figure 11:** Current and field pattern snapshots for a voxel intermediate between the center and the edge of a 20cm-radius cylinder at 1.5T. Ideal patterns are shown on the left, and unperturbed OP patterns are shown on the right, plotted using the same color map. a) Unwrapped current patterns at $\omega_0 t = 110°$. b) Corresponding 3-plane electric field plots with superimposed current at $\omega_0 t = 110°$. c) Unwrapped current patterns at $\omega_0 t = 200°$. d) Corresponding 3-plane electric field plots with superimposed current at $\omega_0 t = 200°$. Whereas the unperturbed OP current patterns maintain non-closed characteristics in some phases, the ideal current patterns are tightly closed, alternating between tightly-packed loop and figure-of-eight configurations analogous to a surface quadrature coil, and thereby not only limiting conservative electric fields but also confining the fields to comparatively small regions of the cylinder.

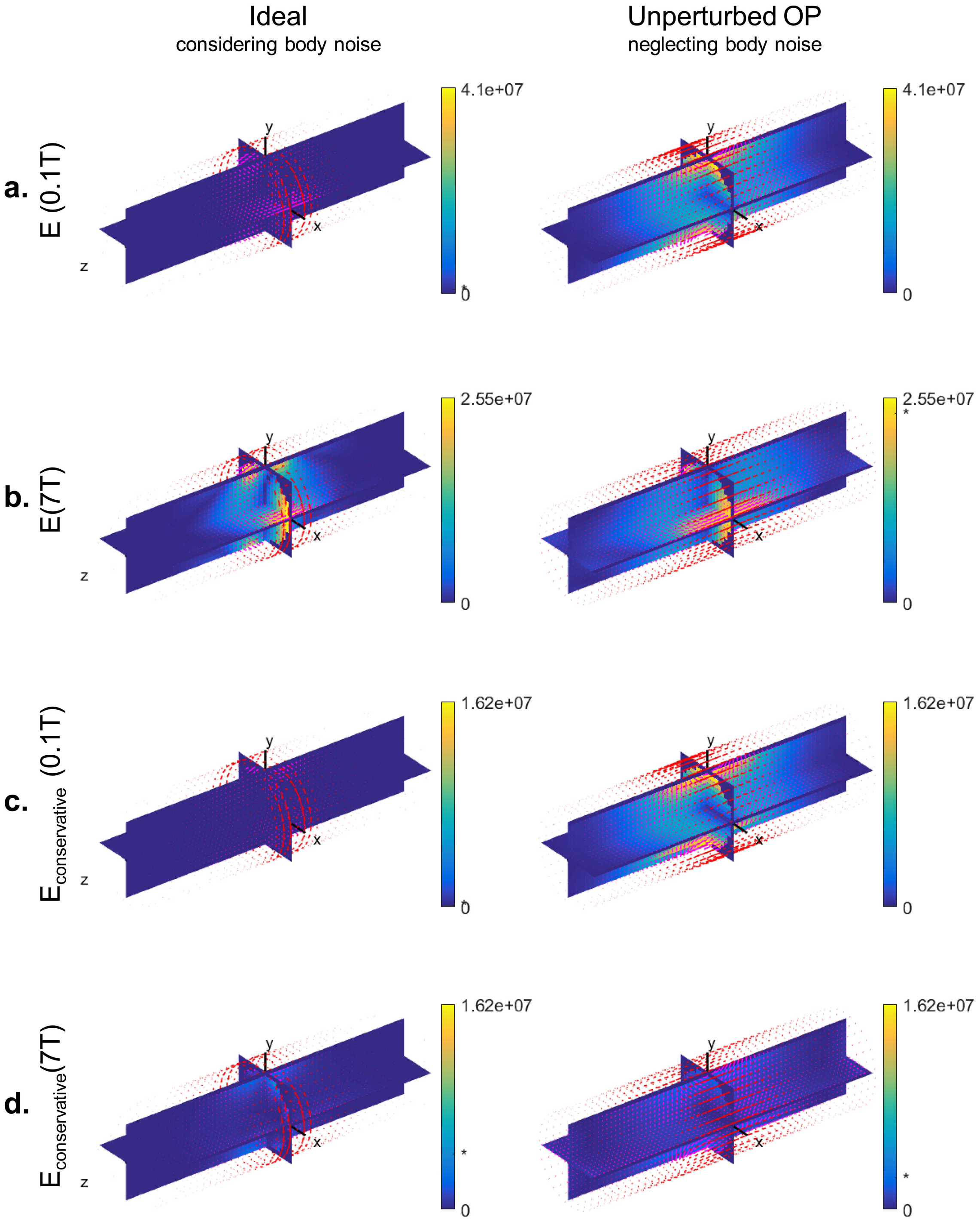

**Figure 12:** Electric field plots in the same format as for Fig 8, corresponding to ideal (left) and unperturbed OP (right) current patterns (shown as encircling red arrows), computed at a radial offset of 1cm from a narrow 5cm-radius cylinder, with properties otherwise the same as for the larger cylinder used in previous simulations (i.e. conductivity $\sigma$ = 0.4 S/m, relative permittivity $\varepsilon_{rel}$ = 39, cylinder length 125cm). a) Full electric field at 0.1T. b) Full electric field at 7T. c) Conservative component of electric field at 0.1T. d) Conservative component of electric field at 7T. As is the case for an intermediate voxel in a larger cylinder, ideal current patterns are closed, in order both to concentrate electric fields and to reduce conservative electric fields, which are otherwise particularly large near the current surface.

a.
b.

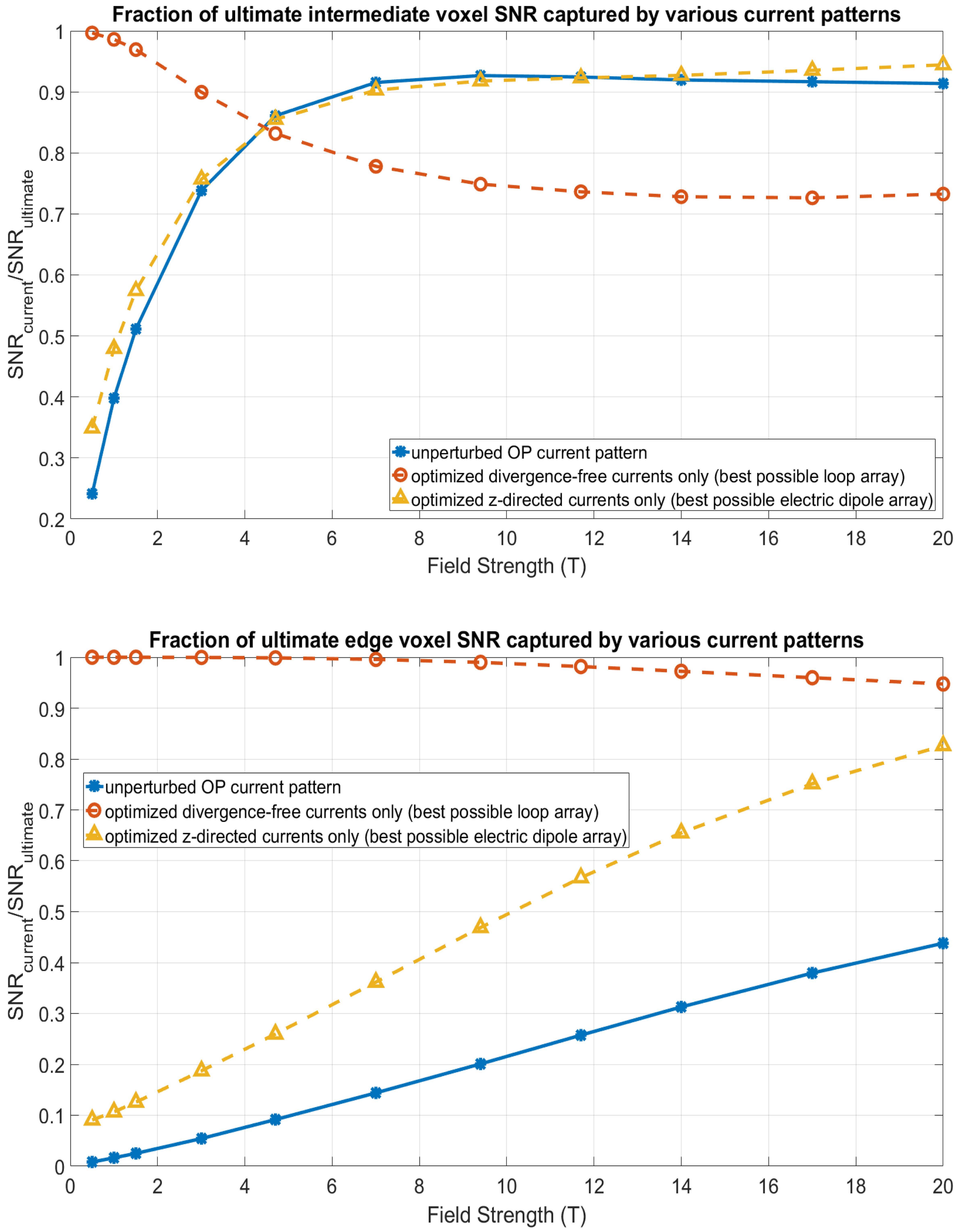

**Figure 13:** SNR contributions as for Fig. 9, but for (a) an intermediate voxel at (x,y,z) = (0,10,0) cm, and (b) an edge voxel at (x,y,z) = (0,19.5,0) cm.

**Figure 14** illustrates a first application of the Optimality Principle for nontrivial surface geometries and realistic body models. All our work in simple cylindrical and spherical geometries has shown that, except in exotic cases such as the traveling wave configuration in Mov. S18, unperturbed OP current patterns approach full SNR-optimizing ideal current patterns at high frequency. Therefore, we used a general purpose electrodynamic solver, CST Microwave Studio (Computer Simulation Technology, Darmstadt, Germany), to compute the electric fields created by a spin precessing at the 7T Larmor frequency in the midbrain of a human head model ("Duke," Virtual Population, IT'IS Foundation, Zurich, Switzerland (37), whole body model at 2 x 2 x 2 $mm^3$ resolution, truncated at the abdomen). These fields were then projected onto a surface composed of a cylinder capped by a half-sphere, to create unperturbed OP current patterns for a close-fitting dome configuration. Snapshots of the resulting patterns (which are animated in **Supplementary Movie S22**) show that the dome combines sphere-like and cylinder-like behavior. On the inferior portion of the cylinder, a chevron pattern dominated by z-directed currents may be discerned (notable for the diagonal blue band indicating the propagation-delayed edge of the chevron), whereas the currents on the hemispherical cap follow what would be closed trajectories on a complete sphere. Note that the computations are highly parsimonious for unperturbed OP – one need only compute the fields once throughout a volume of interest, then one can project these fields onto any surface of interest to derive the OP current pattern for that surface. In other words, computing patterns for a larger dome, or for a longer cylinder or ellipse or custom shape, is a trivial matter of reprojection. (For the full perturbation theory calculation, one must instead perform iterative field calculations based on projected surface currents, so a recomputation for each new surface of interest is required.) Note also that, with current patterns in hand, one can, with a single additional solver call, compute associated magnetic and electric fields in the body to quantitate SNR.

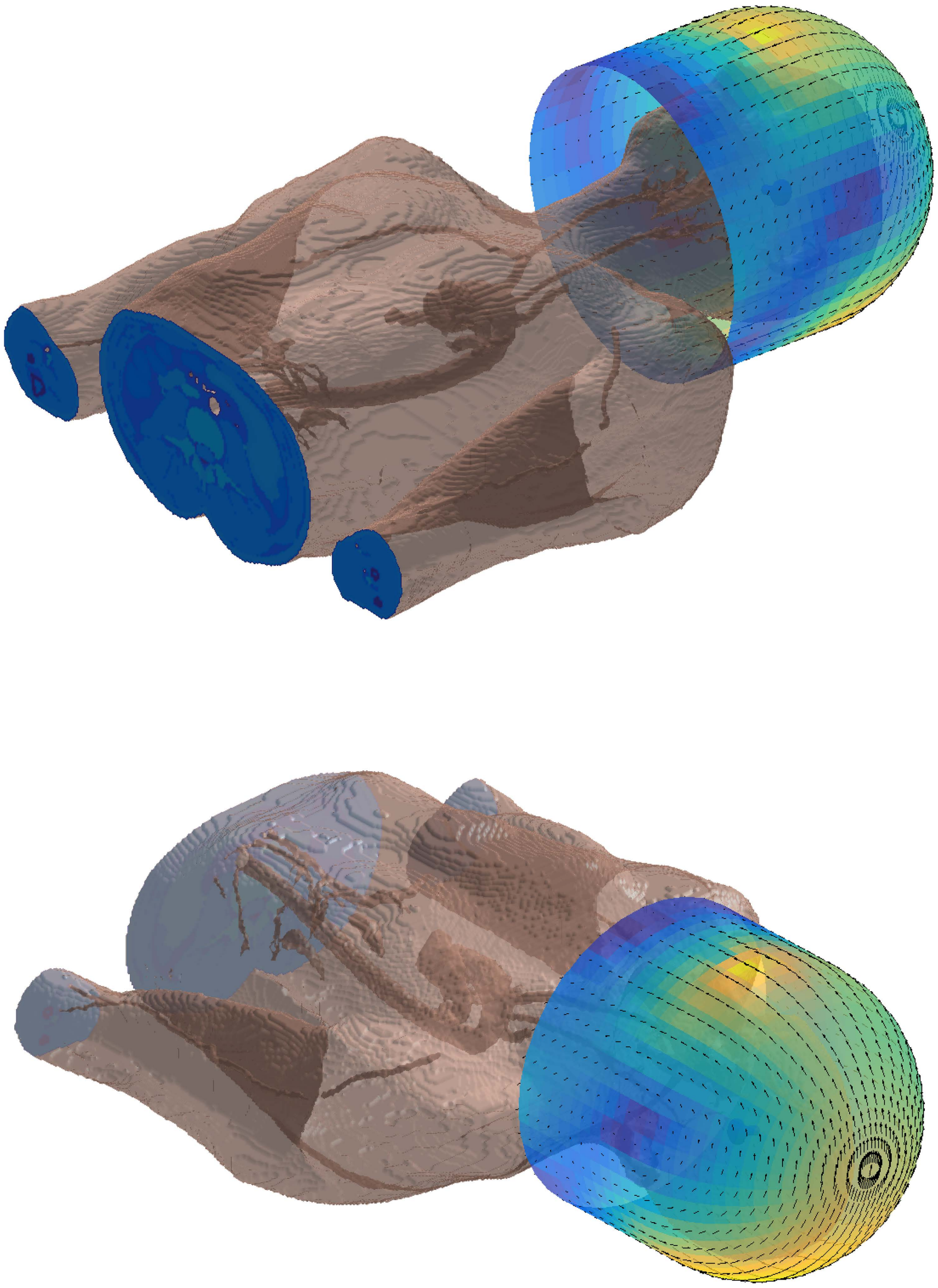

**Figure 14:** Application of the Optimality Principle in a human body model at 7T. Snapshots of the unperturbed OP current pattern are shown on a close-fitting domed surface surrounding the head, emulating the overall geometry of commonly-used helmet coils. Two different views highlight a non-closed chevon-shaped pattern on the cylindrical portion of the dome (top), and a circulating pattern on the spherical cap (bottom). The signal-optimizing current pattern was generated by computing electric fields created by a precessing spin in the midbrain of a Duke body model, and projecting those fields onto a surface composed of a cylinder with 13.5cm length and 13.5cm radius, capped by a hemisphere with the same radius. The body model was truncated at the abdomen, as shown in the figure. Both the current-carrying surface and the body surface are rendered with partial transparency, so as to enable visualization of the spin, shown as a blue arrow within the head. CST Microwave Studio was used for field calculation, as described in the Methods section.

## Discussion

By analogy with the Principle of Reciprocity, the Optimality Principle entails a shift in perspective from the process of signal reception to a reciprocal transmission process, and this shift results in both conceptual and computational shortcuts. Just as reciprocity enables one to understand SNR and receive coil sensitivities intuitively, and to compute them simply, so OP enables one to understand ideal current patterns intuitively, and to compute them simply.

The results highlighted in the previous section illustrate some of the scope of intuitive understanding afforded by the OP perspective and computational framework. Some of these results succinctly address questions that have heretofore been difficult to resolve, such as why electric dipoles perform as well as they do at high field strength. Other results illustrate potentially counterintuitive facts, such as the fact that loops don't necessarily optimize signal even at low field strength, or the fact that signal-optimizing current patterns likely depend more on surface geometry than on body shape or composition. Here we provide some additional commentary on these topics, followed by selected practical considerations and generalizations.

*Electric dipoles and the "emergence" of non-closed current patterns at high frequency*

The current patterns shown in Fig. 8 suggest that signal contributions from electric dipoles do not, in fact, "emerge" at high field strength. They are always present, at least in cylindrical geometries, but are masked by noise-optimizing dark modes at low field strength. These dark modes become ineffective for deep-lying regions at high frequency, and the non-closed signal-optimizing OP patterns dominate.

Indeed, the high-field regime is actually the more intuitive regime from the point of view of the Optimality Principle. It is low field that is, arguably, surprising, and is in any case more laborious to simulate.

*Surprises about familiar loop coils*

The current patterns in Fig. 8 and Fig. 11 also demonstrate that, contrary to some received wisdom, loops do *not* optimize signal in a cylinder: not at the center, not even closer to the periphery; not at low field strength, nor at high field strength. Why, then, do we so commonly use closed loop configurations, such as birdcage structures or surface loops, for our MR coils? They must be minimizing noise.

In fact, ideal current patterns on a cylinder at low frequency are not divergence-free by accident. Divergence-free patterns minimize electric field, as shown in Fig. 9. Meanwhile, for central points surrounded by a spherical surface, divergence-free patterns are in fact the sole contributors to signal (42), and any non-closed elements would only contribute extraneous noise.

*The importance of surface geometry*

It should not, perhaps, be surprising that the geometry of current-carrying surfaces, and the topology of currents on those surfaces, play such a significant role in determining ideal current patterns. Maxwell's equations, after all, are in many ways geometrical and topological equations.

In a cylindrical surface geometry, tangential electric fields from a central spin decay with distance along the cylinder axis, so that unperturbed OP current patterns for central points of interest appear largely z-directed, without appreciable axial return currents. In a spherical geometry, on the other hand, all points on the surface are equidistant from a central spin, and tangential electric fields generated by that spin are closed, creating loop-like OP current patterns at all field strengths. For off-center points of interest in either a cylindrical or a spherical geometry, the Optimality Principle predicts a time-varying mixture of closed and non-closed components, as illustrated in Fig. 11 and Supplementary Movies S3, S7, S19, S20, and S21.

In order to understand these mixed patterns, we may perform a simple thought experiment. As the point of interest approaches either a cylindrical or a spherical surface, that surface begins to appear locally flat. Let us, then, predict the signal-optimizing current pattern for a point at a distance from a flat planar surface – say an infinite half-plane. For such a surface, loops will be optimal when the spin is perpendicular to the surface, since the circularly-symmetric electric field lines of the spin will (after a suitable propagation delay) be parallel to the surface, and points on the surface equidistant from the spin will have the same tangential electric field amplitude. However, when the spin is parallel to the planar surface, a quarter of a cycle further along in its precession, for example, the circles defined by its electric field lines will be perpendicular to that surface. This means that the tangential fields will have no circulation whatsoever – they will, instead, be coaligned along the z direction, with a maximum where the field lines touch the plane, and with minima where the field lines traverse the plane orthogonally. The alternation between closed and non-closed patterns will continue throughout the precession cycle (and will be further complicated at sufficiently high operating frequencies by differential propagation delays to points on the plane at differential distances from the point of interest). This suggests that a combination of loops and z-directed dipoles is required to optimize signal for planar coils, such as spine coil arrays, or even for mildly flexed surface-hugging chest or abdomen arrays. While the imperatives of body noise minimization will tilt the balance toward closed loops for full SNR optimization, we may expect that the loop-dipole combination will still play an important role in optimizing SNR at high frequencies and for deep body regions.

It is clear from this example that the Optimality Principle enables intuitive predictions of coil performance even in the absence of detailed simulations (whose results, of course, will bear out these predictions). For example, one would predict that the use of an irregular current-carrying surface would also require combinations of closed and non-closed components.

Note that most of these predictions make reference only to the shape of the current-carrying surface, rather than to the shape or composition of the body to be imaged. The body, of course, does affect the ideal current pattern, by shaping the electric field generated by the putative spin precessing at the point of interest. To the extent that this perturbed field has a different projection onto the surface, the OP pattern will also be perturbed. However, surface geometry will, in general, have a greater effect than body shape, since different surface projections of the same field can look quite different, as highlighted above with the examples of cylindrical, spherical, and planar surfaces. There are only a few anticipated exceptions to this rule. First, body conductivity can change the relative attenuation of fields at a surface of interest. Since the source for OP calculations is always a point magnetic dipole, however, the usual geometrical falloff of field with distance applies, and additional conductivity-related attenuation may not be a dominant factor. Second, permittivity of the body may change the propagation delay of fields reaching the surface. Since the dielectric constant in tissue can be much larger than that in air, such delays have the potential to be significant. For example, an elliptical dielectric body may change the apparent symmetry of propagation delays, even for a surrounding cylindrical surface. Finally, additional propagation effects within a body may perturb projected fields – effects such as internal traveling waves or dielectric resonances (though these are expected to be supported only in bodies with very low conductivity), as well as reflections and refractions at air-tissue interfaces. In any case, the most significant perturbation away from simple OP patterns results from the integrated electric field defining the noise in a dielectric body, and both body shape and electrical property distribution will clearly affect this quantity.

*Other key determinants of SNR-optimizing ideal current patterns*

Unperturbed OP current patterns maximize signal at the point of interest per unit current on the surface (or, equivalently, minimize surface current for a unit $\mathcal{B}_1^{(-)}$ at the point of interest). Thus, unperturbed OP patterns represent the most parsimonious way to lay out conductor on a surface while maintaining a fixed signal sensitivity. These patterns will also optimize SNR when coil noise, or, more precisely, noise that scales with the quantity of current-carrying material, is dominant. Maximizing SNR in the presence of a conductive body introduces the additional imperative to minimize the quantity $\int_V d^3\mathbf{r}\sigma(\mathbf{r})\boldsymbol{\mathcal{E}}(\mathbf{r})\cdot\boldsymbol{\mathcal{E}}^*(\mathbf{r})$ integrated over the entire volume of the body. For body-noise-dominated cases, this imperative wins out over the pure parsimony of currents, and additional dark-mode currents enter into the patterns. In addition to surface geometry, three key determinants of the shape of resulting ideal current patterns have been identified in the studies presented here:

1. The overall extent of electric fields within the conductive body, which depends upon the extent of current patterns.
2. The magnitude of conservative electric fields resulting from current topology and charge separation.
3. The magnitude and velocity of propagating electric fields resulting from rapidly varying currents.

Let us consider each of these factors in turn.

1. *Extent of electric fields*

For points of interest near the surface, closing the current patterns tends to limit the extent of associated electric fields and, hence, the magnitude of the body noise. The use of closed patterns also tends to cancel electric fields along the central axis of circulating current components, by symmetry. Even in cases with some geometrical preference for closed current patterns at certain time points (such as for intermediate points within spheres or cylinders, or for points adjacent to flat planes), SNR optimization may call for reductions in the size of closed current patterns, as compared with signal-optimizing (i.e., unperturbed OP) patterns (43). When distances and/or operating frequencies become large enough, however, such monolithic electric field cancellation is foiled by wavelength effects (including spatially-dependent constructive and destructive interferences), as well as by field propagation.

*2. Conservative electric fields*

At sufficiently low operating frequency, closed patterns have benefits even for deep-lying points of interest far from the current-carrying surface. This is because not all currents are created equal when it comes to electric field generation at low frequency. Closed currents have no intrinsic charge separation and no conservative electric fields, whereas non-closed currents have significant conservative electric fields. All other things being equal, then, it behooves us to choose closed loop-like current elements, departing from the maximally parsimonious OP patterns, and thereby sacrificing some signal-reception efficiency, in order to reduce the integrated conservative electric field.

Note that, even for closed loops, the presence of lumped-element capacitors will result in some residual charge separation and conservative electric field in practice. The argument for using closed currents in SNR optimization is in fact similar to the practical arguments for distributing capacitance in RF coils: reduced charge separation results in reduced noise-generating conservative electric fields.

*3. Propagating electric fields (and connections to near-field versus far-field interpretations)*

For large objects at high frequency, propagating field components become increasingly dominant for any coil, since such components grow more rapidly with frequency and fall off more slowly with distance from the current-carrying surface than do conservative field components. In this case, the additional currents required to close the patterns would themselves result in new propagating electric field components, which on balance would overwhelm the modest benefits of reduced conservative electric fields.

Whereas conservative electric fields are largely localized, moreover, propagating fields can travel in multiple directions: not only radially within the body but also axially along the scanner bore, in the special case of cylindrical geometries. Axial propagation increases the extent of electric fields, and hence the body noise, with little benefit for localized signal detection, and these components are therefore suppressed in ideal current patterns. Note that the twisting of OP current patterns at high frequency, which is itself a result of propagation delay between the point of interest and the surface, naturally works against purely axial propagation and therefore improves SNR – yet another reason that unperturbed OP patterns approach the full ideal patterns at high field strength. Only in more complex cases, such as in the presence of a bore-lining RF shield at very high frequency, are the OP patterns perturbed more strongly, e.g. in order to cancel traveling-wave cavity modes explicitly (cf. Mov. S18).

The differential effects of conservative as opposed to propagating field components, loosely associated with "near field" and "far field" regimes, have been noted in the literature as an explanation for the frequency-dependent performance of certain RF coil designs such as the radiative dipole antenna (25). These arguments succinctly explain the performance of particular coil elements, but they become less clear when multiple elements are combined into large arrays, for which the balance of signal and noise is more difficult to visualize. The Optimality Principle allows direct application of related physical intuitions not just to individual coil elements, but to the ideal configuration as a whole.

*Dark modes*

The availability of distinct signal-optimizing and SNR-optimizing current patterns also affords new insight, and enables useful thought experiments. If the full ideal pattern differs substantially from the signal-optimizing unperturbed OP pattern, then we know that much of the design is devoted to reducing noise. The noise-reducing elements in ideal current patterns are captured specifically in the dark modes defined in Fig. 8.

Dark modes take advantage of the fact that signal is measured at a point of interest, whereas body-derived noise arises from the entire volume of the imaged body. There are, therefore, numerous degrees of freedom to cancel noise-associated electric field in the volume without affecting the localized magnetic field associated with signal. These degrees of freedom may be limited by symmetry, as in the case of central positions in spherical geometries, for which only a single noise mode survives (42), and dark modes vanish identically. They may also be limited by field propagation, which connects multiple regions and reduces degrees of freedom for noise cancellation at high frequency. In the absence of such limitations, however, dark modes may take various expected or unexpected forms. We have seen that they serve to close otherwise non-closed patterns at low frequency. They also serve to reduce the size of loop-like components in current patterns that optimize signal, say, in planar or gently-curved geometries. The effect of body noise in reducing the optimal radius of a simple loop coil is well known (44). Dark modes that accomplish this reduction in size for full ideal current patterns take the form of two concentric distributed rings with opposite current direction – one to cancel the larger signal-optimizing loop pattern, and the other to impose a smaller SNR-optimizing loop pattern within (45).

*Effects of sample electrical properties and surrounding dielectric materials*

How will spatial variations in tissue conductivity and permittivity affect dark modes and ideal current patterns? At the least, one might expect electric field to be steered away from body regions with high conductivity, in order to minimize integrated noise. If dielectric resonances or other property-dependent perturbations arise that might compromise the balance of signal and noise, one might expect them to be counteracted by suitable dark modes, as in the case of undesired traveling waves in Mov. S18.

Dielectric pads or other deliberately introduced high-permittivity materials have been shown to result in improved transmit and receive efficiency for a range of coil designs (46-50). Both the physical origin of this effect and the extent of gains to be expected may be probed using the Optimality Principle. For example, one may conveniently separate signal and noise effects by looking at changes in ideal and unperturbed OP current patterns (43,45). When it comes to signal, surrounding dielectric materials will perturb tangential electric fields on a current-carrying surface, and will also introduce additional propagation delays. From the standpoint of noise, surrounding dielectrics will confine electric fields in the vicinity of the surface, including conservative electric fields which fall off rapidly with depth, and will also modify the conditions for field propagation, both of which effects can shift the SNR balance. These features of dielectric materials will be investigated in future work.

*Practical considerations for RF coil design*

One additional benefit of ideal current patterns is that they may serve as concrete targets for RF coil design (18,51,52). As may be easily understood from an Optimality Principle perspective, however, such patterns are generally distributed continuously across any surface of choice, and the selection of discrete coil elements to match the target patterns is not always straightforward. The number of available channels is clearly an important practical consideration, since it sets the number of degrees of freedom available, either to match complex patterns or to accommodate multiple positions of interest. Likewise, the topology of individual elements (closed loops versus electric dipoles, or mixtures of the two) plays an important role. It is important to acknowledge, moreover, that not all coil configurations are straightforward to build, tune, decouple, etc. In future, it may be possible to incorporate some practical buildability constraints into a basis-independent perturbation theory framework like the one we have proposed here to account for body noise.

It is also worth noting that the spherical surfaces explored as simple cases in this work are of course an abstraction, since closed spheres are not practical for human imaging. Opening at least one end to allow for placement on a body will introduce distinct topological features, as is illustrated in Fig. 14, and as was also explored recently by Pfrommer and Henning (53).

*Generalization to multiple points of interest*

All of the results we have presented have involved a single point of signal optimization, but practical RF coils are generally designed with more than a single point in mind. The Optimality Principle is still of use in such situations. Either distinct subsets of elements may be devoted to matching distinct patterns separately computed for distinct points of interest, or else composite patterns may be constructed, as weighted combinations of pointwise ideal patterns.

*Generalization to multiple coil layers or volumes*

We have shown that the Optimality Principle applies regardless of the shape of a putative current-carrying surface. It continues to apply if coils are to be placed on multiple layered surfaces, or even if coils are allowed to occupy entire volumes of space. For multiple layers, one need only compute electric fields tangential to each surface of interest. (Formally speaking, the sum over surface current modes in the Theory section is simply extended to a larger set of modes spanning all layers.) For true current-carrying volumes, in which coils are not confined *a priori* to any particular surface, but can traverse space more freely, the ideal current pattern will conform directly to the direction of the precessing spin's electric field within each volume (corresponding to use of volumetric rather than surface current modes).

While the surface equivalence principle suggests that placing coils on multiple surface layers may not improve ultimate intrinsic SNR within the enclosed volume, such an approach may nevertheless yield current patterns that are convenient to emulate in practice, or else it may be useful in generating insight about existing coil designs. For example, the optimality of close-fitting RF shields may be examined directly by positing two concentric surfaces in an Optimality Principle experiment. Unperturbed OP current patterns will clearly have similar currents on both surfaces, if the surfaces are sufficiently close to one another. On the other hand, one would expect to see mirror currents on the outer surface in the full ideal pattern if a shield is, in fact, beneficial for SNR.

*Generalization to include radiation losses*

Speaking of shields, the impact of radiation losses on coil performance may merit a brief comment. As has already been made clear, field propagation, both within a body and in open space, is fully captured in the Optimality Principle, as well as in most of the electrodynamic solvers which might be used to compute OP current patterns. The predictions of the Optimality Principle should therefore be complete and should include the effects of radiation, as long as the conditions for validity of the Principle of Reciprocity continue to hold. One of these conditions requires that appreciable energy not escape from the scanner bore. In the presence of significant net radiative energy flux across the boundaries of a system, the Lorentz reciprocity theorem takes a more general form including surface radiation terms: $\mathbf{J}_1 \cdot \mathbf{E}_2 - \mathbf{E}_1 \cdot \mathbf{J}_2 = \nabla \cdot \left[ \mathbf{E}_1 \times \mathbf{H}_2 - \mathbf{E}_2 \times \mathbf{H}_1 \right]$. (Here, radiative corrections are shown in differential form as divergences of Poynting vectors.) It is straightforward to update Eq.'s [8] and [19] in the Theory section to include these surface terms, which would then appear as an additional factor perturbing ideal current patterns. However, scanner bores are generally surrounded not only by a dedicated RF shield, but also by a shielded room. Any significant outgoing radiation will be reflected back by these shields. In the absence of powerful absorbers in the scanner room, then, net radiative flux across any pertinent surface will be small (54). As a result, radiative perturbations to ideal patterns and ultimate SNR are expected to be negligible, even at comparatively high frequencies. Except in extreme cases, e.g. involving bore-guided traveling waves, it may not even be necessary to include all the relevant shields in the OP calculations.

*Generalization to accelerated parallel imaging*

The ideal current patterns explored so far have not included any accounting for parallel imaging acceleration. Our formulation of the Optimality Principle may be extended to accelerated cases, simply by replacing the coil sensitivity vector **S** in Eq. [21] with a suitable sensitivity matrix. For the case of regular Cartesian undersampling, it may then be shown that the ideal current pattern for any point of interest is simply a weighted combination of the ideal pattern for that point with the ideal patterns for each spatial position connected to that point by aliasing. This property has been noted previously for ultimate intrinsic SNR (55,56), and it holds true just as well for ideal patterns defined by the Optimality Principle.

*Generalization to the transmit case*

Note that a similar derivation as is outlined in the Theory section above may be performed to derive the ideal transmit current pattern, as opposed to the ideal receive current pattern. This derivation is included in Appendix B. In its unperturbed form, the OP transmit pattern results in maximum $\mathcal{B}_1^{(+)}$ at the point of interest per unit applied current on the surface. With perturbation by body-derived losses, the pattern minimizes global SAR while maintaining unit $\mathcal{B}_1^{(+)}$ at the point of interest. Since the resulting current pattern is to be used directly for transmission, the complex conjugation of field and current vectors, whose role required some care to justify in the receive case, is not called for in the transmit case.

*Additional interpretations and new classes of RF coils*

At the beginning of the theory section, we posited (in Eq. [3]) that the detected MR signal corresponds to the EMF measured in a coil as a result of the precessing magnetic dipole of the spin. Applying this perspective to the Optimality Principle results in an interesting potential interpretation of optimized current patterns: these patterns need not, in fact, represent currents at all. They are derived, after all, from tangential electric fields, and we know that integrating electric fields along a suitable path results in a measurable voltage. In other words, OP patterns may be understood directly as target patterns of EMF. In order to capture ideal SNR, we do not need conductors per se, but only suitable structures sensitive to electric field along the right paths. This observation may be particularly relevant in light of the recent introduction of new classes of high-impedance RF receive coils (57). Such coils may, perhaps, be understood as voltage-sensitive elements (as opposed to traditional current-carrying elements, to which high effective impedance is generally added through impedance transformation circuits connected to preamplifiers). High-impedance elements offer the tantalizing possibility of robust and flexible coil design without concerns of inductive coupling. They may also relax traditional constraints related, for example, to self-resonant current distributions along substantial lengths of conductor at high frequency, which can limit the degrees of freedom available to match any arbitrary target current pattern. With such high-impedance structures, one might be able to directly replicate SNR-optimizing OP patterns far more easily than with traditional conductors and lumped-element circuits. The behavior, and optimization, of new classes of coils like this constitutes a rich area for potential future investigation.

## Conclusions

In summary, the Optimality Principle formulated in this work offers the following concrete benefits:

- Simple calculation of signal-optimizing current patterns for any arbitrary surface geometry and body model (via a single call to an electrodynamic solver of choice)
- A perturbative formulation to compute SNR-optimizing ideal current patterns in the presence of body noise, also for any arbitrary surface geometry and body model (via a limited number of calls to an electrodynamic solver)
- Immediate physical intuition about the form of these current patterns
- Ability to separate signal and noise effects in coil optimization

A number of key understandings may be gleaned from the particular cases studied here:

- *Why electric dipoles perform so well at high field strength:* z-directed electric dipoles emulate the unperturbed OP current pattern, which approaches ideality for central positions in large cylinders at high field strength, because propagating rather than conservative electric fields dominate.
- *Why loops don't always optimize signal at low field strength:* For central points in cylindrical geometries, unperturbed OP current patterns are predominantly z-directed even at low frequency, due to the differential distance of points along the cylinder axis from the central point of interest. For more peripheral points and/or for locally flat surfaces, unperturbed OP patterns mix closed loop-like elements with non-closed z-directed elements. Rather than optimizing signal, then, loops minimize integrated electric field at low frequency by symmetry (i.e. by reducing conservative fields for large objects, or by direct cancellation for small objects).
- *Why ideal current patterns depend more on current-carrying surface geometry than on body shape or composition:* The Optimality Principle dictates that signal-optimizing current patterns are proportional to tangential electric fields, which depend strongly on the surface onto which the fields are projected, and only secondarily on body-specific effects such as differential attenuation, refraction, etc. Body-derived noise, and hence body shape and dielectric tissue content, represents an additional frequency-dependent perturbation.

The original goal of ideal current patterns was to provide rational absolute targets for coil design. The Optimality Principle makes those targets easier to access. At the same time, it yields fundamental physical insight into what features are most important in constructing the best possible coil for magnetic resonance.

## Acknowledgments

This work was supported in part by research grants from the US National Institutes of Health (NIH P41 EB017183, R01 EB002568, and R01 EB024536) and National Science Foundation (NSF CAREER 1453675), and was performed under the rubric of the Center for Advanced Imaging Innovation and Research (CAI$^2$R, www.cai2r.net), a NIBIB Biomedical Technology Resource Center (NIH P41 EB017183).

## Appendix A: Ideal current pattern computation as a scattering problem

In order to make the connection between the expression in Eq. [30] and traditional scattering problems, we may identify the free Green's operator $\mathbb{G}_0$ and the interaction potential $\mathbb{V}$ as

$$\begin{aligned} \mathbb{G}_0 &= \mathbf{P} \\ \mathbb{V} &= \left(\mathbf{1} - \tilde{\mathbf{G}}^{*}\bar{\boldsymbol{\sigma}}\tilde{\mathbf{G}}\right), \end{aligned} \qquad [36]$$

and we may write the full Green's operator, $\mathbb{G}$, via a Born series expansion of the Lippmann-Schwinger equation, as

$$\begin{aligned} \mathbb{G} &= \left[\mathbb{I} - \mathbb{G}_0\mathbb{V}\right]^{-1}\mathbb{G}_0 = \mathbb{G}_0 + \mathbb{G}_0\mathbb{V}\mathbb{G}_0 + \left(\mathbb{G}_0\mathbb{V}\right)^2\mathbb{G}_0 + \ldots = \left\{\sum_{m=0}^{\infty}\left[\mathbb{G}_0\mathbb{V}\right]^m\right\}\mathbb{G}_0 \\ &= \left\{\sum_{m=0}^{\infty}\left[\mathbf{P}\left(\mathbf{1} - \tilde{\mathbf{G}}^{*}\bar{\boldsymbol{\sigma}}\tilde{\mathbf{G}}\right)\right]^m\right\}\mathbf{P} \end{aligned} \qquad [37]$$

Substituting this operator into Eq. [30] yields the simple formal expression

$$\left|\boldsymbol{J}^{\text{ideal}}(\mathbf{r})\right\rangle \propto \mathbb{G}\left|\boldsymbol{\mathcal{E}}_{\text{spin}}(\mathbf{r})\right\rangle \qquad [38]$$

The zeroth-order Born approximation, which corresponds to taking $\mathbb{G} \to \mathbb{G}_0$ in Eq. [38], then results in unperturbed OP current patterns obtained by projection of the dipole field onto the surface. Additional iterations involve successive calls to an electrodynamic solver, as specified earlier. In other words, the simple zeroth order OP pattern is "scattered" by body noise contributions to yield the full ideal pattern.

In general, the speed and radius of convergence of Born series are known to be related to the eigenvalues of the operator $\mathbb{G}_0\mathbb{V}$. The well-established condition that this operator must have eigenvalues of modulus less than 1 is straightforward to meet. Since the noise covariance matrix $\boldsymbol{\Psi}$ is Hermitian, its eigenvalues are nonnegative. If we do a Taylor expansion around a suitable multiple *c* of the identity, i.e. $\boldsymbol{\Psi}^{-1} = \left[\mathbf{1} - \left(\mathbf{1} - \boldsymbol{\Psi}\right)\right]^{-1} = \left[c\mathbf{1} - \left(c\mathbf{1} - \boldsymbol{\Psi}\right)\right]^{-1} = c\mathbf{1} + \left(c\mathbf{1} - \boldsymbol{\Psi}\right) + \left(c\mathbf{1} - \boldsymbol{\Psi}\right)^2 + \ldots$, then we can guarantee a suitable eigenvalue spectrum in any chosen basis. The rate of convergence of the series under various particular conditions is explored in Fig. 11.

One of benefits of the scattering analogy is that we can bring to bear a wide range of known techniques for the generation of intuition, or for parsimonious computation (58). For example, we may represent subsequent expansion orders by Feynman diagrams. Moreover, we may consider modifying the scattering problem to reduce computational burden. We could, if desired, include analytic electrodynamics for simple geometries and uniform electrical properties (e.g. DGF operations) in the free Green's operator, and treat both body noise and deviations from simple body geometry as a joint perturbation, updating both effects in each iteration. Such an approach would take advantage of a common motivation for Born expansions – the fact that operation with full Green's operators is typically difficult or impossible, whereas the free Green's operator is selected to be more tractable. Such approaches will be the subject of future investigations.

## Appendix B: Ideal current patterns for optimal transmit efficiency

Eliminating the complex conjugation of the ideal current pattern in Eq. [18] yields

$$\left|\boldsymbol{\mathcal{E}}_{\text{spin, tangential}}(\mathbf{r})\right\rangle = \mathbf{P}\left|\boldsymbol{\mathcal{E}}_{\text{spin}}(\mathbf{r})\right\rangle = \sum_n \frac{\left|\boldsymbol{J}_n(\mathbf{r})\right\rangle\left\langle \boldsymbol{J}_n(\mathbf{r})\middle|\boldsymbol{\mathcal{E}}_{\text{spin}}(\mathbf{r})\right\rangle}{\left\langle \boldsymbol{J}_n(\mathbf{r})\middle|\boldsymbol{J}_n(\mathbf{r})\right\rangle}, \quad [39]$$

and

$$\begin{aligned}\left\langle \boldsymbol{J}_n(\mathbf{r})\middle|\boldsymbol{\mathcal{E}}_{\text{spin}}(\mathbf{r})\right\rangle &= -\left\langle \boldsymbol{\mathcal{E}}_n(\mathbf{r})\middle|\boldsymbol{J}_{\text{spin}}(\mathbf{r})\right\rangle = i\omega\left\langle \boldsymbol{\mathcal{B}}_n(\mathbf{r})\middle|\boldsymbol{\mathcal{M}}_{\text{spin}}(\mathbf{r})\right\rangle \\ &= i\omega M_0\left(\boldsymbol{\mathcal{B}}_n^*(\mathbf{r}_{\text{spin}})\cdot(\hat{\mathbf{x}}-i\hat{\mathbf{y}})\right) = 2i\omega M_0\left(\mathcal{B}_{1,\text{n}}^{(+)}(\mathbf{r}_{\text{spin}})\right)^*.\end{aligned} \quad [40]$$

A similar derivation as for the receive case yields the following result:

$$\left|\boldsymbol{J}_{\text{ideal}}^{\text{transmit}}(\mathbf{r})\right\rangle \propto \sum_n\sum_{n'}\left|\boldsymbol{J}_n(\mathbf{r})\right\rangle \Psi_{nn'}^{-1*}\left\langle \boldsymbol{J}_{n'}(\mathbf{r})\middle|\boldsymbol{\mathcal{E}}_{\text{spin}}(\mathbf{r})\right\rangle, \quad [41]$$

In this case, the matrix $\boldsymbol{\Psi}$ represents the power covariance matrix, sometimes denoted $\boldsymbol{\Phi}$, associated with transfer of power between transmit coils, which is equal, by reciprocity, to the noise covariance matrix. Note that the expression in Eq. [41] applies for maximizing transmit efficiency at a single point of interest. This is not a common use-case for transmit optimization, which usually aims for a particular multi-point transmission pattern (e.g. emphasizing homogeneity or selectivity), or else optimizes efficiency over extended regions. The special case with negligible body losses corresponds to maximizing pointwise transmit field per unit applied current, or per unit input power lost in the coils. The case including body losses corresponds to maximizing pointwise transmit field per unit global SAR.

## Supplementary Movies

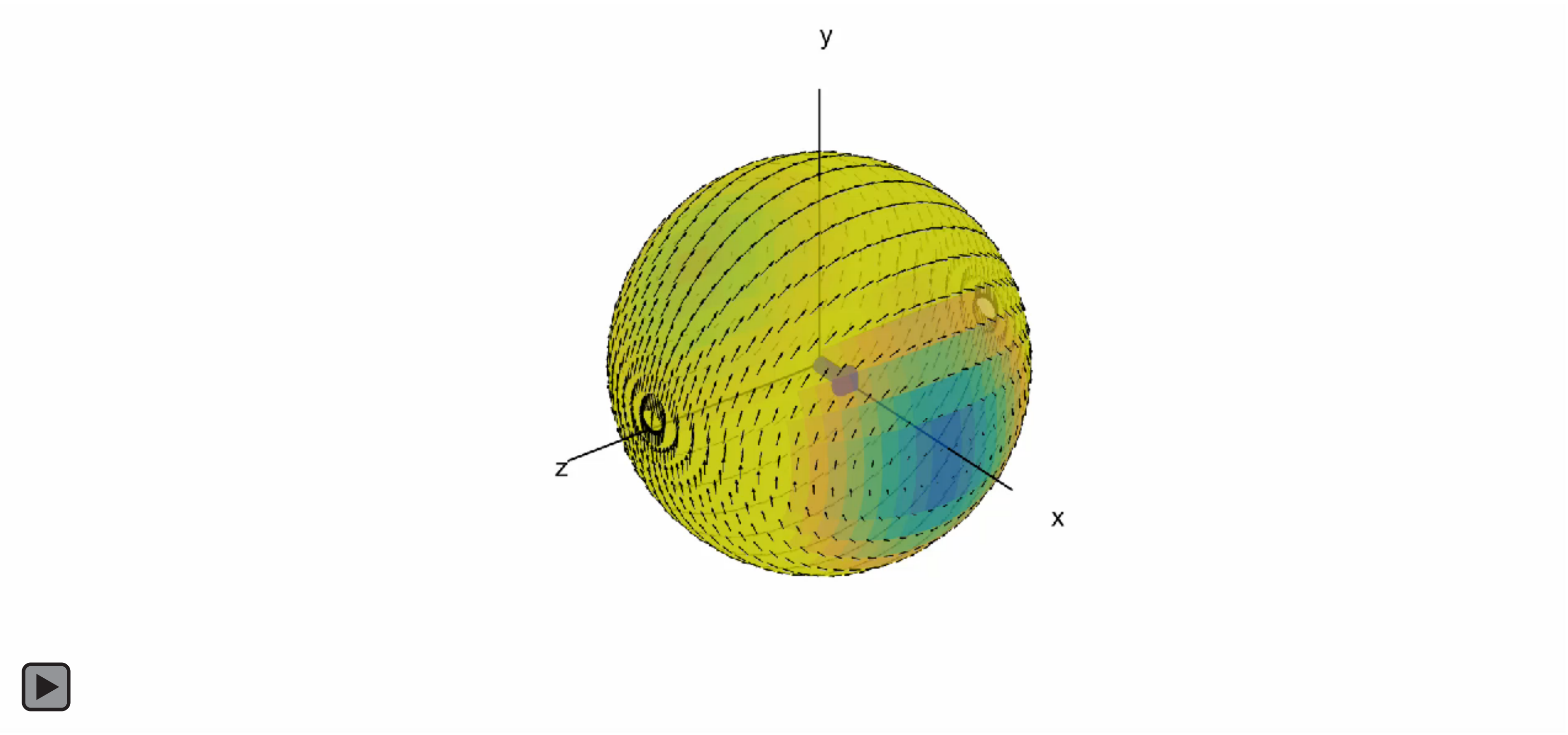

**Supplementary Movie S1:** Animation of a central spin precessing at 0.1T Larmor frequency, along with its tangential electric field pattern (corresponding to the signal-optimizing current pattern) on the spherical surface shown in snapshots on the right-hand side of Fig 2. Note the closed loop character of the signal-optimizing pattern, resulting from the fact that all points on the surface of the sphere are equidistant from the central spin position. This animation was created with DGF calculations using the default spherical object and surface properties identified in Methods (i.e. conductivity $\sigma$ = 0.4 S/m, relative permittivity $\varepsilon_{rel}$ = 39, dielectric sphere radius $\rho_{sphere}$ = 20cm, and spherical current-carrying surface radius $\rho_{surface}$ = 21cm).

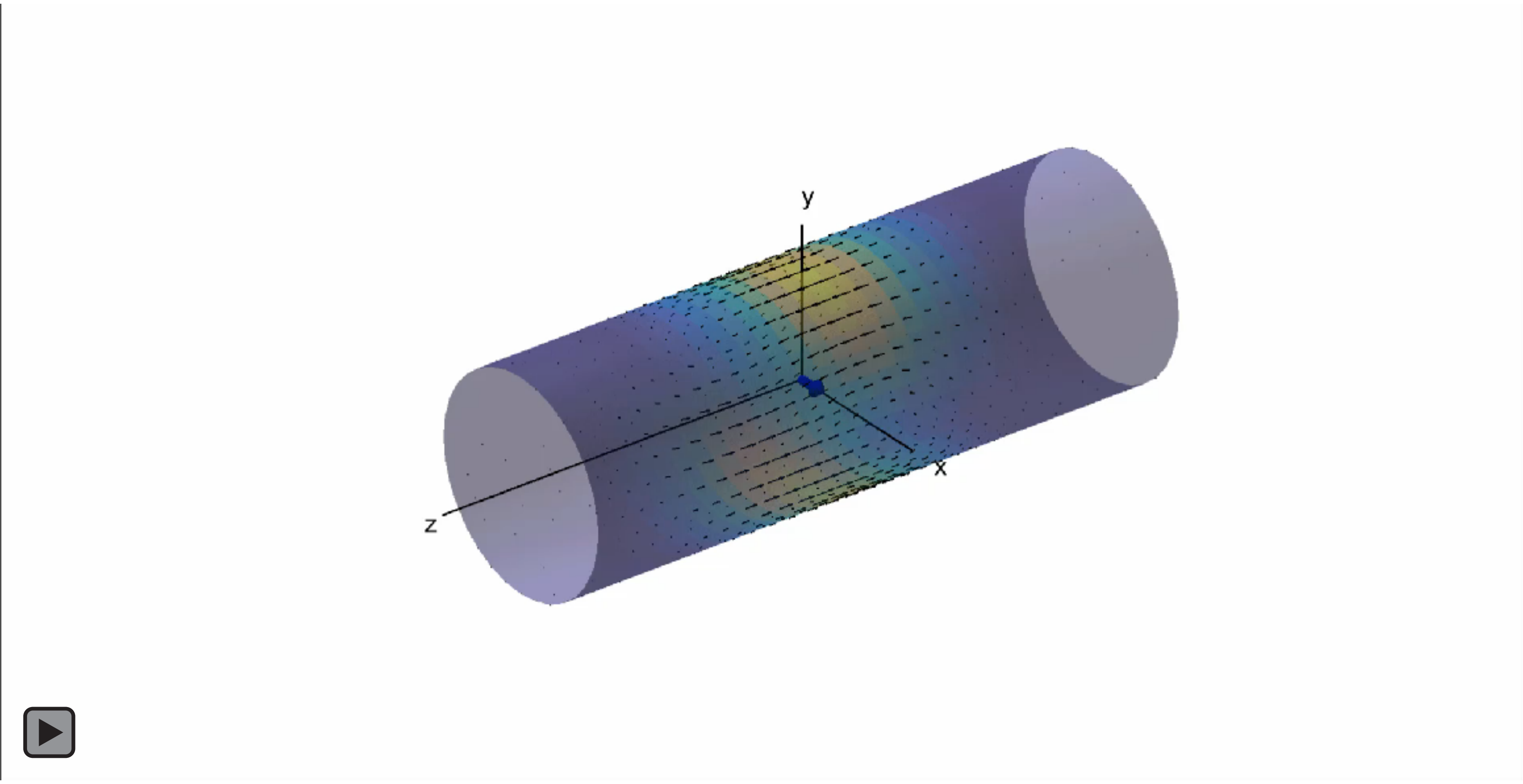

**Supplementary Movie S2:** Animation of a central spin precessing at 0.1T Larmor frequency, along with its tangential electric field pattern (corresponding to the signal-optimizing current pattern) on the cylindrical surface shown in snapshots on the left-hand side of Fig 2. Note the largely z-directed, non-closed character of the signal-optimizing pattern, resulting from attenuation of tangential electric fields with increasing distance from the center. This animation was created with DGF calculations using the default cylindrical object and surface properties identified in Methods (i.e. conductivity $\sigma$ = 0.4 S/m, relative permittivity $\varepsilon_{rel}$ = 39, cylinder length 125cm, cylinder radius $\rho_{body}$ = 20 cm, and cylindrical current-carrying surface radius $\rho_{surface}$ = 21 cm).

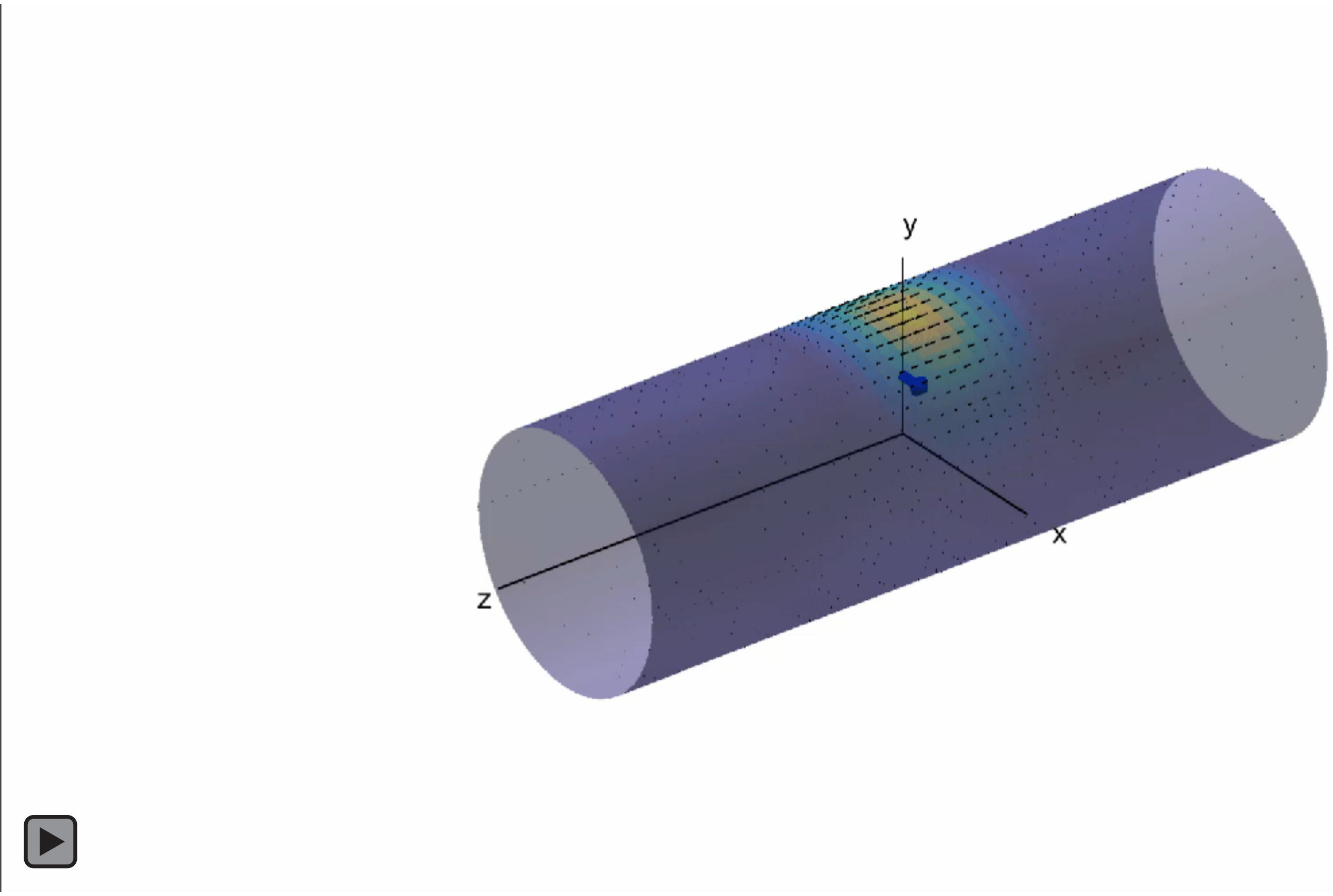

**Supplementary Movie S3:** Similar animation for a spin location (and a corresponding point of interest for signal optimization) intermediate between the center and the surface of the cylinder, i.e. at position (x, y, z) = (0, 10, 0) cm. Note that the z-directed character of the current pattern is still pronounced, but some more circulatory behavior appears at certain temporal phases, and the pattern is now more localized to the side of the cylinder closest to the spin. Two sample snapshots of this animation are shown in Fig. 3.

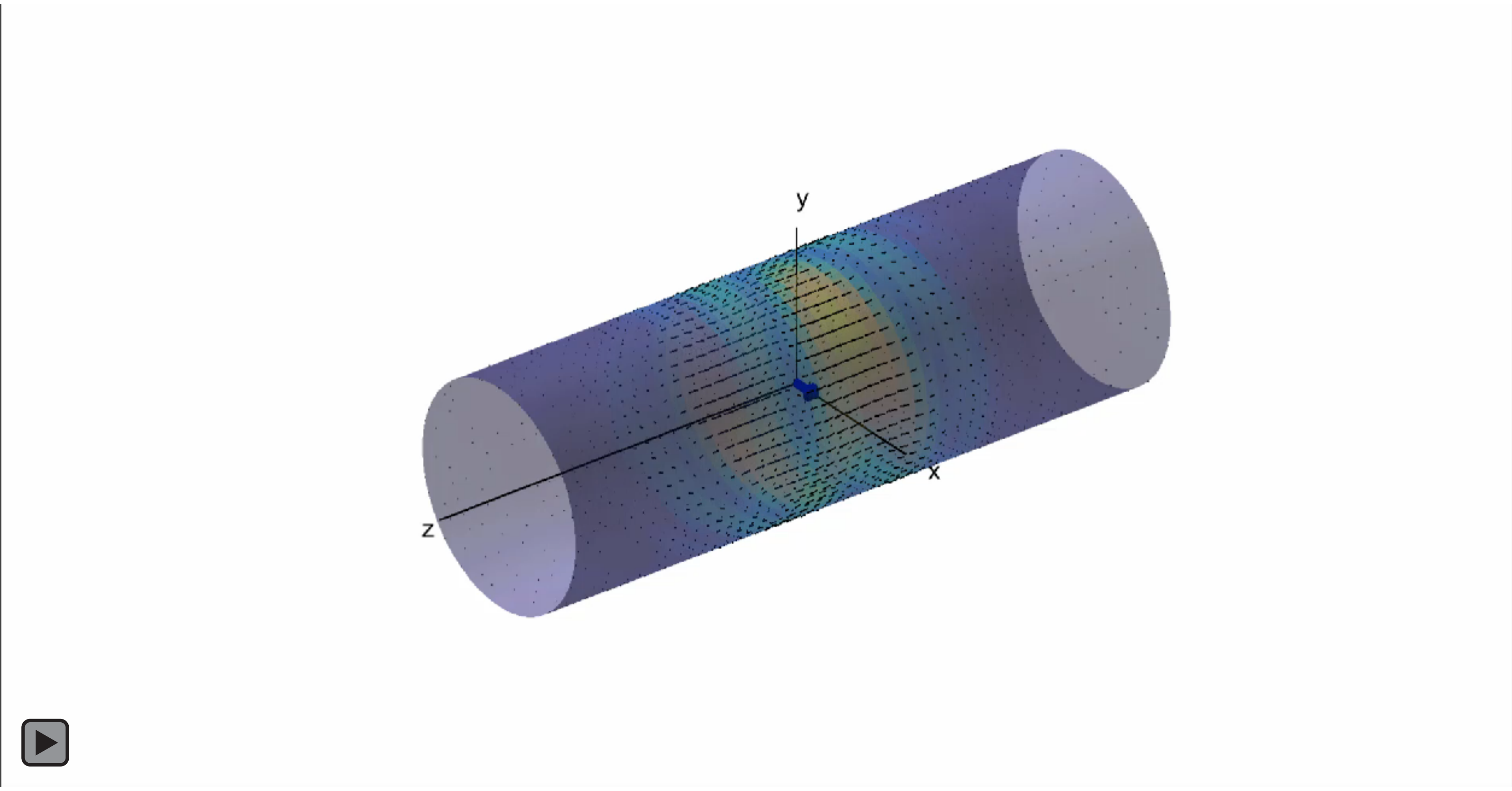

**Supplementary Movie S4:** Animated spin and signal-optimizing current pattern for a central position at 7T Larmor frequency (with simulation parameters otherwise the same as for Movs. S2 and S3). z-directed character is maintained, but note the distortion of the pattern at ±z, as compared with low-field patterns, resulting in a characteristic “chevron” structure. This distortion is a result of differential relativistic time delay associated with the spin’s field reaching different locations on the surface. At the high frequency (298.1 MHz) associated with 7T field strength, the spin has already precessed appreciably by the time the electric field reaches the more distant regions on the surface. Two sample snapshots of this animation are shown in Fig. 4.

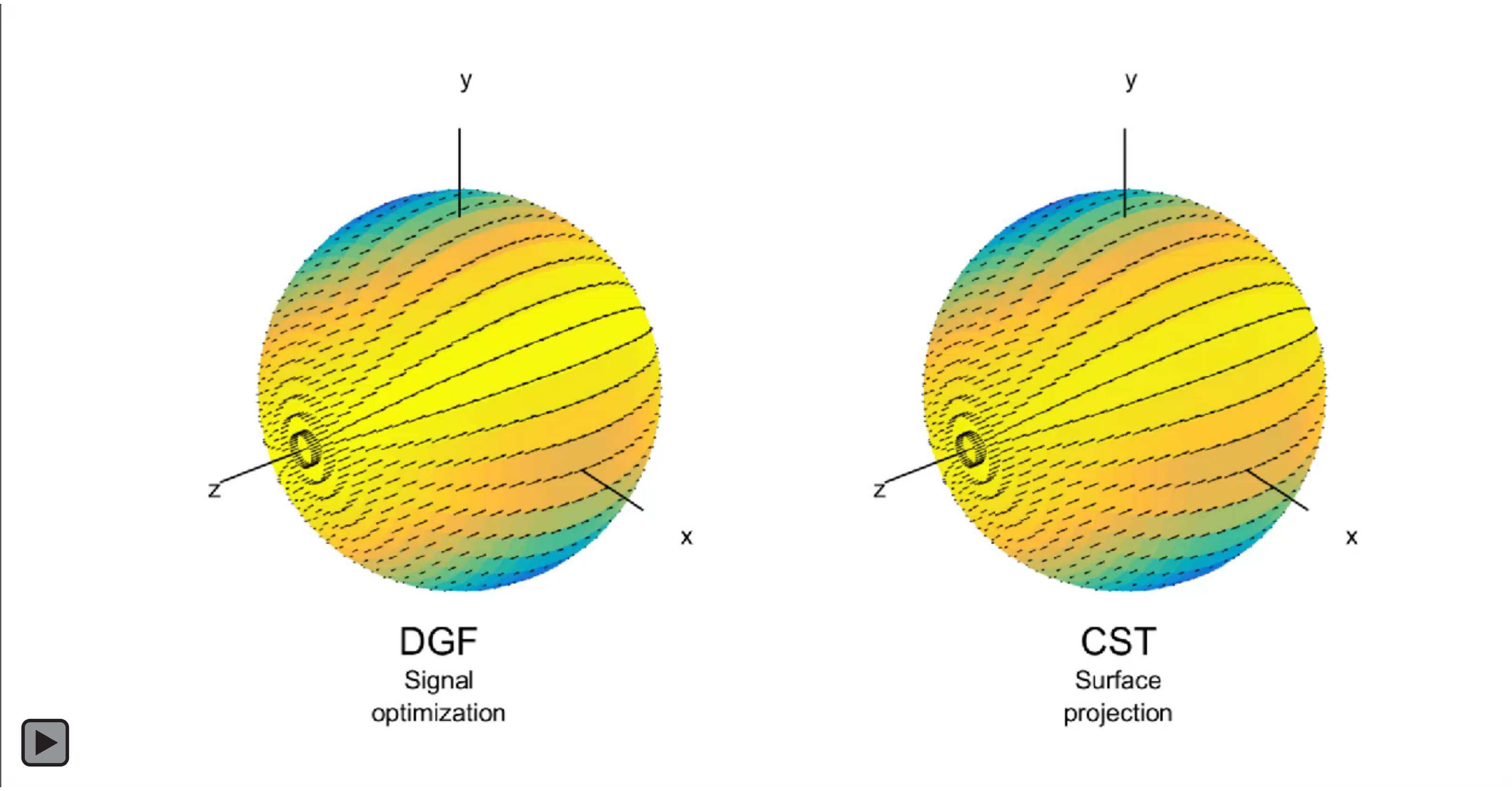

**Supplementary Movie S5:** Animated version of Fig. 5, showing patterns precessing at the 1.0T Larmor frequency.

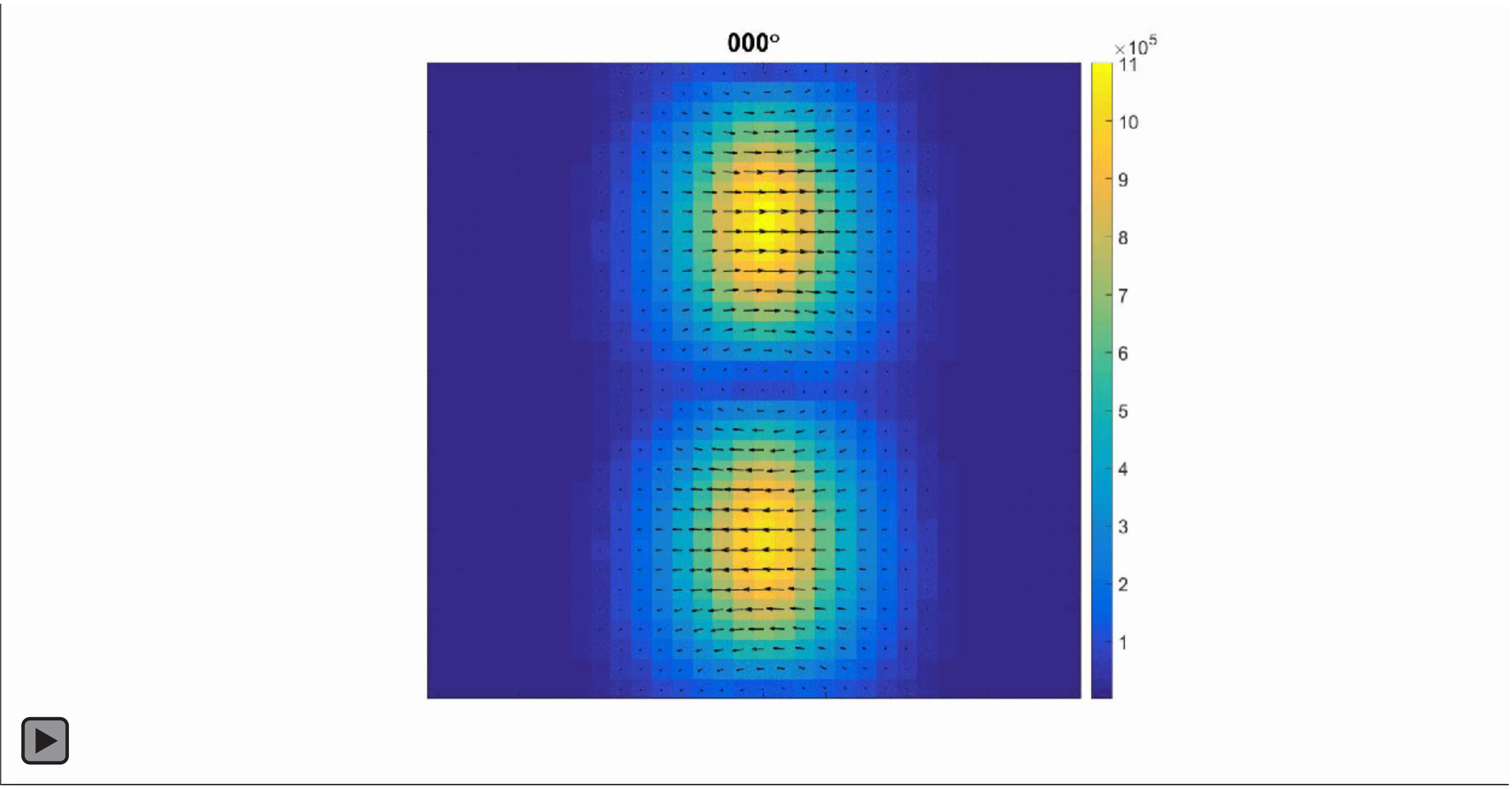

**Supplementary Movie S6** Animation of the unwrapped pattern from Fig. 7 (i.e., the signal-optimizing pattern for a central position at 0.1T, corresponding to the pattern shown in Mov. S2). Precession of the pattern around the cylinder appears in this format as a cyclical upward motion within the two-dimensional frame.

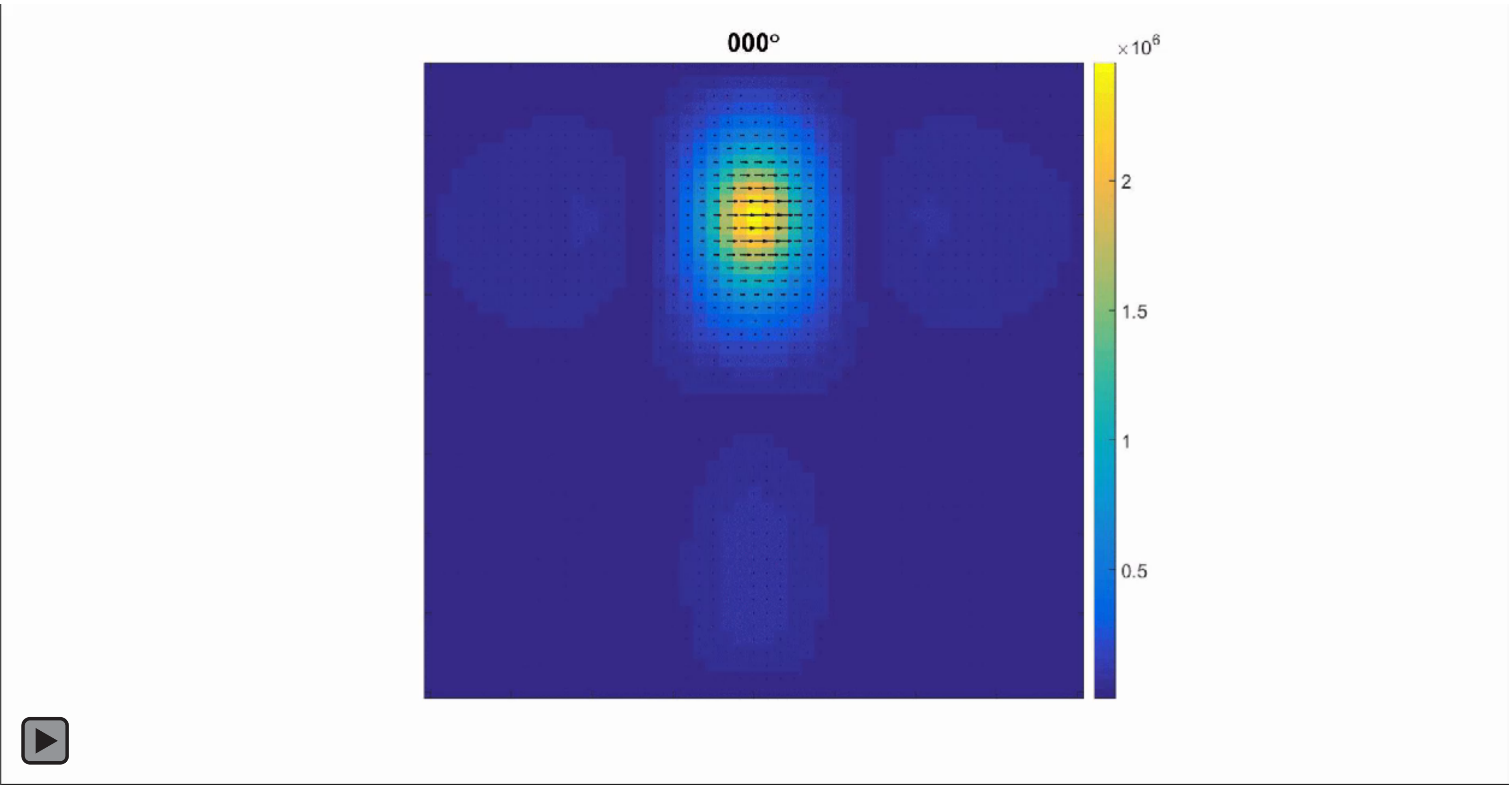

**Supplementary Movie S7:** Animation of an unwrapped signal-optimizing current pattern for an intermediate position at 0.1T, corresponding to the pattern shown in Mov. S3.

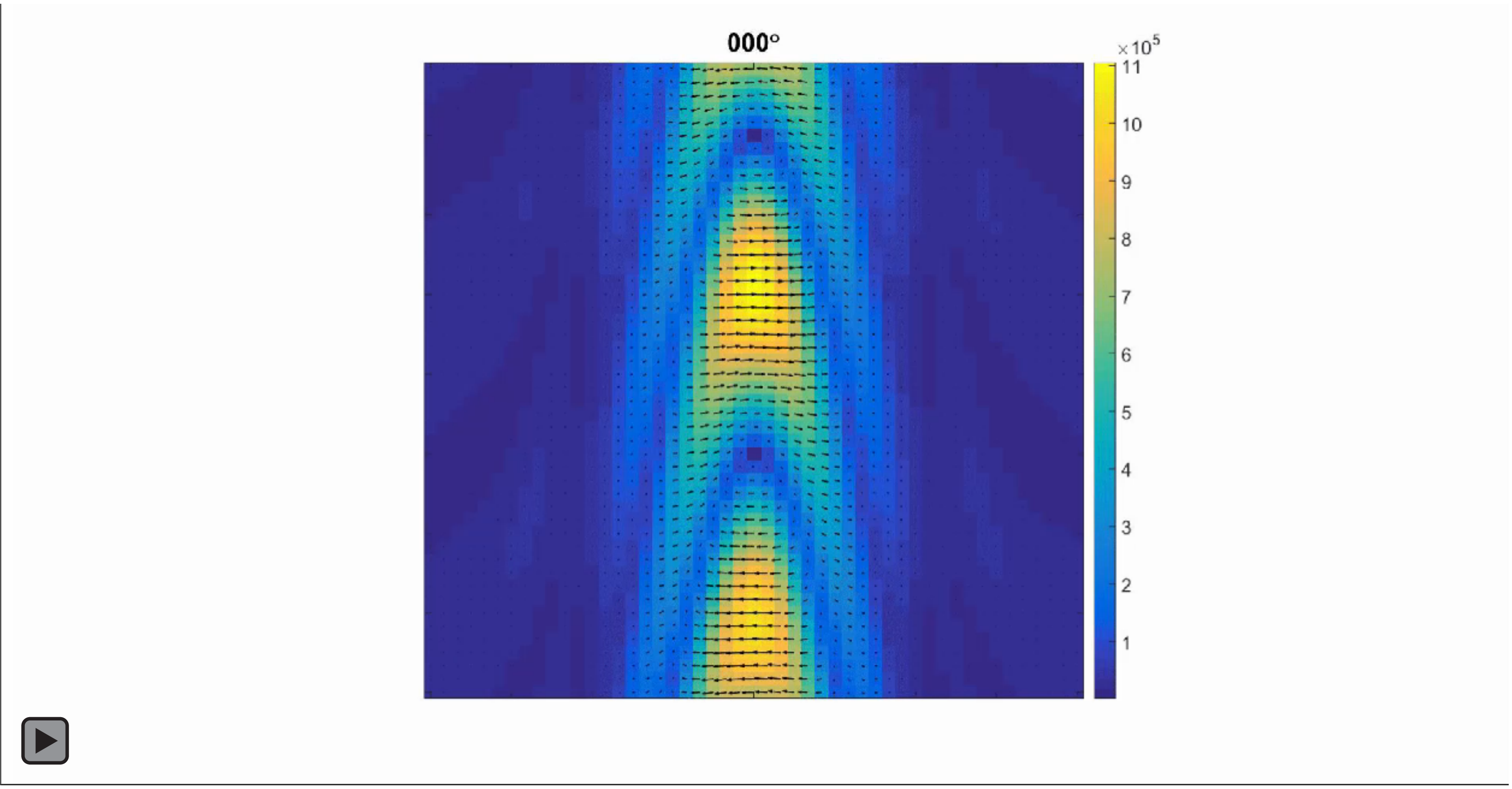

**Supplementary Movie S8:** Animation of an unwrapped signal-optimizing current pattern for a central position at 7T, corresponding to the pattern shown in Mov. S4.

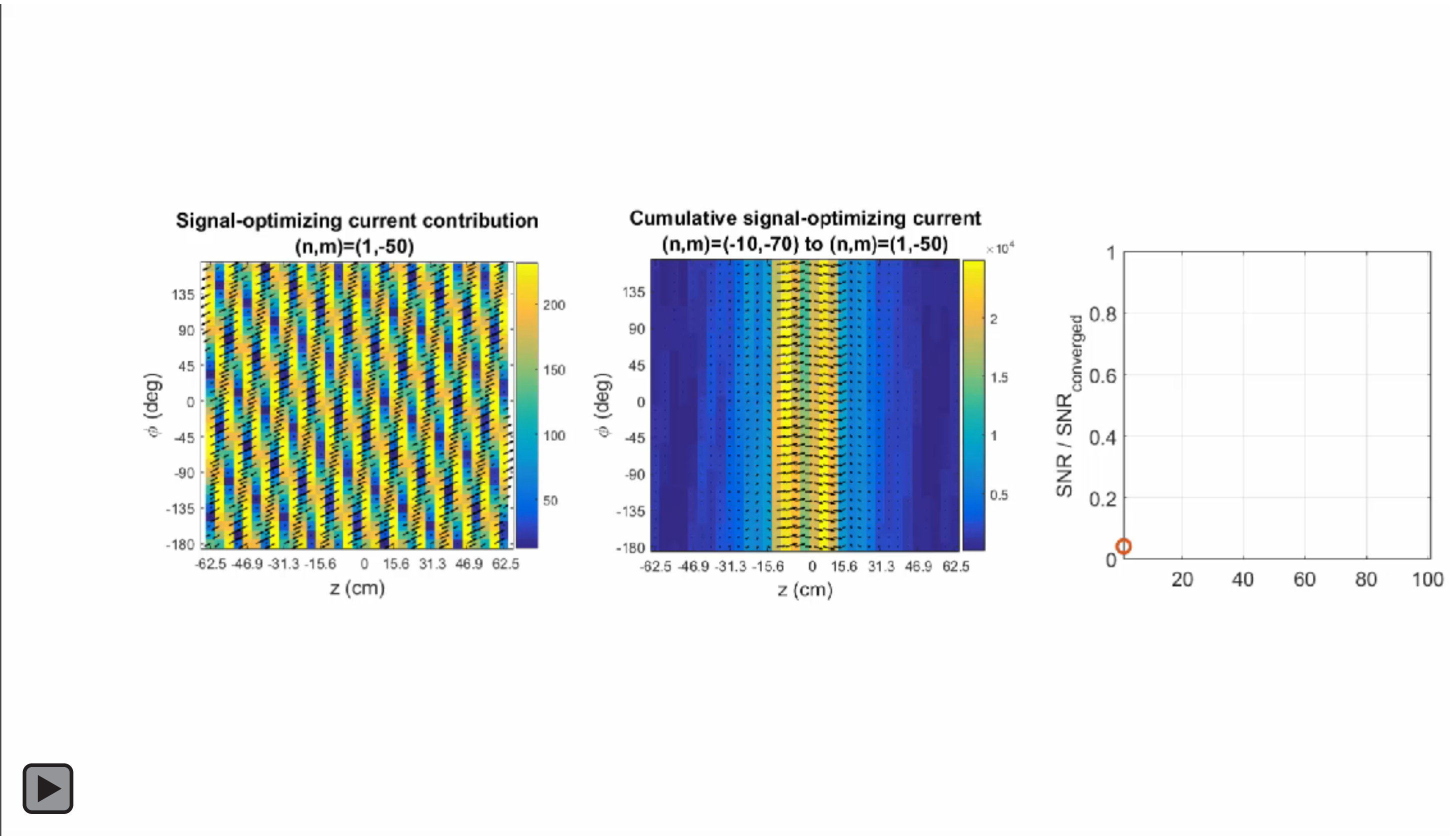

**Supplementary Movie S9:** Animation of convergence of the unwrapped current pattern from Mov. S8, with progressive addition of weighted contributions from various surface current mode orders. Left: snapshot of the current pattern associated with each cylindrical harmonic mode in turn. Modes are identified by circumferential harmonic order *n* and axial harmonic order *m.* Center: cumulative current pattern resulting from the weighted combination of all modes in the sequence so far. Colormap is autoscaled for the maximum current amplitude at each point in the sequence. Right: Fraction of the converged SNR achieved so far in the sequence.

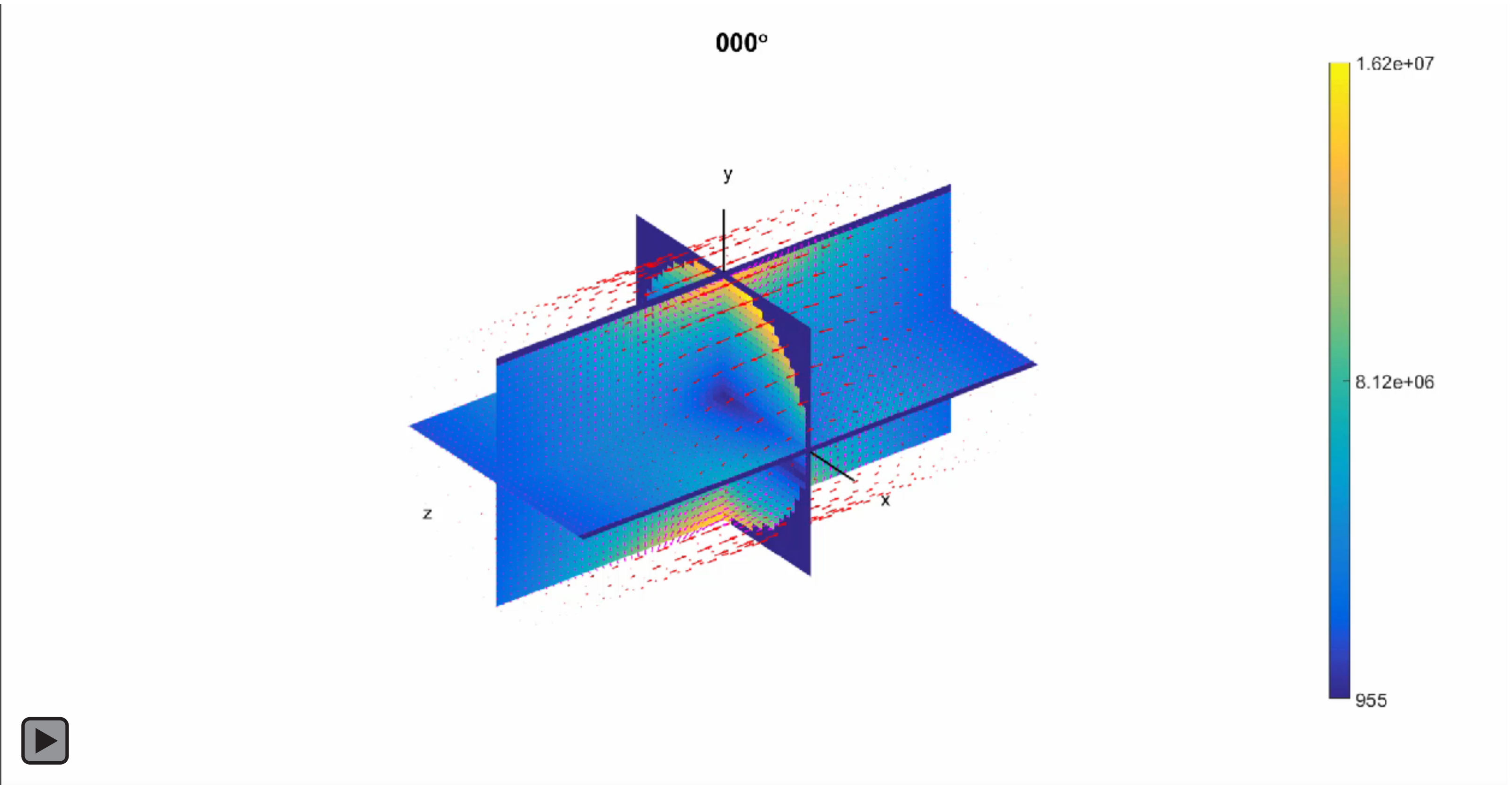

**Supplementary Movie S10:** Animation of the electric field pattern at 0.1T, associated with the snapshot on the right-hand side of Fig. 9a, with the corresponding unperturbed OP current pattern (associated with the unwrapped snapshot from the center of Fig. 8a) shown as red arrows encircling the field plot.

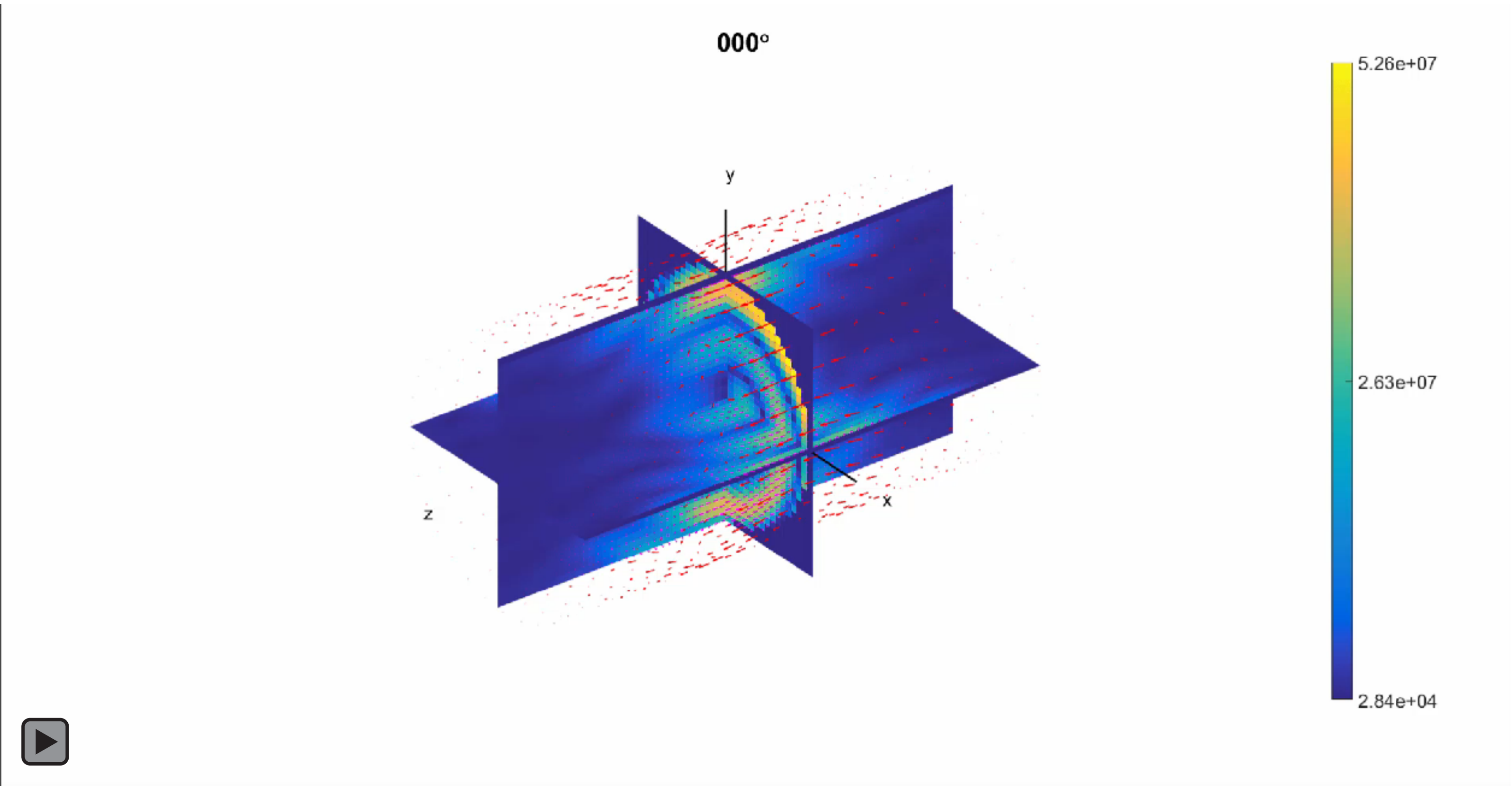

**Supplementary Movie S11:** Animation of the electric field pattern at 7.0T, associated with the snapshot on the right-hand side of Fig. 9b, with the corresponding unperturbed OP current pattern (associated with the unwrapped snapshot from the center of Fig. 8d) shown as red arrows encircling the field plot. Note the spiral twisting pattern associated with field propagation outward towards the current-carrying surface as the entire pattern precesses at high frequency.

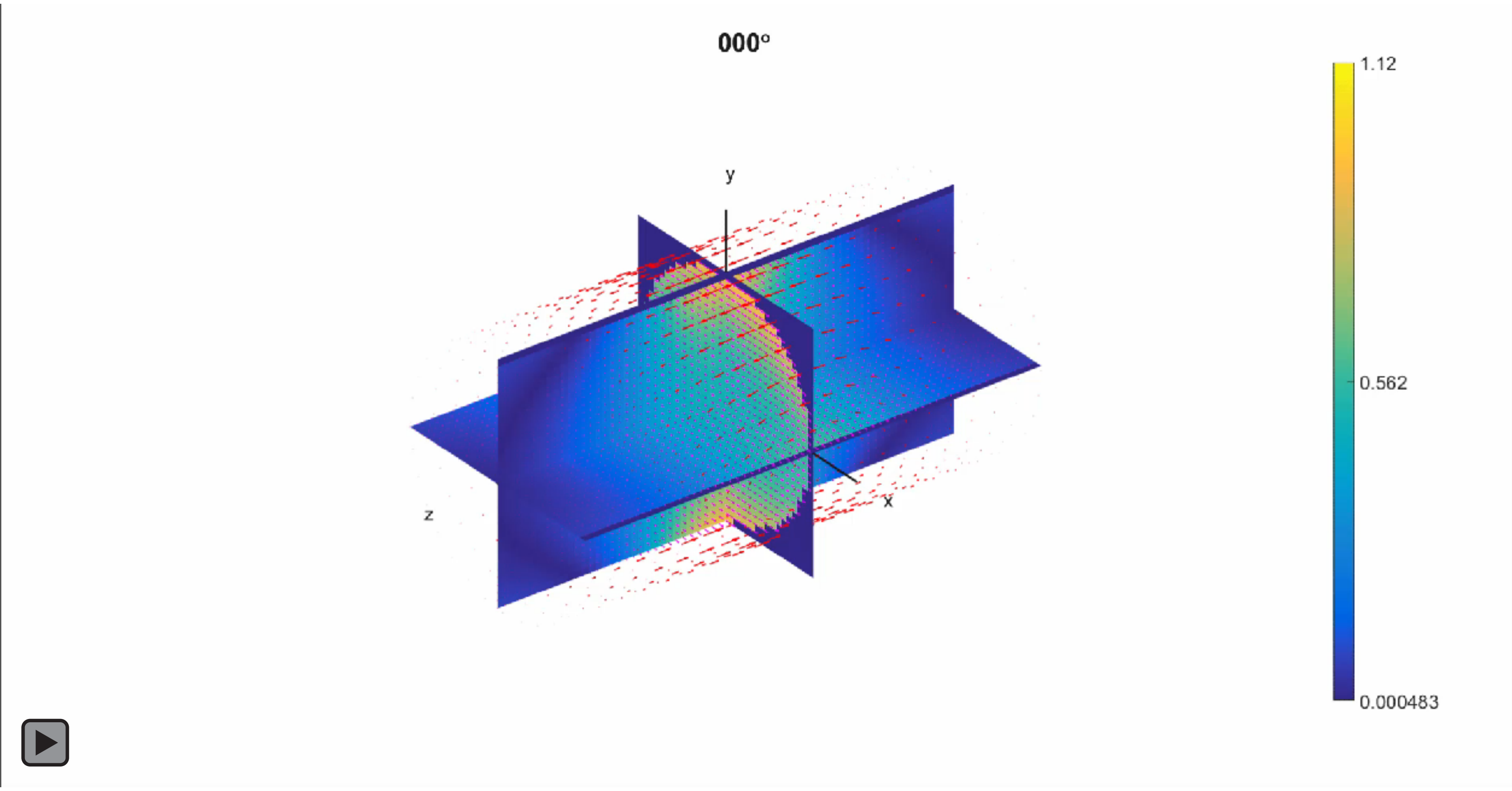

**Supplementary Movie S12:** Animation of the magnetic field pattern at 0.1T, associated with the electric field in Mov. S10. Note the precessing character of the magnetic field, which assumes a unit amplitude at the central position of interest, corresponding to the prescribed value of $B_1^{(-)} = 1$.

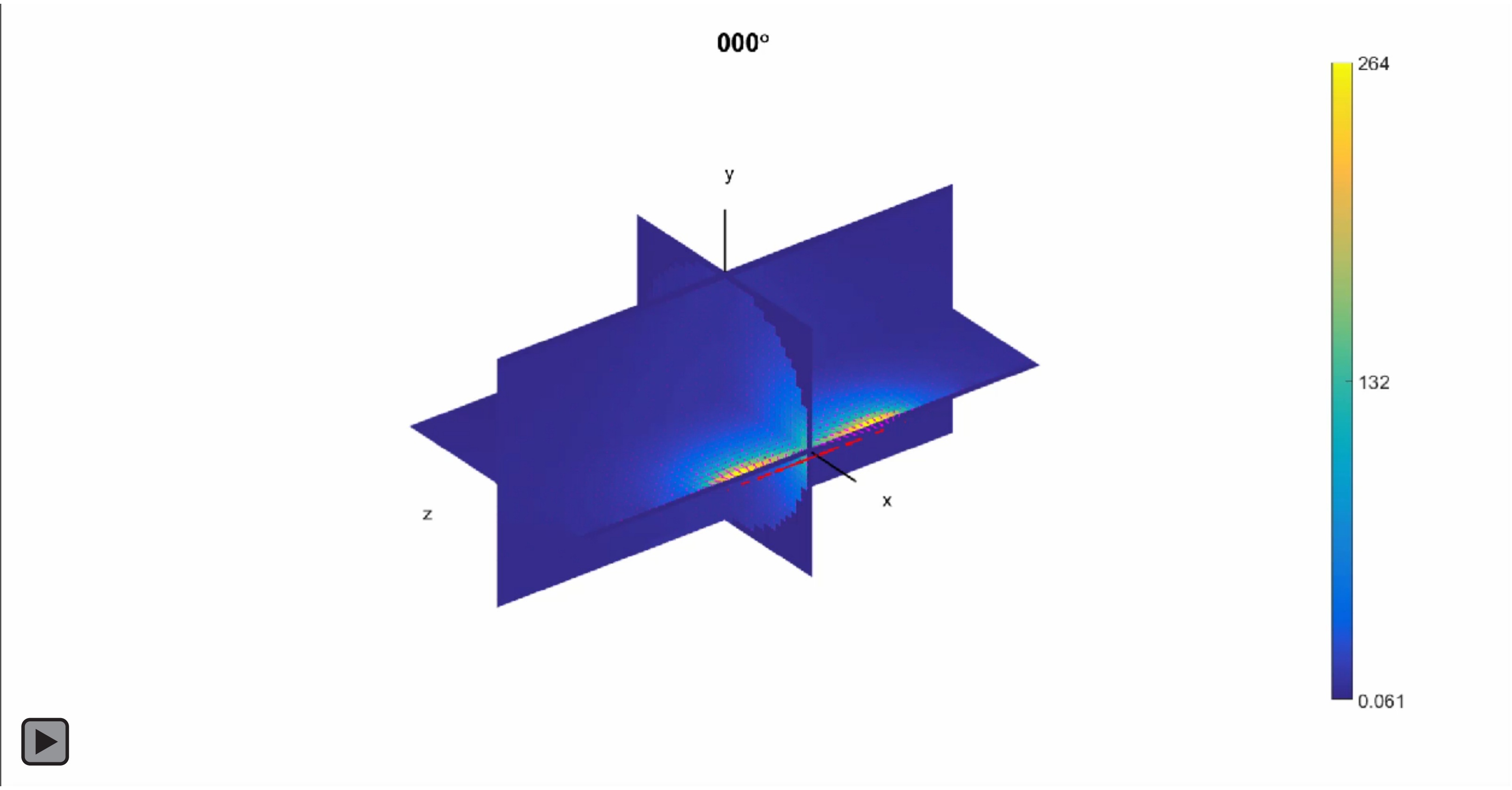

**Supplementary Movie S13:** Animation of the electric field pattern for a single filamentary z-directed electric dipole adjacent to the same cylinder at 0.1T. The current of the dipole is shown as red arrows that reverse direction over time, following a sinusoidal amplitude variation typical of a center-fed dipole antenna. The field assumes a classic quasistatic electric dipole field pattern, reversing direction in sync with the dipole current.

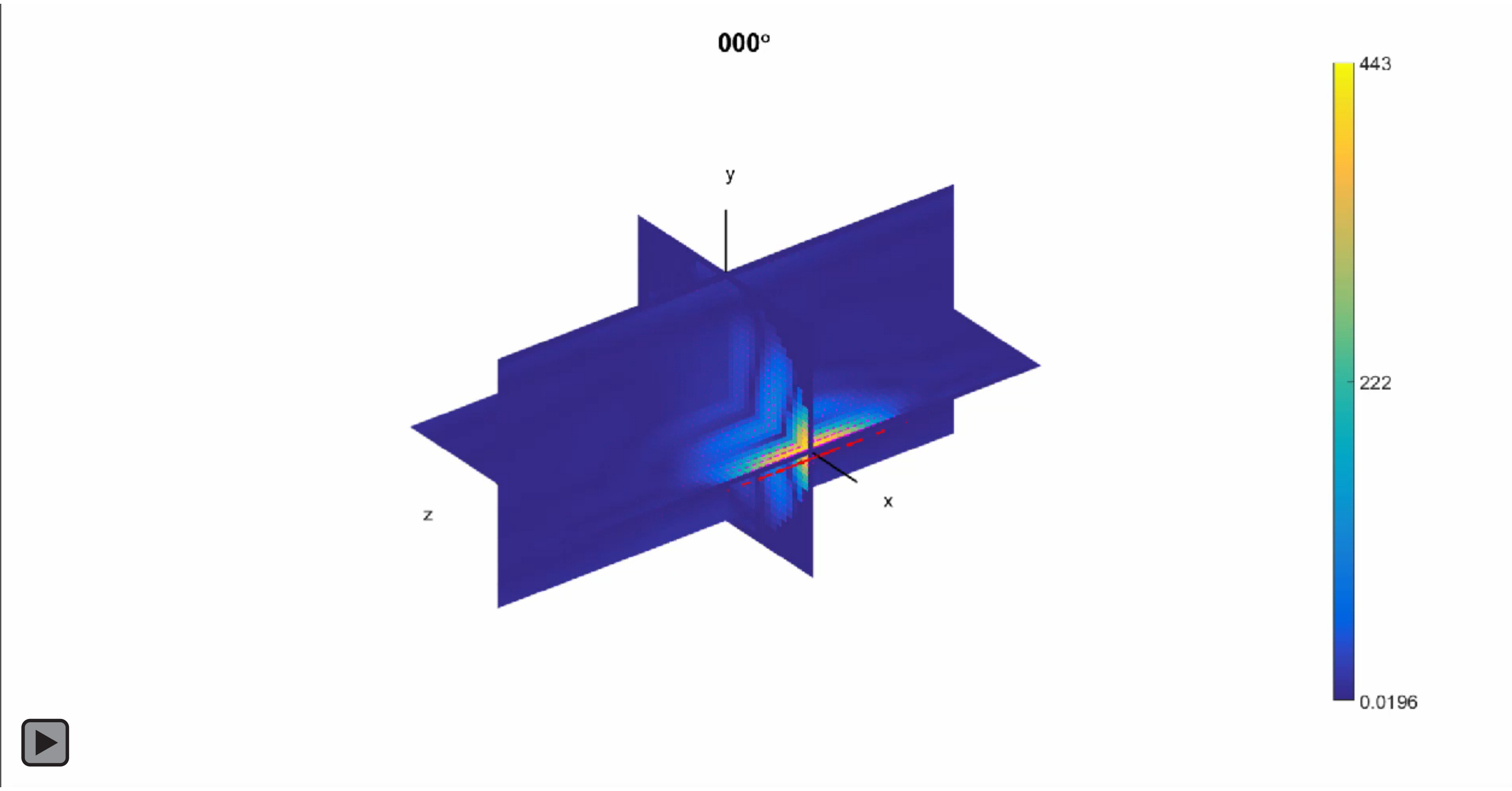

**Supplementary Movie S14:** Animation of the electric field pattern for the same single filamentary z-directed electric dipole as in Mov. S13, but at a frequency corresponding to the 7.0T Larmor frequency. The propagation behavior associated with this efficient radiator at high frequency is evident. Note that the direction of propagation is from the center outward to the current dipole rather than from the dipole inward, since it is the electric field associated with signal reception that is shown here, as opposed to the field associated with transmission by the current dipole.

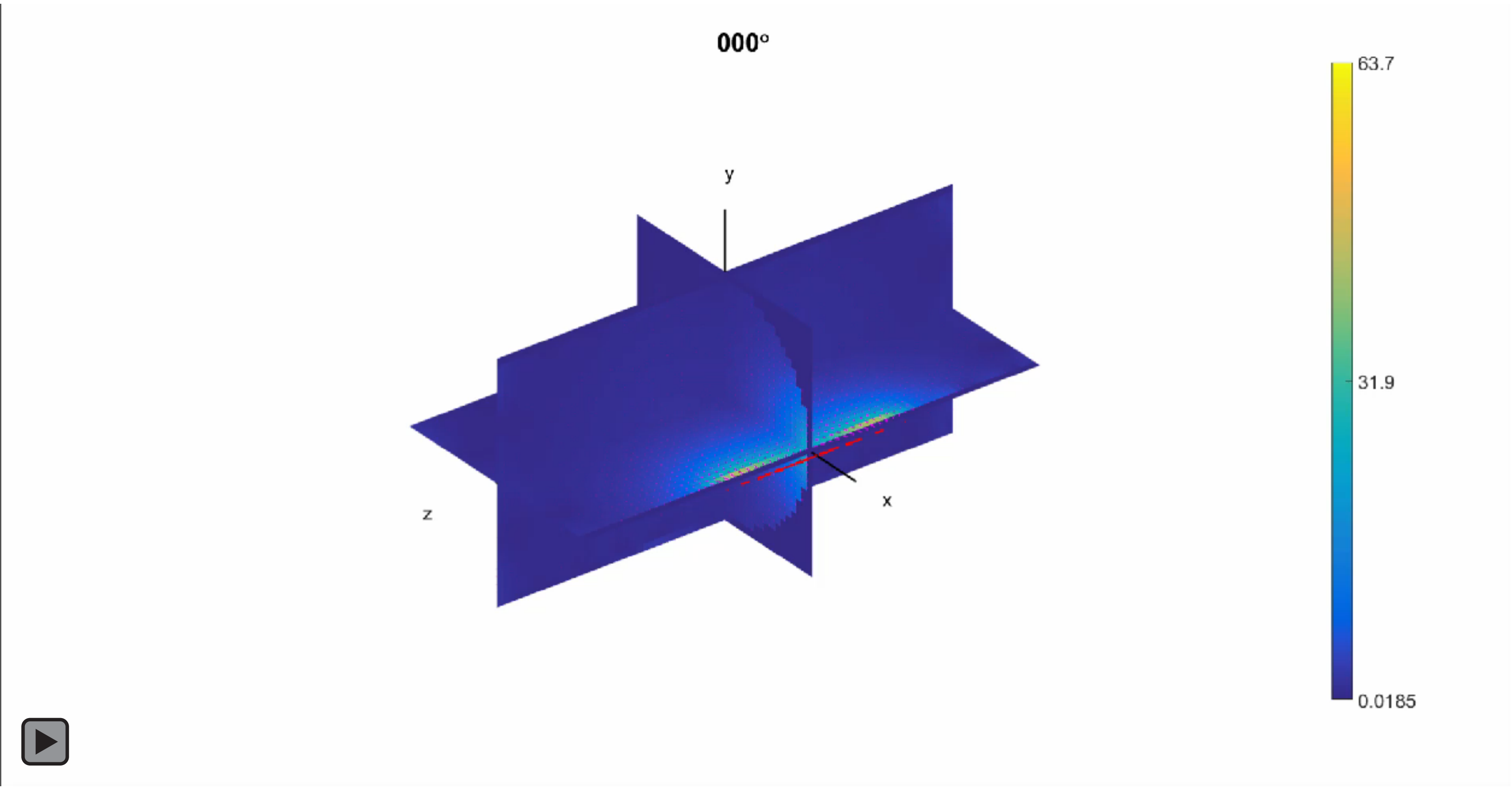

**Supplementary Movie S15:** Animation of the conservative component of the electric field pattern from Supplementary Movie S14, for the same single filamentary z-directed electric dipole at 7.0T. The conservative electric field component at high frequency strongly resembles the full electric field pattern at low frequency, but at a small fraction of the amplitude of the full propagating E field (i.e., with a maximum amplitude of nearly 64, as opposed to the maximum amplitude of 443 in Mov. S14).

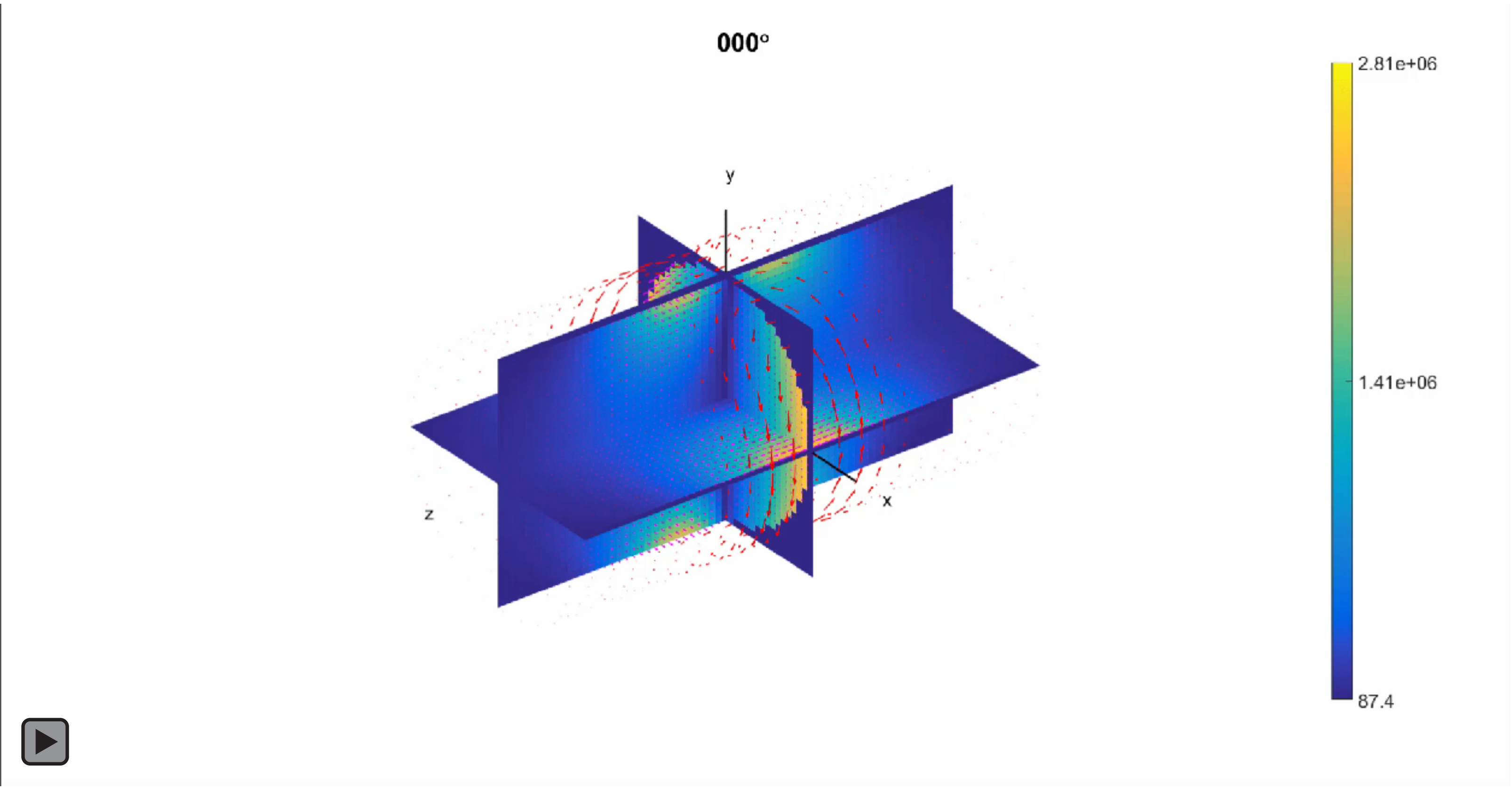

**Supplementary Movie S16:** Animation of the electric field pattern (associated with the snapshot on the left-hand side of Fig. 9a) corresponding to the full ideal current pattern optimizing SNR, including body noise, for a central point at 0.1T. The surrounding ideal current pattern (shown in red, associated with the unwrapped pattern on the left-hand side of Fig. 8a) is closed.

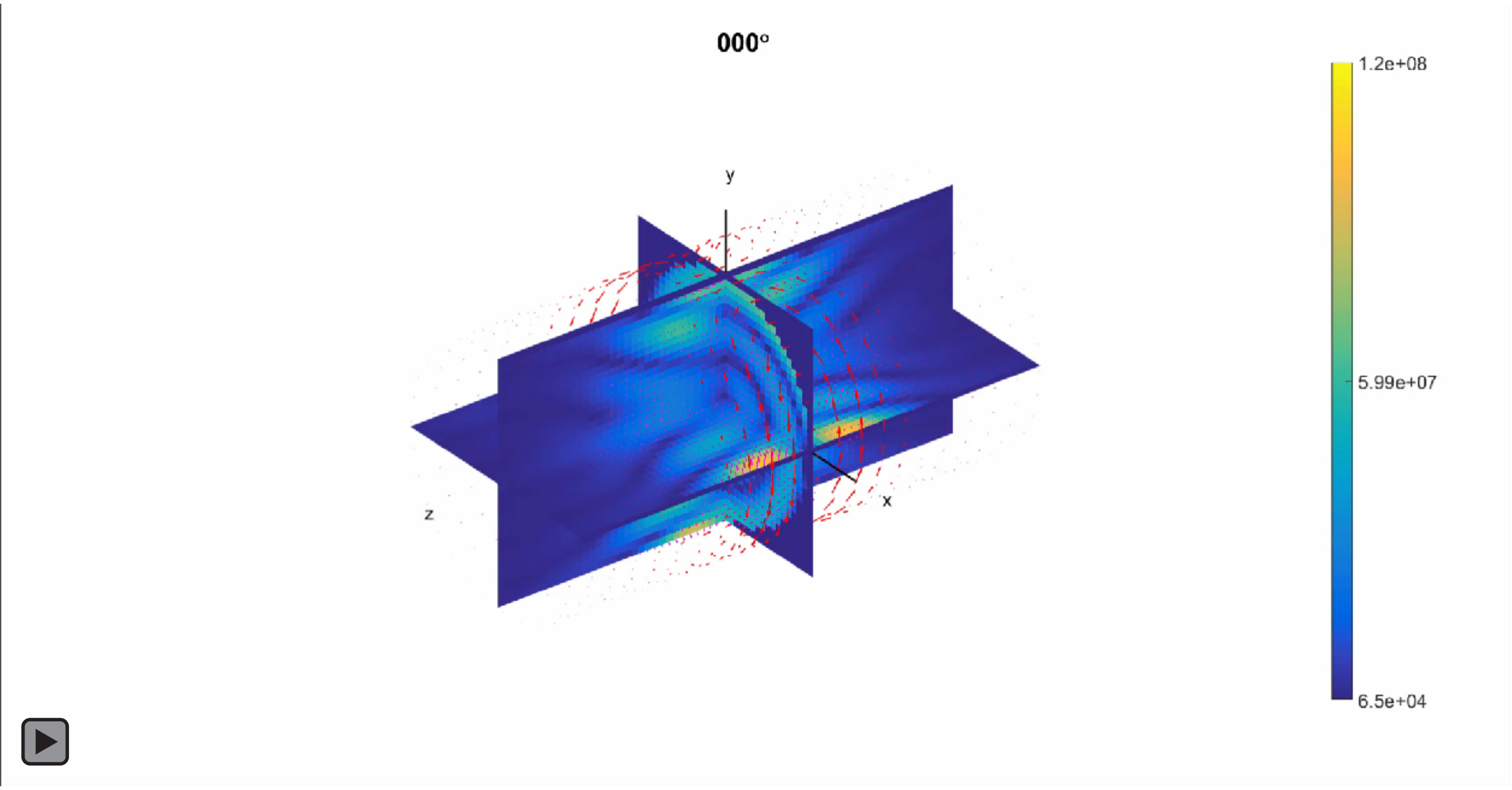

**Supplementary Movie S17:** Animation of the electric field pattern corresponding to the same closed current pattern from Supplementary Movie S16 that is optimal at 0.1T, but imposed artificially at 7.0T. Note the substantial propagating field components that extend along ±z. The benefits of closing the current pattern are lost at high frequency, and the ideal pattern more closely resembles the unperturbed OP pattern (see Fig 8d).

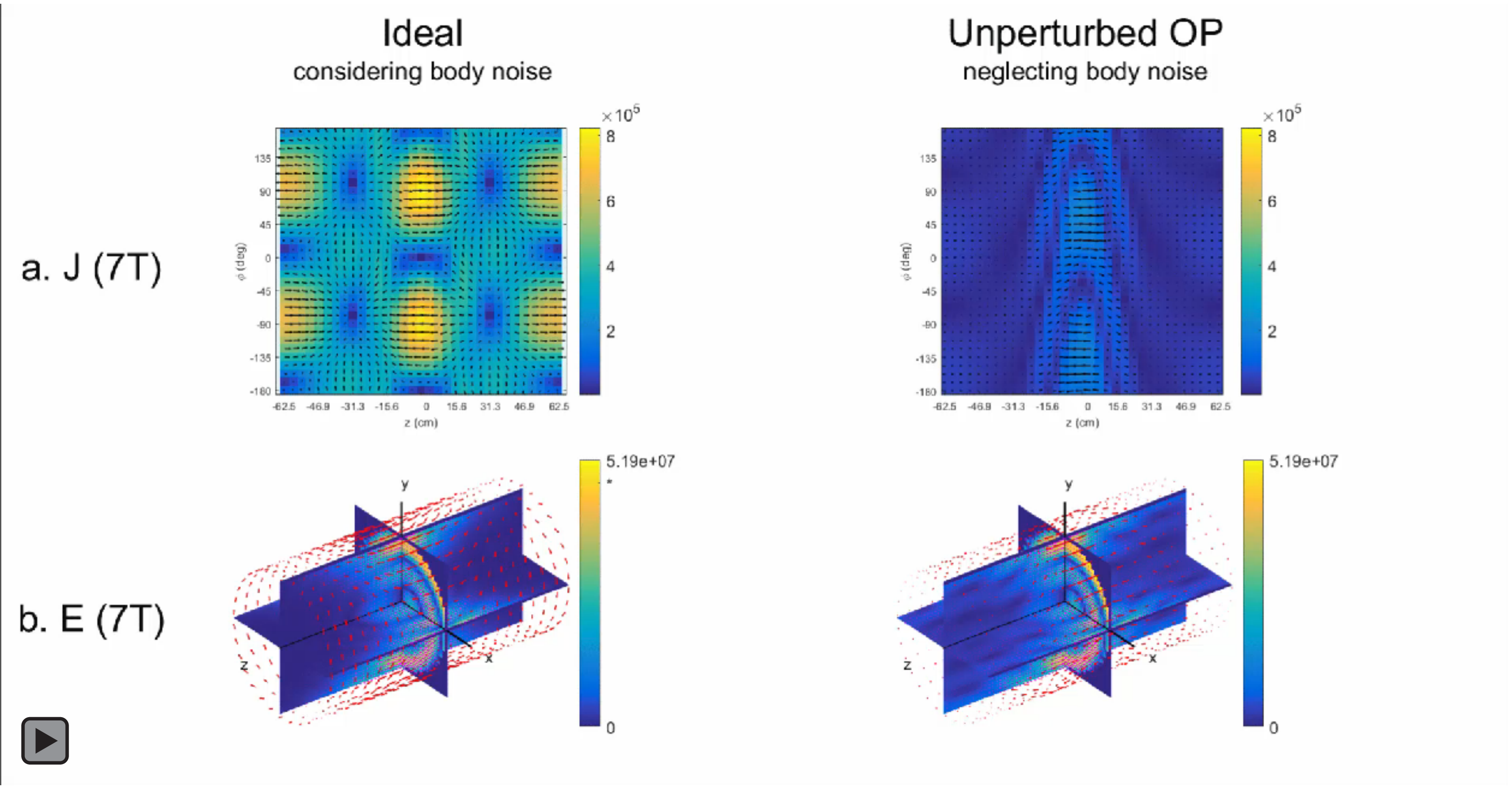

**Supplementary Movie S18:** Animations of surface current patterns and internal electric field patterns for a configuration at 7T in which traveling wave modes are supported by a cylindrical RF shield with radius 37.5cm surrounding a 20cm dielectric cylinder (the same cylinder that was used in previous simulations without an encircling shield). a: Unwrapped ideal (left) and unperturbed OP (right) current patterns. b: ideal (left) and unperturbed OP (right) electric field patterns, with surrounding surface current patterns shown in red. As compared with the corresponding patterns without an encircling shield at 7T, e.g. in Fig. 8d and Mov. S8, the unperturbed OP current patterns here maintain a similar chevron pattern but are slightly more distributed – i.e., they have lower maximum amplitude but extend farther along the cylinder axis, reflecting outward axial propagation of the tangential electric field created by a precessing spin at the center. The full ideal current patterns in the presence of the shield, on the other hand, are markedly different from the corresponding shield-free ideal current patterns. The chevron pattern associated with radial propagation delay is lost, and paired current densities out of phase with the central current density appear symmetrically along ±z. As may be seen in the electric field patterns (b), the net result is a marked reduction in axially-propagating electric fields in the body. In other words, rather than generating traveling wave modes, which are SNR inefficient for a single central point of interest, the ideal current pattern actually arranges for cancellation of these modes.

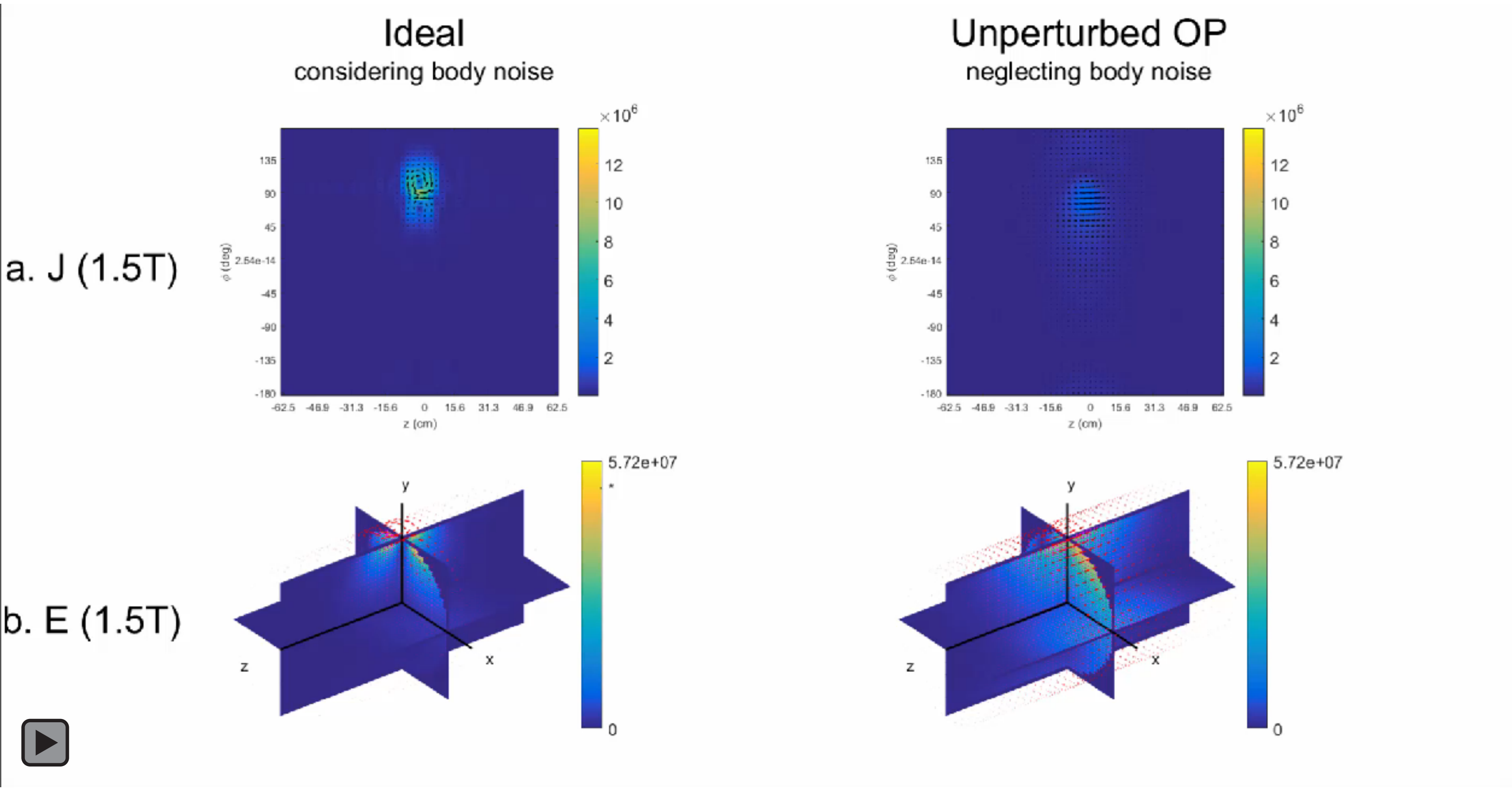

**Supplementary Movie S19:** Animated version of Fig. 11, showing ideal (left) and unperturbed OP (right) currents (a) and electric fields (b) looping over a full cycle of phases.

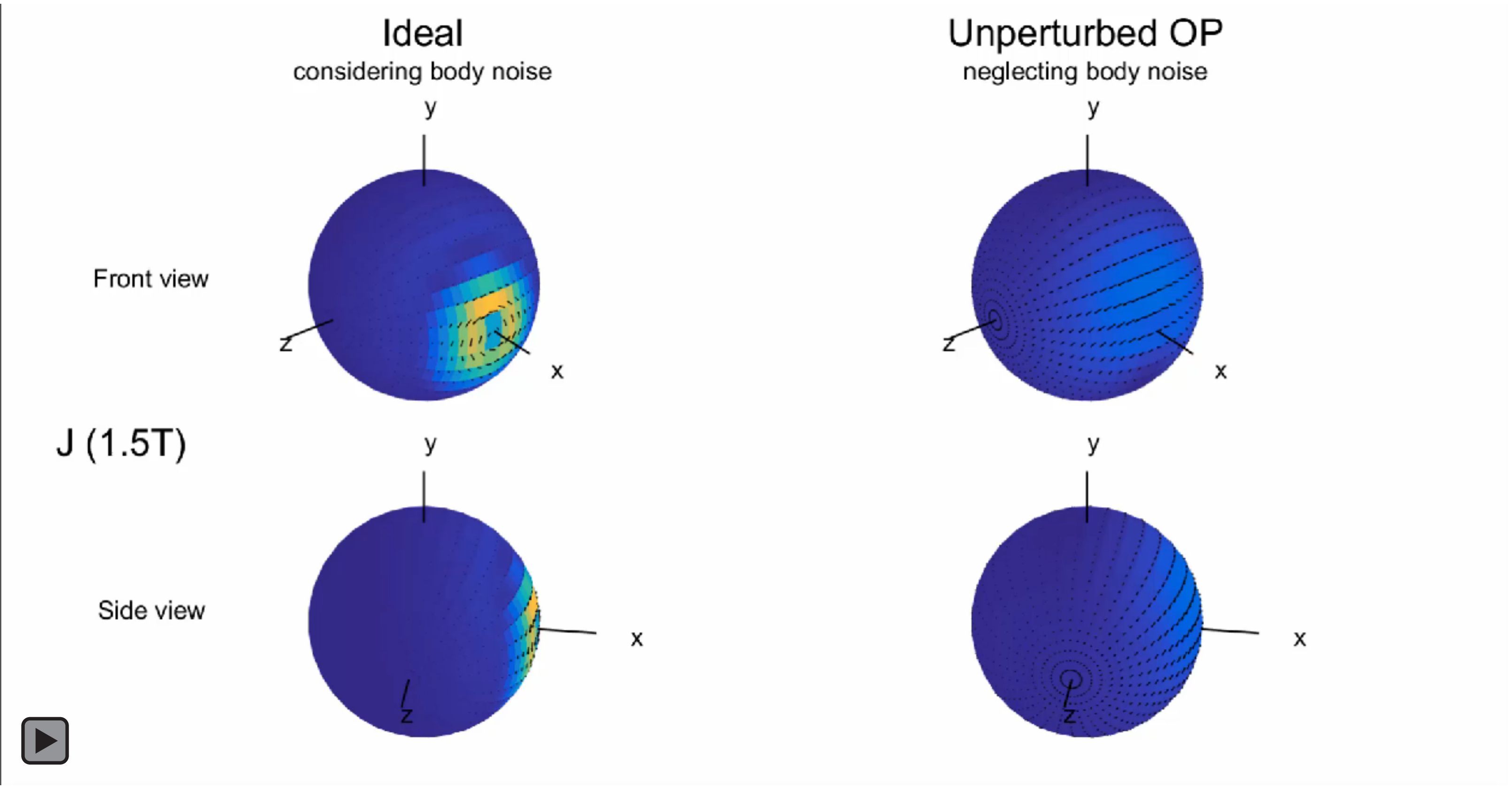

**Supplementary Movie S20:** Animated current patterns for a voxel intermediate between the center and the edge of a 20cm-radius sphere surrounded by a concentric 21cm-radius spherical surface at 1.5T. The voxel of interest was located at (x,y,z) = (10,0,0) cm. The sphere had a conductivity of $\sigma$ = 0.4 S/m, and a relative permittivity $\varepsilon_{rel}$ = 39. Left: ideal current patterns considering body noise. Right: unperturbed OP patterns neglecting body noise. Top: Front view. Bottom: Side view. The patterns display characteristics similar to the case of an intermediate voxel in a cylinder (Fig. 11 and Mov. S19). Ideal patterns are closed and tightly concentrated near the voxel of interest, whereas unperturbed OP patterns are more broadly distributed, and are non-closed in some phases, as a result of the differential distance of the voxel of interest from different regions on the surface.

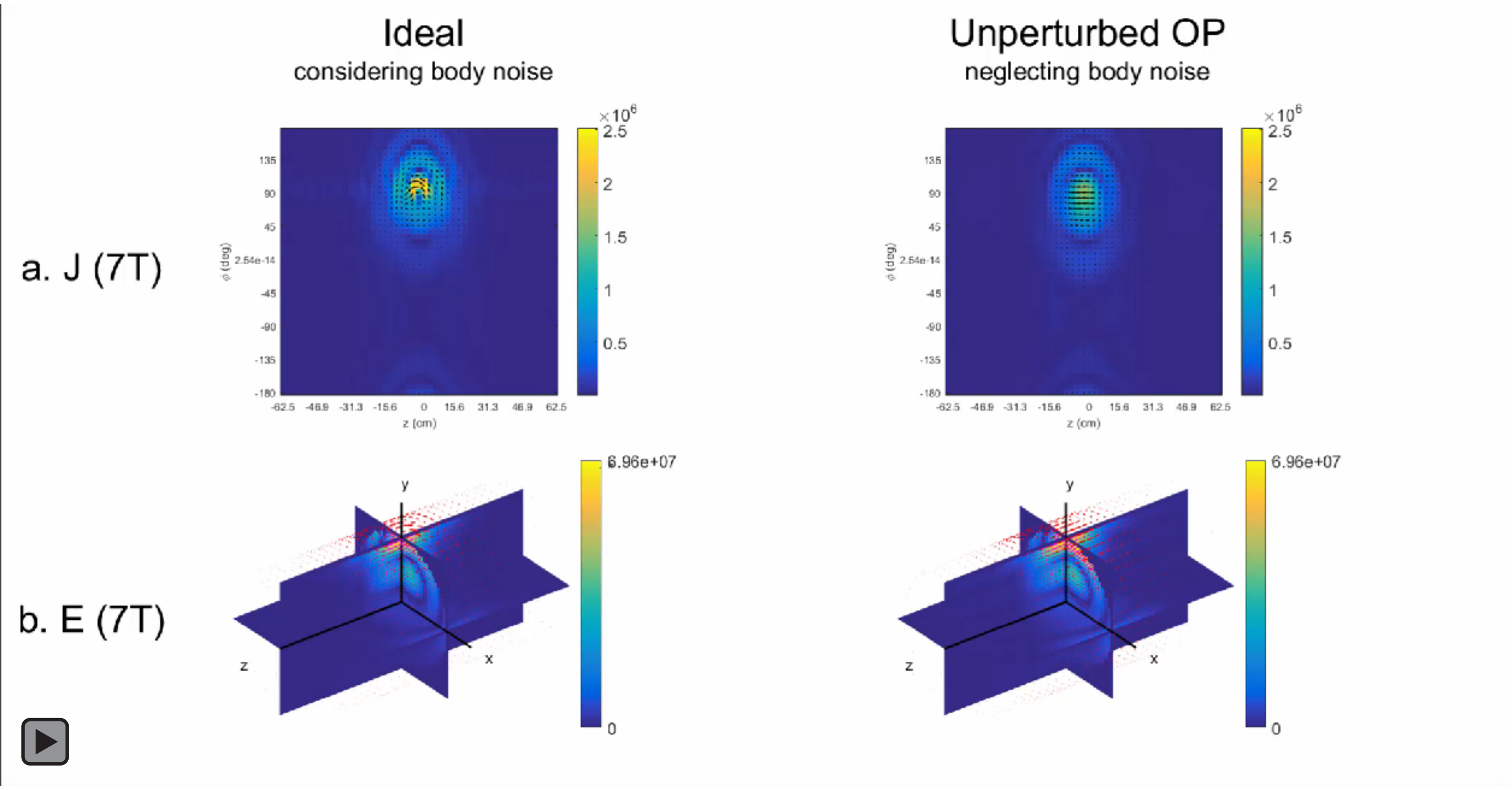

**Supplementary Movie S21:** The same animations as in Mov. S19, but for 7T rather than 1.5T field strength. Note that, in this case, ideal patterns more closely resembles OP patterns, and electric field patterns contain a significant propagating component.

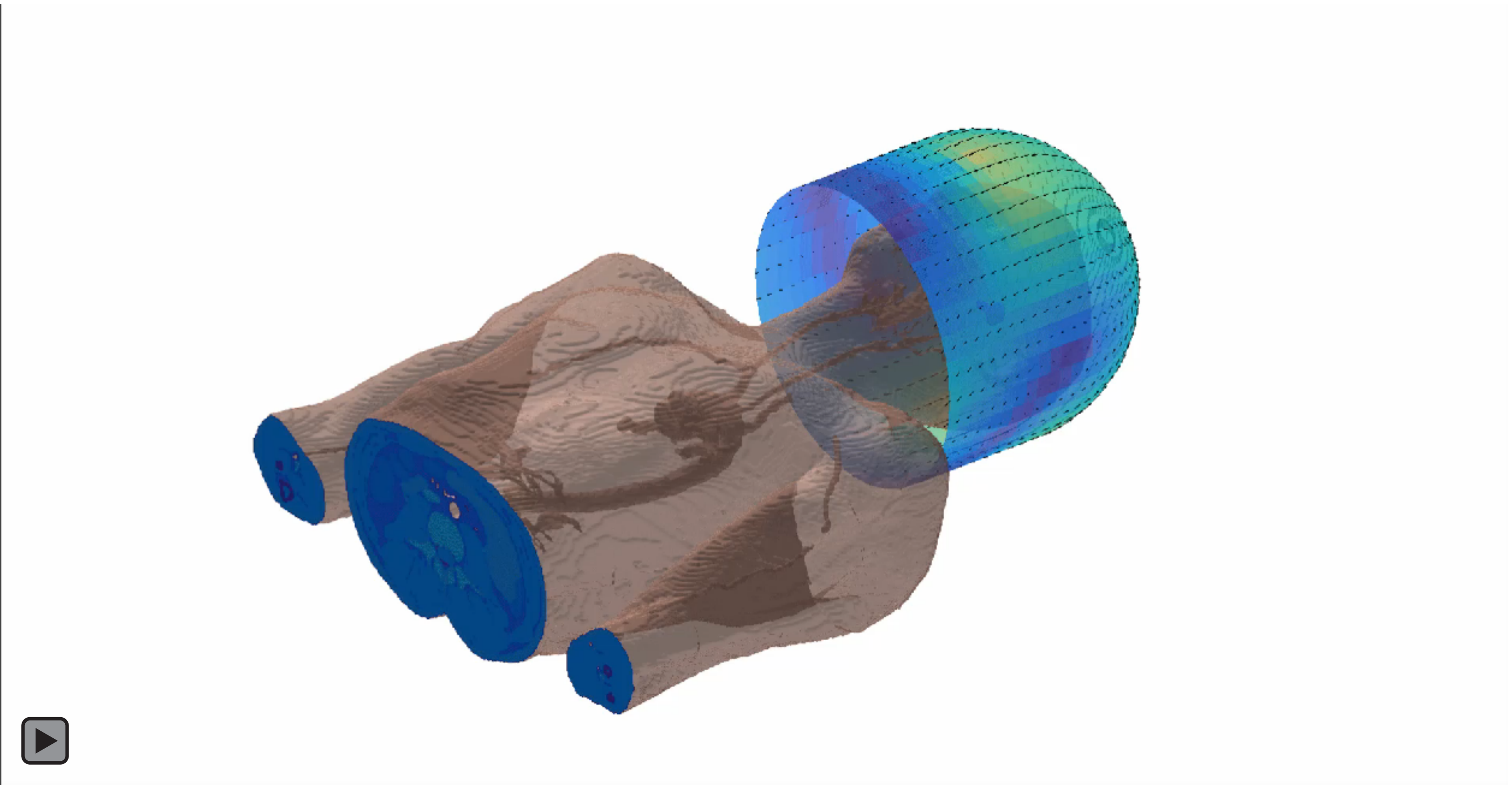

**Supplementary Movie S22:** Animation of the current pattern in Fig. 14. Three cycles are shown from a feet-first viewpoint matching that at the top of Fig. 14, followed by three cycles from a head-first viewpoint matching that at the bottom of Fig. 14.